\documentclass[prd,twocolumn,nofootinbib,letter,superscriptaddress]{revtex4-1}

\usepackage{graphicx}
\usepackage{amssymb}
\usepackage{textcomp}
\usepackage{amsmath}
\usepackage{bm}
\usepackage{times}
\usepackage{epsfig}
\usepackage{color}
\usepackage{graphics}
\usepackage{hyperref}
\usepackage{setspace,ulem}

\allowdisplaybreaks

\hypersetup{
    pdfnewwindow=true,      
    colorlinks=true,       
    linkcolor=black,          
    citecolor=blue,        
    filecolor=blue,      
    urlcolor=blue           
}

\def\ba{\begin{array}}
\def\ea{\end{array}}
\def\non{\nonumber\\}

\newcommand{\GeV}      {~\mathrm{GeV}}
\def\hf{\frac{1}{2}}

\def\mI{\mathcal{I}}

\def\mL{\mathcal{L}}
\def\mM{\mathcal{M}}

\def\mO{\mathcal{O}}

\def\mR{\mathcal{R}}

\def\bea{\begin{eqnarray}}
\def\eea{\end{eqnarray}}

\begin{document}

\begin{flushright}
ULB-TH/17-14
\end{flushright}

\title{CP violation effects in the diphoton spectrum of heavy scalars}

\author{Ligong Bian}
\email{lgbycl@cqu.edu.cn}
\affiliation{Department of Physics, Chongqing University, Chongqing 401331, China}
\affiliation{Department of Physics, Chung-Ang University, Seoul 06974, Korea}

\author{Ning Chen}
\email{ustc0204.chenning@gmail.com}
\affiliation{Department of Physics, University of Science and Technology Beijing, Beijing 100083, China}
\affiliation{CAS Center for Excellence in Particle Physics, Beijing 100049, China}

\author{Yongchao Zhang}
\email{yongchao.zhang@physics.wustl.edu}
\affiliation{Service de Physique Th\'{e}orique, Universit\'{e} Libre de Bruxelles, Boulevard du Triomphe, CP225, 1050 Brussels, Belgium}
\affiliation{Department of Physics and McDonnell Center for the Space Sciences, \\  Washington University, St. Louis, MO 63130, USA}

\date{\today}

\begin{abstract}
  In a class of new physics models, an extended Higgs sector and new CP-violating sources are simultaneously present in order to explain the baryon asymmetry in the Universe. The aim of this work is to study the implications of beyond the Standard Model (SM) CP violation for the searches of heavy scalars at the LHC. In particular, we focus on the diphoton channel searches in the CP-violating two-Higgs-doublet model (CPV 2HDM). To have a sizable CPV in the scalar sector, the two heavy neutral scalars in 2HDM tend to be nearly degenerate. The theoretical constraints of unitarity, perturbativity and vacuum stability are considered, which requires that the heavy scalars $M_H \lesssim 1$ TeV in a large region of the parameter space. The experimental limits are also taken into account, including the direct searches of heavy neutral scalars in the final state of the SM $h$, $W$ and $Z$ bosons, the differential $t\bar{t}$ data, those from the charged scalar sector which is implied by the oblique $T$ parameter, as well as the precise measurements of the electric dipole moments of electron and mercury. The quantum interference effects between the resonances and the SM background are crucially important for the diphoton signals, and the CPV mixing of the quasi-degenerate heavy scalars could enhance significantly the resonance peak. With an integrated luminosity of 3000 fb$^{-1}$ at the LHC, almost the whole parameter space of CPV 2HDM could be probed in the diphoton channel, and the CPV could also be directly detected via the diphoton spectrum.
\end{abstract}
\maketitle

\tableofcontents

\section{Introduction}
\label{section:intro}

The discovery of the 125 GeV  Standard Model (SM)  Higgs boson opens a new era in particle physics. But the hierarchy problem, neutrino masses, dark matter and the origin of matter-antimatter asymmetry still need to be addressed in new physics models, such as supersymmetry~\cite{Flores:1982pr,Gunion:1984yn,Djouadi:2005gj}, composite Higgs models~\cite{Gripaios:2009pe}, and two-Higgs-doublet models (2HDM)~\cite{Branco:2011iw},  as well as variants of these models, etc.
To explain the baryon asymmetry of the universe (BAU), three Sakharov conditions needs to be accomplished in the new physics models beyond the SM~\cite{Sakharov:1967dj}. One of the most attractive mechanisms to explain BAU is the electroweak baryogenesis (EWBG), we refer to Ref.~\cite{Morrissey:2012db} for a recent review. To realize the mechanism, the SM Higgs sector need to be extended in order to obtain a strong first order electroweak (EW) phase transition, which might induce deviation of triple scalar coupling from the SM prediction to be detected at high energy hadron colliders~\cite{Arkani-Hamed:2015vfh}. The CP violation (CPV) beyond the SM is one of the three ingredients of Sakharov conditions. The 2HDM with complex parameters in the scalar potential offers a most economical possibility to introduce new CPV sources in the extended Higgs sector~\cite{Lee:1973iz}, and provides one economical renormalizable framework to address the BAU using EWBG~\cite{Dorsch:2016nrg, Dorsch:2014qja,Dorsch:2013wja,Shu:2013uua,Bian:2014zka}.

To probe the CPV effects with extended Higgs sector, the previous literatures focus on two distinct categories of methods. (i) One may rely on direct measurement of the CPV couplings of the $125\,\GeV$ SM-like Higgs boson at colliders. A lot of efforts have been made to study the physics opportunities in measuring the parity of the SM-like Higgs boson at the LHC~\cite{Harnik:2013aja, Berge:2013jra, Sun:2013yra, Anderson:2013afp, Chen:2013waa, Askew:2015mda, Belyaev:2015xwa, Buckley:2015vsa, Berge:2015nua, Rindani:2016scj} and the future colliders~\cite{Li:2015kxc, Hagiwara:2016rdv, Chen:2017bff, Chen:2015fca, Gori:2016zto}.
(ii) One can look for the indirect CPV effects in particular processes, e.g. the precise measurements of electric dipole moments (EDMs).
There have been incredible progresses in improving the upper bounds on the EDMs of electrons~\cite{Baron:2013eja} and mercury~\cite{Graner:2016ses}.
CPV beyond the SM is severely constrained by the increasing precision of the EDM measurements, see e.g.~\cite{Inoue:2014nva}, except for the case where the SM-like Higgs boson has a sizable CPV mixing of $\mO(0.1)$ and a sizable cancellation exists in evaluating the Barr-Zee diagrams for electron EDM~\cite{Shu:2013uua, Bian:2014zka}.
The recent electron and mercury EDM measurements could provide considerable constraints on the parameter space of CPV 2HDM, as will be explored in this paper, which is largely complementary to the direct searches of CPV in the scalar sector of 2HDM at the LHC in the diphoton decay mode.

At hadron colliders, the dominant production and decay mode of the CP even/odd heavy scalar is the $gg\rightarrow H/A\rightarrow t\bar{t}$ process, that makes the $t\bar{t}$ final states to be an important channel to probe heavy scalars, which, however, suffers from large systematic uncertainties and smearing effects~\cite{Craig:2015jba}. As a result of the clean and well understood background, the diphoton decay mode is very likely one of the main channels to search for beyond SM heavy scalars or even probe directly new source of CPV in the framework of 2HDM, as for the SM Higgs, though the branching ratio (BR) to diphoton is usually very small~\cite{Strumia:2016wys}.


In general, for resonant particles at high energy colliders, e.g. the heavy neutral scalar in 2HDM, 
the interference of resonances with the continuum SM background may have non-negligible impacts on the shape and size of resonant signals~\cite{Dicus:1987fk, Dixon:2003yb, Martin:2012xc, Martin:2013ula, Jung:2015sna, Jung:2015gta, Carena:2016npr}, which could help us to probe the spin and production mode of the resonance particles.
In the framework of 2HDM, due to the relative phase between the resonance and continuum background, various resonance shape can be obtained in the channels of $gg \rightarrow H_i \rightarrow t\bar{t}$, $\gamma\gamma$ ($H_i$ being the heavy scalars in 2HDM)~\cite{Martin:2012xc, Gaemers:1984sj, Dicus:1994bm, Dicus:1987fk, Dixon:2003yb, Djouadi:2016ack, Carena:2016npr}. The relative phase can be generated by either the (fermion) loop diagrams or CP-violating interactions.





In the diphoton decay mode of heavy scalars in CPV 2HDM $gg \to H_i \to \gamma\gamma$, a large interference effect can be expected, since the continuum background $gg \to \gamma\gamma$ is one-loop process while the resonance is via two-loop.\footnote{Generally, one can expect relatively small interference in ZZ channel since both the $gg\rightarrow ZZ$ continuum background and the resonance signal processes $gg\rightarrow H_i \rightarrow ZZ$ are both dominated by one-loop diagrams~\cite{Jung:2015sna}.}
In particular, for heavy scalar masses above the mass threshold of $2m_{t}$, the top loop in the production process of $gg\rightarrow H_i$ and the decay process $H_i \to \gamma\gamma$ can induce imaginary parts in the amplitude $gg \rightarrow H_i \rightarrow \gamma\gamma$ and then change the magnitude of the cross section,\footnote{It should be noted that for a relatively large $\tan\beta$, the bottom quark loop can also induce sizable imaginary parts~\cite{Jung:2015sna,Djouadi:2016ack}. However, this possibility is not favored by the CPV 2HDM scenario explored in this paper.} with the real parts serve to shift the resonance shape~\cite{Djouadi:2016ack}. When the two heavy scalars are nearly degenerate with CP-violating mixing~\cite{Bian:2016zba, Carena:2015uoe, Carena:2016npr},
there will be additional interference effects in the heavy scalar sector, which tends to amplify the nearly-degenerate scalar resonance (as shown in Fig.~\ref{fig:example_CP}). The amplification could in principle be directly probed at the high energy colliders, which is largely complementary to the indirect constraints from the measurements of EDMs.
One should note that, the CPV in the scalar sector of 2HDM depend non-trivially on the heavy scalar masses, which is significant only when two heavy scalars are quasi-degenerate~\cite{Bian:2016zba}. That is the reason why we focus in this paper only on the 2HDM scenarios with a small mass splitting for the two heavy scalars.


The layout of this paper is described as follows. In Section~\ref{section:CPV2HDM}, we review the setup of CPV 2HDM, with emphasis on the neutral scalars and their CPV mixings. For our discussions, we set up the formalism for processes involving two quasi-degenerate heavy scalars. Their propagators are written in the form a $2\times2$ matrix, with the off-diagonal elements might be non-negligible for certain parameter sets. All the theoretical and experimental constraints on the CPV 2HDM are collected in Section~\ref{sec:limitsall}, including the requirements of unitarity, perturbativity and vacuum stability of the scalar potential, the direct searches of heavy scalars decaying into $WW/ZZ$, $hZ$ and $hZ$, and the consistency of differential $t\bar{t}$ data with the SM predictions. The neutral-charged scalar mass splitting $|M_H - M_\pm|$ is tightly constrained by the oblique $T$ parameter which could ``transfer'' the charged scalar limits onto the neutral scalars. All these limits are exemplified in Figs.~\ref{fig:limits1} to \ref{fig:limits5} with the small mass splitting of heavy neutral scalars set to be 1 GeV or 10 GeV. The electron and mercury EDM limits are also considered, which exclude large region in the parameter space of 2HDM, as expected. The diphoton signal in the CPV 2HDM is discussed in Section~\ref{sec:diphoton}, where we show the full parton-level (differential) cross sections including both the resonance and interference contributions. By scanning the parameter space, it turns out that the diphoton signals could probe almost all the regions we considered at the high-luminosity LHC (HL-LHC) 14 TeV run with an integrated luminosity of 3000 fb$^{-1}$, at least at the 95\%, that are allowed by the limit above. The relation between the diphoton line shapes and CPV mixing in the scalar sector is also discussed. It turns out that the CPV can be directly probed in the diphoton events, if the scalar masses are $\lesssim 600$ GeV, when $\tan\beta = 0.5$. The conclusions are given in Section~\ref{section:conclusion}. The formulas for differential $H_i\to t\bar{t}$ cross sections in the CPV 2HDM, the oblique parameters and the details of evaluating the EDMs are respectively summarized in the appendices. 



\section{Quasi-degenerate heavy neutral scalars in CPV 2HDM}
\label{section:CPV2HDM}

\subsection{The CPV 2HDM}

There are two scalar doublets $\Phi_{1,2}$ in the general 2HDM. For simplicity, we introduce a discrete $Z_2$ symmetry to avoid tree-level flavor-changing neutral currents, under which two scalar doublets transform as $(\Phi_1\,,\Phi_2)\to (-\Phi_1\,, \Phi_2)$. When the $Z_2$ symmetry soft-breaking term is introduced, one can obtain CPV in the Higgs sectors~\cite{Inoue:2014nva,Accomando:2006ga}. The general scalar potential can be written as
\begin{eqnarray}
\label{eqn:potential}
&& V(\Phi_1\,,\Phi_2) =
m_{11}^2|\Phi_1|^2+m_{22}^2|\Phi_2|^2-(m_{12}^2 \Phi_1^\dag\Phi_2+ {\rm H.c.} ) \nonumber \\
&& \quad +\hf\lambda_1 |\Phi_{1}|^{4} +\hf\lambda_2|\Phi_{2}|^{4}
+ \lambda_3|\Phi_1|^2 |\Phi_2|^2+\lambda_4 |\Phi_1^\dag \Phi_2|^2 \nonumber \\
&& \quad + \hf \Big[ \lambda_5 (\Phi_1^\dag\Phi_2)^2 + {\rm H.c.} \Big] \,,
\end{eqnarray}
where $\lambda_5$ and the $Z_2$ soft-breaking term $m_{12}^2$ are complex and all other mass and quartic parameters are real.
The imaginary components of $m_{12}^2$ and $\lambda_5$ are the source of CP violation, which leads to mixings of all the three neutral states, as shown bellow.
For the ease of discussions below, we define the soft mass parameter as the real part of $m_{12}^2$, i.e.
\begin{eqnarray}
\label{eqn:msoft}
m_{\rm soft}^2\equiv {\rm Re} \, m_{12}^2 \,,
\end{eqnarray}
which is assumed to be positive. After the EW symmetry breaking, the two scalar doublets obtain non-vanishing vacuum expectation values (VEVs) 
\begin{eqnarray}
\label{eqn:doublets}
&&\Phi_1=\left(
\ba{c}  -s_\beta\, H^+ \\
 \frac{1}{\sqrt{2}} ( v_1 + H_1^0 - i s_\beta A^0)
 \ea  \right)\,,\nonumber \\
&& \Phi_2= \left(
\ba{c}  c_\beta\, H^+ \\
 \frac{1}{\sqrt{2}} ( v_2  + H_2^0 + i c_\beta A^0 )   \,,
 \ea  \right) \,,
\end{eqnarray}
where we have neglected the relative phase between the two VEVs, $v^2 = v_1^2+v_2^2 = (\sqrt{2}\, G_F)^{-1}$ with $G_F$ the Fermi constant, $s_\beta \equiv \sin\beta$ and $c_\beta \equiv \cos\beta$ with the angle defined as the VEV ratio $t_\beta \equiv \tan\beta = v_2/v_1$.

It is straightforward to obtain the mass square matrix for the three neutral scalars:
\begin{eqnarray}
\label{eqn:matrix}
\mM_0^2&=& \left(
\begin{array}{ccc}
  \lambda_1 c_\beta^2 + \nu s_\beta^2 &
 (\lambda_{345} -\nu ) s_\beta c_\beta  &
  -\frac12 {\rm Im} \lambda_5 \, s_\beta   \\
  ( \lambda_{345} -\nu ) s_\beta c_\beta  &
  \lambda_2 s_\beta^2 + \nu c_\beta^2   &
  -\frac12 {\rm Im} \lambda_5 \, c_\beta    \\
  - \frac12 {\rm Im}\lambda_5  \, s_\beta &
  -\frac12 {\rm Im} \lambda_5 \, c_\beta  &
  - {\rm Re} \lambda_5 + \nu  \\
\end{array}  \right) v^2\,, \nonumber \\
\end{eqnarray}
with the short-handed notation of
\begin{eqnarray}
\label{eqn:lambda345}
\lambda_{345} &\equiv& \lambda_3+\lambda_4+ {\rm Re} \lambda_5 \,, \nonumber \\
\nu &\equiv&  \frac{m_{\rm soft}^2}{v^2 s_\beta c_\beta} \,.
\end{eqnarray}
In the limits of CP conservation ${\rm Im} \, m_{12}^2 =0$ and ${\rm Im} \, \lambda_5 = 0$, the first two scalars $H_{1,2}^0$ are CP-even while the third one $A^0$ is CP-odd. The nonzero imaginary components of $m_{12}^2$ and $\lambda_5$ lead to mixings of all the three neutral states via the rotation matrix ${\cal R}$:
\begin{eqnarray}
\mM_0^2&=& \mR^T \, {\rm diag}(M_1^2\,, M_2^2\,, M_3^2 )\, \mR \,,
\end{eqnarray}
where $M_i$ the mass eigenvalues for the three physical scalars $H_i$. For concreteness we assume the first scalar is SM-like ($H_1 = h$) with a mass of $M_1 = m_h = 125$ GeV, while the other two scalars $H_{2,3}$ are heavier with (nearly-degenerate) masses $M_{2,3}$. 
The $3\times 3$ mixing matrix can be parameterized explicitly as~\cite{Khater:2003wq}
\begin{eqnarray}
\label{eqn:matrix2}
\mR&=&\mR_{23}(\alpha_c) \mR_{13}(\alpha_b) \mR_{12}(\alpha+\frac{\pi}{2})\non
&=&\left( \ba{ccc}
-s_\alpha c_{\alpha_b} & c_\alpha c_{\alpha_b} & s_{\alpha_b} \\
  s_\alpha s_{\alpha_b}s_{\alpha_c} - c_\alpha c_{\alpha_c} & -s_\alpha c_{\alpha_c} - c_\alpha s_{\alpha_b} s_{\alpha_c} & c_{\alpha_b}s_{\alpha_c}  \\
  s_\alpha s_{\alpha_b} c_{\alpha_c} + c_\alpha s_{\alpha_c}  &  s_\alpha s_{\alpha_c} - c_\alpha s_{\alpha_b} c_{\alpha_c}  &  c_{\alpha_b} c_{\alpha_c}  \\
  \ea  \right)\,, \nonumber \\
\end{eqnarray}
where the angle $\alpha$ parameterizes the mixing between the two CP-even states, and the other two $\alpha_{b,c}$ determining the CP violating mixing of the scalars. 

The physical neutral scalar masses $M_{1,2,3}$ and the charged scalar mass $M_{\pm}$ in the scalar spectrum are directly related to the mass parameters and quartic couplings in the potential (\ref{eqn:potential}).
In practice, one often trade the quartic scalar self-couplings into the physical inputs, i.e. the EW VEV $v$, the scalar masses and the mixing angles:
\begin{eqnarray}
\label{eqn:lambda1}
\lambda_1 &=& - \nu\, t_\beta^2 +
\frac{1}{v^2 c_\beta^2} \sum_i M_i^2 {\cal R}_{i1}^2  \ , \\
\label{eqn:lambda2}
\lambda_2 &=& - \frac{\nu}{t_\beta^2} +
\frac{1}{v^2 s_\beta^2} \sum_i M_i^2 {\cal R}_{i2}^2 \ , \\
\label{eqn:lambda3}
\lambda_3&=&- \nu +
\frac{1}{v^2 s_\beta c_\beta } \sum_i M_i^2 {\cal R}_{i1} R_{i2} +
\frac{2 M_\pm^2 }{v^2}  \,,\\
\label{eqn:lambda4}
\lambda_4 &=& 2\nu -{\rm Re} \lambda_5 - \frac{2 M_\pm^2 }{v^2}   \,, \\
\label{eqn:lambda5Re}
{\rm Re} \lambda_5 &=& \nu -
\frac{1}{v^2} \sum_i M_i^2 {\cal R}_{i1}^2 \ , \\
\label{eqn:lambda5Im}
{\rm Im} \lambda_5 &=& -\frac{1}{v^2 s_\beta c_\beta} \Big[
c_\beta \sum_i M_i^2 {\cal R}_{i1} {\cal R}_{i3} +
s_\beta \sum_i M_i^2 {\cal R}_{i2} {\cal R}_{i3} \Big] \,, \nonumber \\
\end{eqnarray}
where all the $i$ run from 1 to 3. These relations are very useful to apply the perturbative, unitarity and stability constraints on the quartic couplings, which would imply also limits on the physical parameters such as the mass ranges of heavy neutral scalar in the quasi-degenerate case (cf. the following discussions in Section~\ref{sec:theoretical} and Figs.~\ref{fig:sensitivity1} to \ref{fig:sensitivity7}).


\subsection{Couplings in the CPV 2HDM}

We collect here all the couplings of heavy scalars in the CPV 2HDM, which are important for examining the theoretical and experimental constraints as well as the diphoton prospects at the LHC.

With the discrete ${Z}_2$ symmetry, the scalar doublets $\Phi_{1,2}$ couple only to the up-type quarks or the down-type quarks and charged leptons, which is sufficient to suppress the dangerous tree-level flavor changing neutral couplings of the scalars. The type-I and type-II Yukawa couplings are, respectively, with the quark mixing suppressed,
\begin{eqnarray}
\label{eq:2HDM_Yuk}
\mathcal{L}= \left\{\begin{array}{ll}
-\biggl( \displaystyle{ c_\alpha\over s_\beta}{m_u\over v} \biggr)\overline Q_L \tilde \Phi_2 u_R  -\biggl(  {  c_\alpha\over s_\beta}{m_d\over v}\biggr) \overline Q_L \Phi_2 d_R + {\rm h.c.} & \\ 
-\biggl(  \displaystyle{ c_\alpha\over s_\beta}{m_u\over v} \biggr)\overline Q_L \tilde \Phi_2 u_R
+\biggl( { s_\alpha\over c_\beta}{m_d\over v} \biggr)\overline Q_L \Phi_1 d_R
+ {\rm h.c.} & \\ 
\end{array} \right. \nonumber 
\end{eqnarray}
where $Q_L^T=(u_L,d_L)$ is the SM quark doublet, $\tilde \Phi_2 \equiv i \sigma_2 \Phi_2^\ast$ with $\sigma_2$ being the second Pauli matrix, and the Yukawa couplings to the charged leptons are of the same form as that of the down-type quarks in both the two cases. After the rotation ${\cal R}$, the couplings of physical scalars $H_i$ to the SM fermions and $W$ and $Z$ gauge bosons can be parameterized as
\begin{eqnarray}
\label{eqn:couplings}
\mL&=& \sum_{i=1}^3 \Big[ -m_f\left( c_{f,i} \bar f f+ \tilde c_{f,i} \bar f i\gamma_5 f  \right) \nonumber \\ && + a_i \left( 2  m_W^2 W_\mu W^\mu +  m_Z^2 Z_\mu Z^\mu \right)  \Big] \frac{H_i}{v}  \,,
\end{eqnarray}
with the coefficients $c_{f,i}$, $\tilde{c}_{f,i}$ and $a_i$ collected in Table~\ref{tab:couplings}, as functions of the ${\cal R}$ matrix elements in Eq.~(\ref{eqn:matrix2})~\cite{Shu:2013uua, Inoue:2014nva, Chen:2015gaa}. In the CP conserving limit of $\alpha_{b,c}=0$, it is clear that the ${\cal R}$ matrix is block-diagonal, and the first two scalars $H_{1,2}$ have the purely CP-even Yukawa couplings of $c_{f\,,i}$ while the couplings of the third one $H_3$ are purely CP-odd $\tilde c_{f\,,i}$. In the most general case, when $c_{f,i}\tilde c_{f,i}\neq 0$ or $a_{i}\tilde c_{f,i}\neq 0$, all the three mass eigenstate $H_i$ couples to both CP-even and CP-odd currents, and the CP symmetry is violated.

The Yukawa couplings of the charged scalar $H^\pm$ is
\begin{widetext}
\begin{eqnarray}
\label{eq:2HDM_Yuk2}
\mathcal{L}= \left\{\begin{array}{ll}
- \frac{\sqrt2}{v} H^+ \bar{u}_i V_{ij} \Big[
\cot\beta \, m_{u_i} (1-\gamma_5) +
\cot\beta \, m_{d_j} (1+\gamma_5)
\Big] d_j + {\rm h.c.} &
\hspace{0.5cm} {\rm 2HDM-I}\vspace{0.2cm} \\
- \frac{1}{\sqrt2} H^+ \bar{u}_i V_{ij} \Big[
\cot\beta \, m_{u_i} (1-\gamma_5) -
\tan\beta \, m_{d_j} (1+\gamma_5)
\Big] d_j + {\rm h.c.} &
\hspace{0.5cm} {\rm 2HDM-II} \, ,
\end{array} \right.
\end{eqnarray}
\end{widetext}
with $V_{ij}$ the CKM quark mixing matrix. These couplings can be used to interpret the LHC limits on the charged scalars in terms of the 2HDMs, which would be applied to the neutral scalar sector, due to the oblique constraints on the heavy scalar mass splitting, as shown in Section~\ref{sec:chargedscalar}.

\begin{table*}[!t]
  \centering{
  \caption{Yukawa and gauge couplings of the physical neutral scalars in CPV 2HDM, in terms of the corresponding SM couplings.}
  \label{tab:couplings}
  \begin{tabular}{cccccc}
  \hline\hline
  &  $c_{u,i}$ & $\tilde c_{u,i}$ & $c_{d,i}$ &  $\tilde c_{d,i}$ & $a_i$ \\
  \hline
  Type I & ${\cal R}_{i2}/\sin\beta$ &
  $-{\cal R}_{i3}\cot\beta$ & ${\cal R}_{i2}/\sin\beta$ & ${\cal R}_{i3}\cot\beta$ &
  ${\cal R}_{i2}\sin\beta+{\cal R}_{i1}\cos\beta$ \\
  \hline
  Type II & ${\cal R}_{i2}/\sin\beta$ &
  $-{\cal R}_{i3}\cot\beta$ & ${\cal R}_{i1}/\cos\beta$ & $-{\cal R}_{i3}\tan\beta$ &
  ${\cal R}_{i2}\sin\beta+{\cal R}_{i1}\cos\beta$ \\
  \hline\hline
  \end{tabular}}
\end{table*}

In evaluations of the decay widths and propagator matrix for the (quasi-degenerate) heavy neutral scalars $H_{2,3}$ below, we also need the Higgs-gauge couplings $g_{1iZ}$ involving two different physical scalars and the SM $Z$ boson, which is of form
\begin{eqnarray}
\label{eqn:g1iZ}
g_{1iZ}&=& \frac{e}{2s_{W} c_W} \Big[  ( - s_\beta \mR_{11} + c_\beta \mR_{12}) \mR_{i3} \nonumber \\
&&- ( - s_\beta \mR_{i1} + c_\beta \mR_{i2} ) \mR_{13}  \Big]\,,
\end{eqnarray}
and can be significantly simplified in the parameter set of $\alpha = \beta - \pi/2$ and $\alpha_{b,c}\neq0$:
\begin{eqnarray}
g_{12Z} &=&- \frac{e}{2s_{W} c_W} \, s_{\alpha_b} c_{\alpha_c} \,,\nonumber \\
g_{13Z} &=& \frac{e}{2s_{W} c_W} \,  s_{\alpha_b} s_{\alpha_c} \,.
\end{eqnarray}
The trilinear scalar coupling is relevant to the decay of $H_{2,3} \to hh$, and can be extracted from the Higgs potential as
\begin{eqnarray}
\label{eqn:lambda11i}
\lambda_{11i} \equiv
\frac12 \frac{\partial^3 {\cal L}_{3s}}{ \partial^2 H_1\, \partial^2 H_i} =
- \frac{v}{2} \sum_{m,n,k} {\cal R}_{1m} {\cal R}_{1n} {\cal R}_{ik} \, a_{mnk} \,, \nonumber \\
\end{eqnarray}
with $i = 2,\,3$, ${\cal L}_{3s}$ the original Lagrangian for the cubic scalar couplings in the basis of $(H_1^0, H_2^0, A^0)$ before the rotation ${\cal R}$,
and the coefficients~\cite{Osland:2008aw}
\begin{eqnarray}
a_{111} &=& \frac12 c_\beta \lambda_1 \,, \nonumber \\
a_{112} &=& \frac12 s_\beta \lambda_{345} \,, \nonumber \\
a_{113} &=& -\frac12 s_\beta c_\beta {\rm Im} \lambda_5 \,, \nonumber \\
a_{122} &=& \frac12 c_\beta \lambda_{345} \,, \nonumber \\
a_{123} &=& - {\rm Im} \lambda_{5} \,, \nonumber \\
a_{133} &=& \frac12 c_\beta (s^2_\beta \lambda_1 + c^2_\beta {\rm Re} \lambda_{345}
- 2 {\rm Re} \lambda_5 ) \,, \nonumber \\
a_{222} &=& \frac12 s_\beta \lambda_{2} \,, \nonumber \\
a_{223} &=& - \frac12 s_\beta c_\beta {\rm Im} \lambda_{5} \,, \nonumber \\
a_{233} &=& \frac12 s_\beta (c^2_\beta \lambda_2 + s^2_\beta {\rm Re} \lambda_{345}
- 2 {\rm Re} \lambda_5 ) \,, \nonumber \\
a_{333} &=& \frac12 s_\beta c_\beta {\rm Im} \lambda_5 \,,
\end{eqnarray}
with $\lambda_{345}$ defined in Eq.~(\ref{eqn:lambda345}). A general derivation of the scalar cubic and quartic self couplings in CPV 2HDM can be found in Ref.~\cite{Pilaftsis:1999qt,Carena:2002bb,Osland:2008aw}.

We list here also the trilinear couplings of neutral scalars to the charged scalar $H^\pm$, which could, in principle, enter the $H_i \gamma\gamma$ coupling through the $H^\pm$ loop:
\begin{eqnarray}
\lambda_{i +-} = \sum_j {\cal R}_{ij} \tilde{\lambda}_{j+-} \,,
\end{eqnarray}
with the coefficients $\tilde{\lambda}$ written in the basis of $(H_1^0, H_2^0, A^0)$~\cite{Carena:2002bb,Osland:2008aw}
\begin{eqnarray}
\label{eqn:lambda+-1}
\tilde{\lambda}_{1+-} &=& -v\cos\beta \left[ \sin^2\beta
( \lambda_1 - \lambda_4 - {\rm Re} \lambda_5 ) + \cos^2 \beta \lambda_3 \right] \,, \nonumber \\ && \\
\label{eqn:lambda+-2}
\tilde{\lambda}_{2+-} &=& -v\sin\beta \left[ \cos^2\beta
( \lambda_2 - \lambda_4 - {\rm Re} \lambda_5 ) + \sin^2 \beta \lambda_3 \right] \,, \nonumber \\ && \\
\label{eqn:lambda+-3}
\tilde{\lambda}_{3+-} &=& -v\sin\beta \cos\beta \, {\rm Im} \lambda_5 \,.
\end{eqnarray}




\subsection{Quasi-degenerate heavy scalars and CP violation}
\label{sec:degenerate}

The $(1,3)$ and $(2,3)$ elements the mass square matrix ${\cal M}_0^2$ in Eq.~(\ref{eqn:matrix}) are the source of CPV in the 2HDM, and they are correlated via
\begin{eqnarray}
(\mM_0^2)_{13} = (\mM_0^2)_{23} \, t_\beta \,,
\end{eqnarray}
which relates the scalar masses to the CPV angles as follows~\cite{Khater:2003wq}
\begin{eqnarray}
\label{eqn:relation0}
&&( M_1^2 - M_2^2 s_{\alpha_c}^2  - M_3^2 c_{\alpha_c}^2 ) s_{\alpha_b} (1 + t_\alpha) \nonumber \\
&&= (M_2^2 - M_3^2 ) (t_\alpha t_\beta - 1) s_{\alpha_c} c_{\alpha_c}\,.
\end{eqnarray}
In particular, the magnitudes of CPV is very sensitive to the mass splitting $\Delta M_H \equiv M_3- M_2$ (here for simplicity we assume the scalar $H_3$ is heavier than $H_2$, i.e. $M_3 - M_2 >0$). Given the relation in Eq.~(\ref{eqn:relation0}), with larger deviation of $t_\beta$ from $1$ and smaller mass splitting of $\Delta M$
\begin{eqnarray}
\Delta M_H \ll M_H \equiv \frac{M_2 + M_3}{2} \,,
\end{eqnarray}
one gets larger CPV mixing of $|\alpha_c|$ in the heavy Higgs sector. The non-trivial dependence of CPV in 2HDM on the mass splitting $\Delta M_H$ (and other parameters such as the heavy scalar mass $M_H$ and $\tan\beta$) is crucially important for the couplings of the heavy scalars to the SM particles, for a transparent physical picture we refer to Fig.4 of Ref.~\cite{Bian:2016zba}. On the phenomenological side, this is intimately related to the theoretical, collider and EDM constraints on the heavy neutral scalars in Section~\ref{sec:limitsall}.
This is also the strongest motivation in this work for us to study in great detail the phenomenologies of CPV in the degenerate limit.


In the scalar sector of CPV 2HDM, we have the mass parameters $m_{ij}^2$, and the quartic couplings $\lambda_i$ in the potential~(\ref{eqn:potential}), including also the two phases of $m_{12}^2$ and $\lambda_5$. After spontaneous symmetry breaking, these are related to the phenomenological parameters, of which some are already known and some others are measurable at the LHC: the EW VEV $v_{}$ and ratio $\tan\beta$, the neutral scalar masses $M_{i}$, the charged scalar mass $M_{\pm}$, the (CPV) mixing angles $\alpha$, $\alpha_b$, $\alpha_c$, and the soft $Z_2$ breaking parameter $m_{\rm soft}$.
To simplify the numerical calculations below in the high-dimensional parameter space and obtain some physically meaningful results, we will not scan the whole parameter space, but rather make the following reasonable assumptions, which are applied to all the numerical calculations below and suffice to demonstrate the non-trivial features in the scalar sector of CPV 2HDM:
\begin{itemize}

  \item
  The {\it alignment limit} requires both $\alpha = \beta - \pi/2$ and $\alpha_{b}=0$~\cite{Haber:2013mia,Grzadkowski:2014ada}, i.e., no CPV in the $Z_2$ symmetric model. However, a small deviation from the exact alignment limits is still allowed by current LHC Higgs data, i.e. the couplings of SM Higgs to the gauge bosons $a_1 = -\cos\alpha_b \sin (\alpha-\beta) \simeq 1$; in other words, $\alpha$ may deviate from $\beta-\pi/2$ and/or $\alpha_b$ may be non-zero. In this paper, we consider a particular direction in vicinity of the exact alignment limit, i.e., $\alpha=\beta-\pi/2$ and $\alpha_b \neq 0$, which is also adopted in Refs.~\cite{Shu:2013uua, Bian:2014zka, Inoue:2014nva}. It is found that in this direction the mixing angle $\alpha_b$ can be allowed up to $\sim\mathcal{O}(10^{-1})$ for $\tan\beta\sim 1$, and, as aforementioned, the scenario can generate abundantly the BAU through the EWBG mechanism due to the existence of beyond SM CPV and the feasibility of strong first order phase transition~\cite{Dorsch:2013wja, Cline:2011mm, Shu:2013uua, Bian:2014zka}. This motivates us to investigate further the CPV effects in the upcoming LHC data in this scenario, which is largely complementary to the EDMs experiments in the direct searches of CP violation~\cite{Chen:2015gaa,Chen:2017com}. It turns out that, as seen in the following sections, the amplified CP violating interference effects in the diphoton spectrum in comparison with the CP conserving cases are able to be detected at the HL-LHC 14 TeV run~\cite{Jung:2015gta,Jung:2015sna,Djouadi:2016ack}.

  \item As illustrating examples, we will focus on two specific value of the small mass splitting $\Delta M_H = 1$ and $10$ GeV; for larger values, say 50 GeV, the two heavy scalar $H_{2,3}$ are significantly separated apart, with much weaker correlations between them. An even smaller mass splitting $\Delta M_H$ is, on the other hand, in practice possible, but would not change too much the qualitative features and might need mild tuning of the parameters in the potential.
  \item The $Z_2$ breaking parameter $m_{\rm soft}^2 = {\rm Re} \, m_{12}^2$ is directly related to the quartic couplings $\lambda_i$, see Eqs.~(\ref{eqn:lambda1}) to (\ref{eqn:lambda5Im}). Its impact on the diphoton signal is two folded: On one hand, it will enter the trilinear couplings of the neutral and charged scalars $\lambda_{i+-} H_i H^+ H^-$ in Eqs.~(\ref{eqn:lambda+-1}) to (\ref{eqn:lambda+-3}), and contribute to the $H^\pm$ loop for the effective $H_i \gamma\gamma$ interaction. However, as long as the quartic couplings $\lambda_{i}$ are within the perturbative ranges, the $H^\pm$ loop contribution to the diphoton signal is always subdominant to the fermion loops. On the other hand, the perturbativity, stability and unitarity bounds on the quartic couplings would also set limits on $m_{\rm soft}$, depending on $\tan\beta$, the heavy scalar masses and the mixing angles. For the scalar masses below 1 TeV, the theoretical limits require that the soft breaking mass parameter to be of the few hundred GeV. To be specific, throughout the numerical calculations below we set $m_{\rm soft} = 300$ GeV.
  \item We will not apply any ``artificial'' constraints on the heavy scalar mass $M_H > m_h$, $\tan\beta$, and the CPV mixing angles $-\frac{\pi}{2} < \alpha_{b,c} < \frac{\pi}{2}$, besides the  correlation obtained from Eq.~(\ref{eqn:relation0}):
      \begin{eqnarray}
      \label{eqn:relation}
      \sin \alpha_b = \frac12
      \frac{(M_{3}^2 - M_{2}^2) \sin2\alpha_c \tan2\beta}
      {\left( M_{2}^2 \sin^2 \alpha_c + M_{3}^2 \cos^2 \alpha_c \right) - m_h^2 } \,,
      \end{eqnarray}
      or, alternatively,
      \begin{eqnarray}
      \label{eqn:relation2}
      \tan2\beta = \left[ \cos2\alpha_c
      +\frac{\left( M_{2}^2+M_{3}^2\right) - 2 m_h^2}{M_{3}^2 - M_{2}^2} \right]
      \frac{\sin\alpha_b}{\sin2\alpha_c} \,. \nonumber \\
      \end{eqnarray}
      In the calculations below we will exclude the unphysical regions in which the phase $\alpha_b$ or $\alpha_c$ does not have a real solution.
\end{itemize}

\subsection{Heavy neutral scalar decay and propagator matrix}
\label{sec:decay}



Compared to the SM-like scalar $H_1 = h$, the couplings of $H_{2,3}$ to the SM fermions and massive gauge bosons are rescaled respectively by the $c_{f,i}$ ($\tilde{c}_{f,i}$) and $a_i$ coefficients in Table~\ref{tab:couplings}, and therefore the decay widths into SM fermion pairs, $WW$ and $ZZ$ are respectively proportional to the linear combinations of $(c_{f,i})^2$ and $(\tilde{c}_{f,i})^2$ and $a_i^2$. The beyond SM decay channels $H_i \to hh$, $hZ$ are dictated respectively by the couplings $\lambda_{11i}$ in Eq.~(\ref{eqn:lambda11i}) and $g_{1iZ}$ in Eq.~(\ref{eqn:g1iZ}). In the CP conserving limit of $\alpha_{b,c} = 0$, both the two couplings, and thus the two decay modes, are vanishing. At loop level, the heavy scalars could decay into two gluons and two photons, as in the SM, mediated by the SM fermion loops, with subleading contribution from the $W^\pm$ and $H^\pm$ loops for the diphoton channel (as long as the quartic couplings $\lambda_i$ in the potential~(\ref{eqn:potential}) are within the perturbative range). The partial decay widths in CPV 2HDM for these channels are respectively, at the leading order,
\begin{widetext}
\begin{eqnarray}
\label{eqn:Gammaff}
\Gamma (H_i \to f \bar{f}) 
&=& \frac{N_C^f G_{F} M_i m_f^2}{4\sqrt2 \pi}
\left[ \left( 1 - \frac{4m_f^2}{M_{i}^2} \right)
(c_{f,i})^2 + (\tilde{c}_{f,i})^2 \right]
\left( 1 - \frac{4m_f^2}{M_{i}^2} \right)^{1/2} \,, \\
\label{eqn:GammaVV}
\Gamma (H_i \to VV) &=&
\Gamma (H_{\rm SM} \to VV) \times
(a_i)^2 \nonumber \\
&=& \frac{G_F \delta_V M_{i}^2 (a_i)^2}{16\sqrt2 \pi}
\left( 1- \frac{4m_V^2}{M_{i}^2} \right)^{1/2}
\left( 1- \frac{4m_V^2}{M_{i}^2} + \frac{12m_V^4}{M_{i}^4} \right) \,, \\
\label{eqn:Gammahh}
\Gamma (H_i \to hh) &=&
\frac{|\lambda_{11i}|^2}{4\pi M_{i}}
\left( 1- \frac{4m_h^2}{M_{i}^2} \right)^{1/2} \,, \\
\label{eqn:GammahZ}
\Gamma[H_i \to h Z]&=& \frac{  | g_{1iZ} |^2 m_Z^2 }{16\pi M_i }
\left[
\left( 1 - \frac{(m_h + m_Z)^2}{M_{i}^2}  \right)
\left(  1 - \frac{(m_h - m_Z)^2}{M_{i}^2}  \right)  \right]^{1/2} \nonumber \\
&& \times \left[  1 - \frac{2 (M_{i}^2 + m_h^2)}{m_Z^2}
+ \frac{( M_i^2  - m_h^2 )^2}{m_Z^4}  \right] \,, \\
\label{eqn:Gammagg}
\Gamma (H_i \to gg) &=&
\frac{G_F \alpha_s^2 (M_i) M_{i}^3}{64 \sqrt2 \pi^3}
\left[ \left| \sum_q c_{q,i} A_{1/2}^{H} (\tau_q) \right|^2 +
\left| \sum_q \tilde{c}_{q,i} A_{1/2}^{A} (\tau_q) \right|^2  \right] \,, \\
\label{eqn:GammaAA}
\Gamma (H_i \to \gamma\gamma) &=&
\frac{G_F \alpha_{\rm EM}^2 M_{i}^3}{128 \sqrt2 \pi^3}
\left[ \left| \sum_{j=1,2} \frac{R_{ij} \tilde{\lambda}_{j+-}v}{2M_\pm^2} A_{0} (\tau_\pm) +
\sum_f N_C^f Q_f^2 c_{f,i} A_{1/2}^{H} (\tau_f)
+ a_i A_{1} (\tau_W) \right|^2 \right. \nonumber \\
&& \qquad\qquad\qquad + \left.
\left| \frac{R_{i3} \tilde{\lambda}_{3+-}v}{2M_\pm^2} A_{0} (\tau_\pm) +
\sum_f N_C^f Q_f^2 \tilde{c}_{f,i} A_{1/2}^{A} (\tau_f)  \right|^2  \right] \,,
\end{eqnarray}
\end{widetext}
where $\tau_X = M_H^2 / 4m_X^2$ (with $\tau_\pm = M_H^2 / 4 M_\pm^2$), $\delta_V = 1$ for the $Z$ boson and $2$ for the $W$ boson, $\alpha_{\rm EM}$ the fine-structure constant, the strong coupling $\alpha_s$ evaluated at the scale $M_H$, the trilinear scalar couplings $\tilde\lambda$ given in Eqs.~(\ref{eqn:lambda+-1}) to (\ref{eqn:lambda+-3}), the loop functions $A(\tau)$ for the charged scalars, fermions and vectors are the same as in~\cite{Djouadi:2005gi, Djouadi:2005gj}.

\begin{figure*}[!t]
  \centering
  \includegraphics[width=0.4\textwidth]{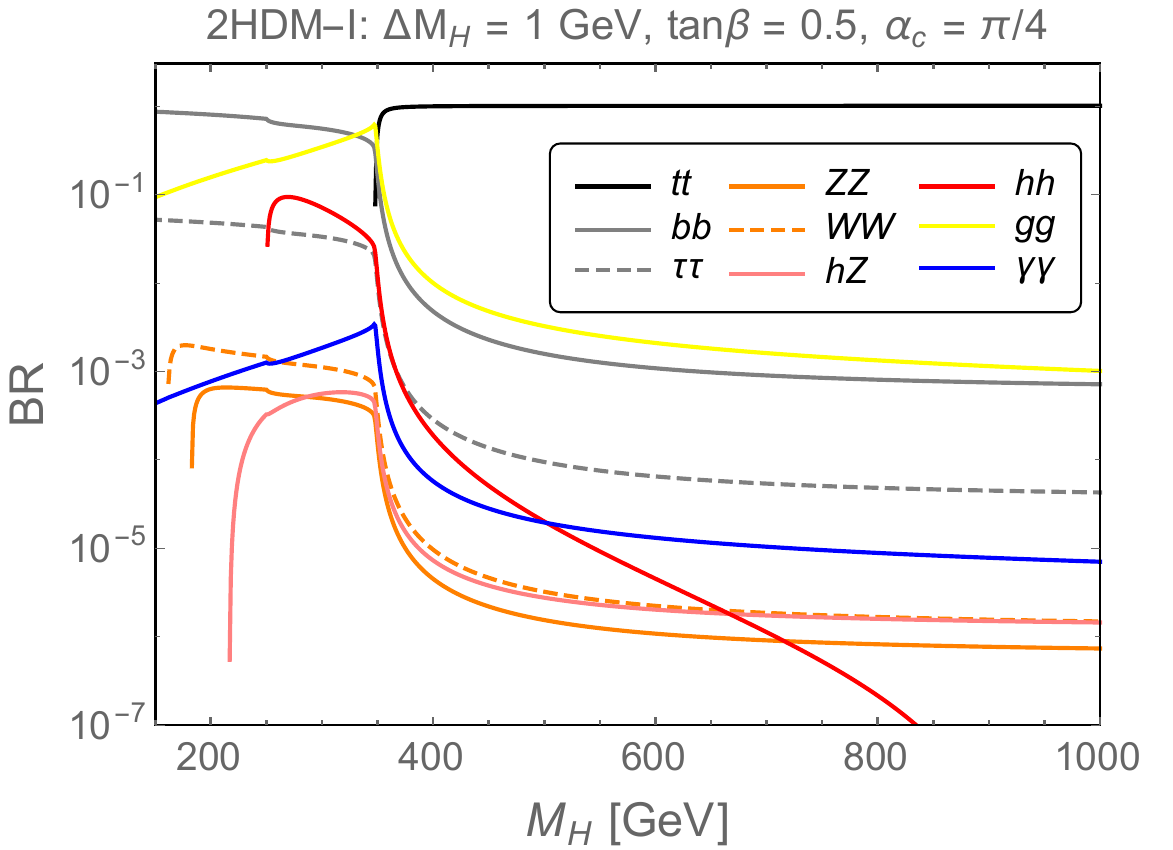} \hspace{-10pt}
  \includegraphics[width=0.4\textwidth]{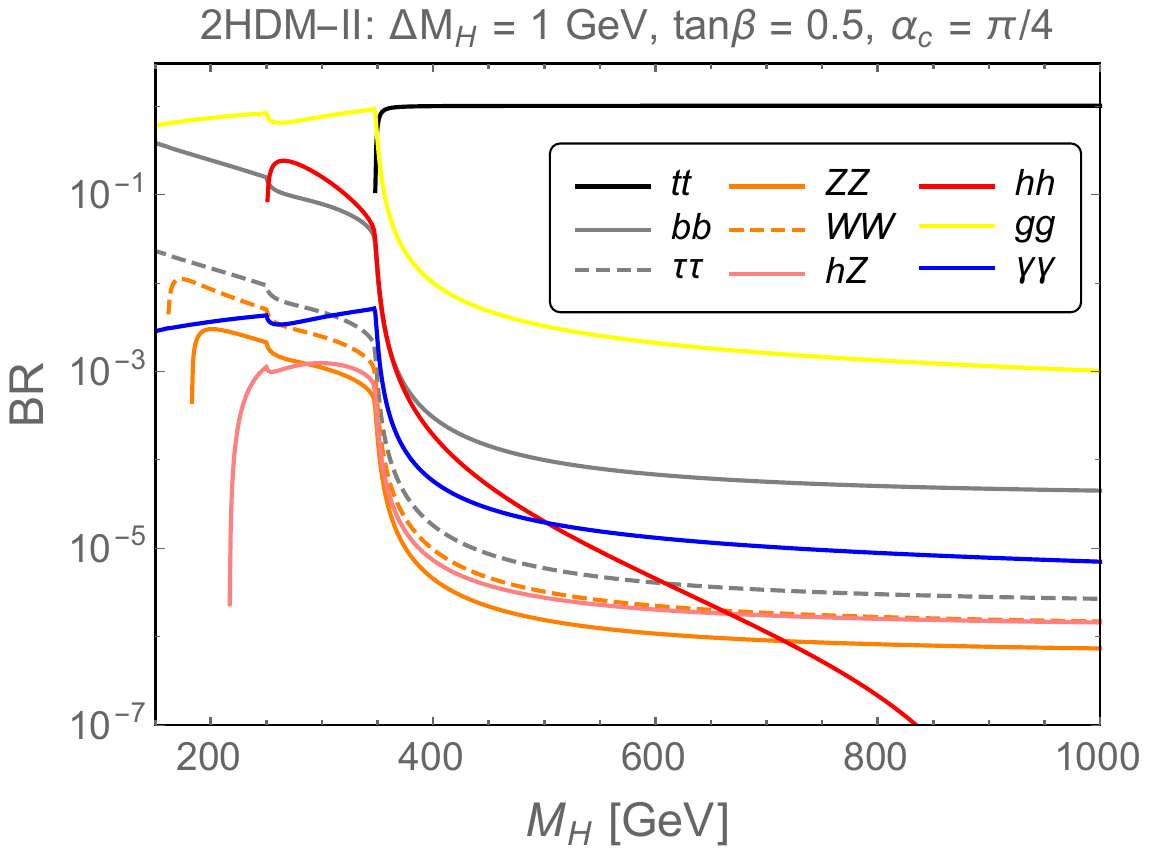} \\
  \includegraphics[width=0.4\textwidth]{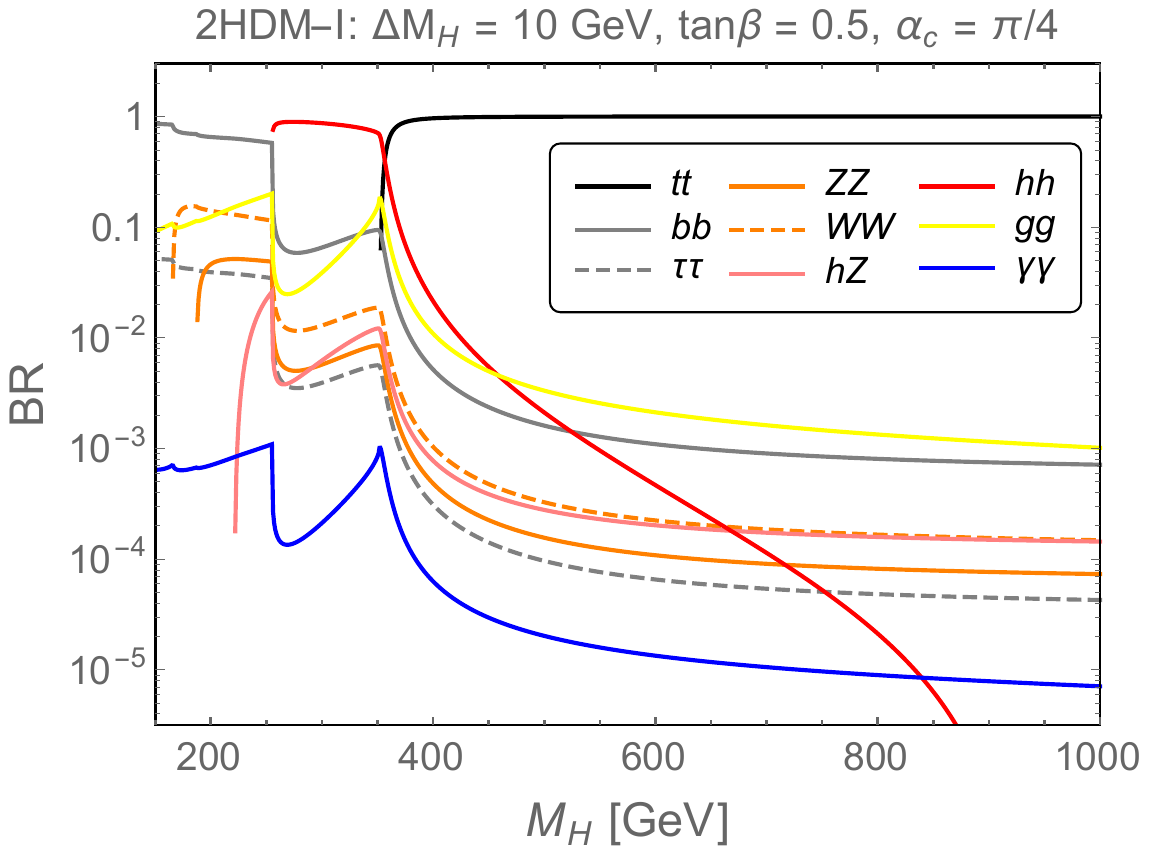} \hspace{-10pt}
  \includegraphics[width=0.4\textwidth]{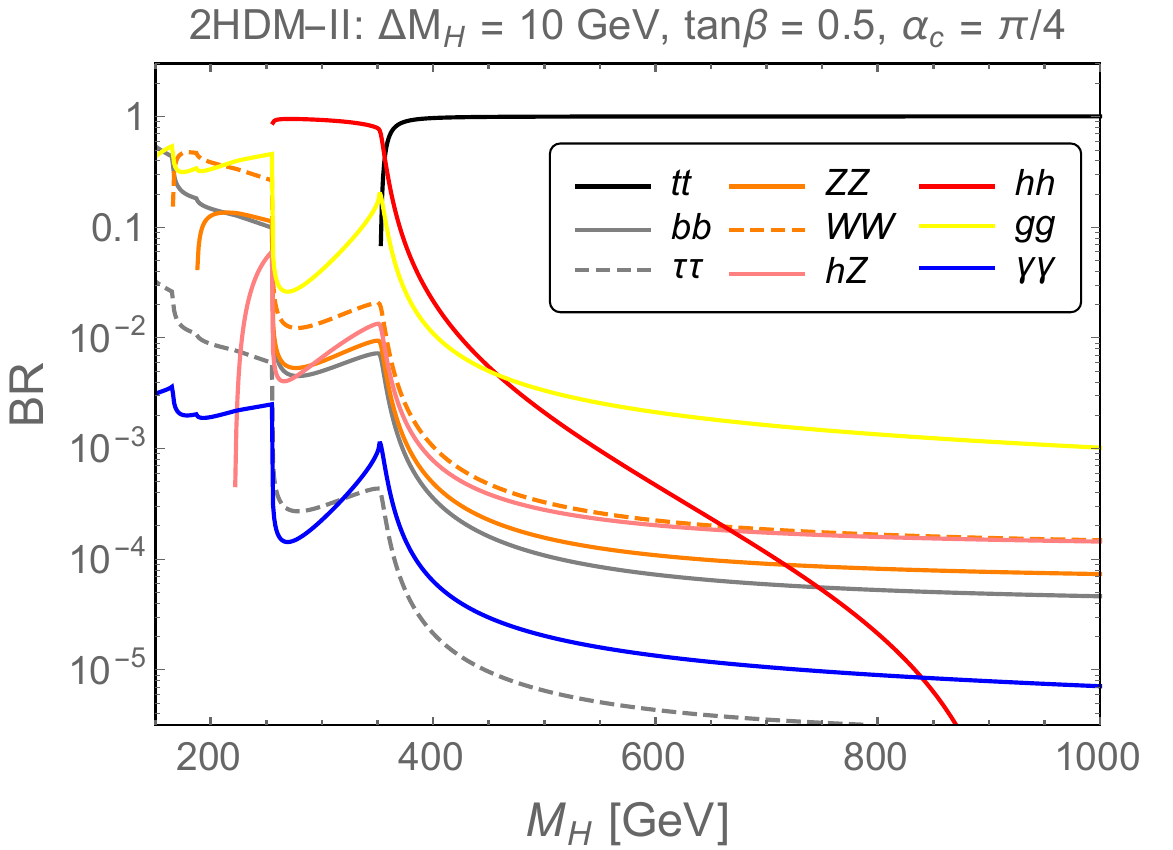}
  \caption{Representative examples of BRs of the heavy scalars $H_{2,3}$ in CPV 2HDM of type-I (left) and type-II (right), as functions of scalar mass $M_H$, with a small mass splitting of $\Delta M_H = 1$ GeV (upper) or 10 GeV (lower), and $\tan\beta = 0.5$. With the maximal mixing $\alpha_c = \pi/4$, the BRs of the two heavy scalars $H_{2,3}$ are almost the same.}
  \label{fig:BR}
\end{figure*}

\begin{figure*}[!t]
  \centering
  \includegraphics[width=0.4\textwidth]{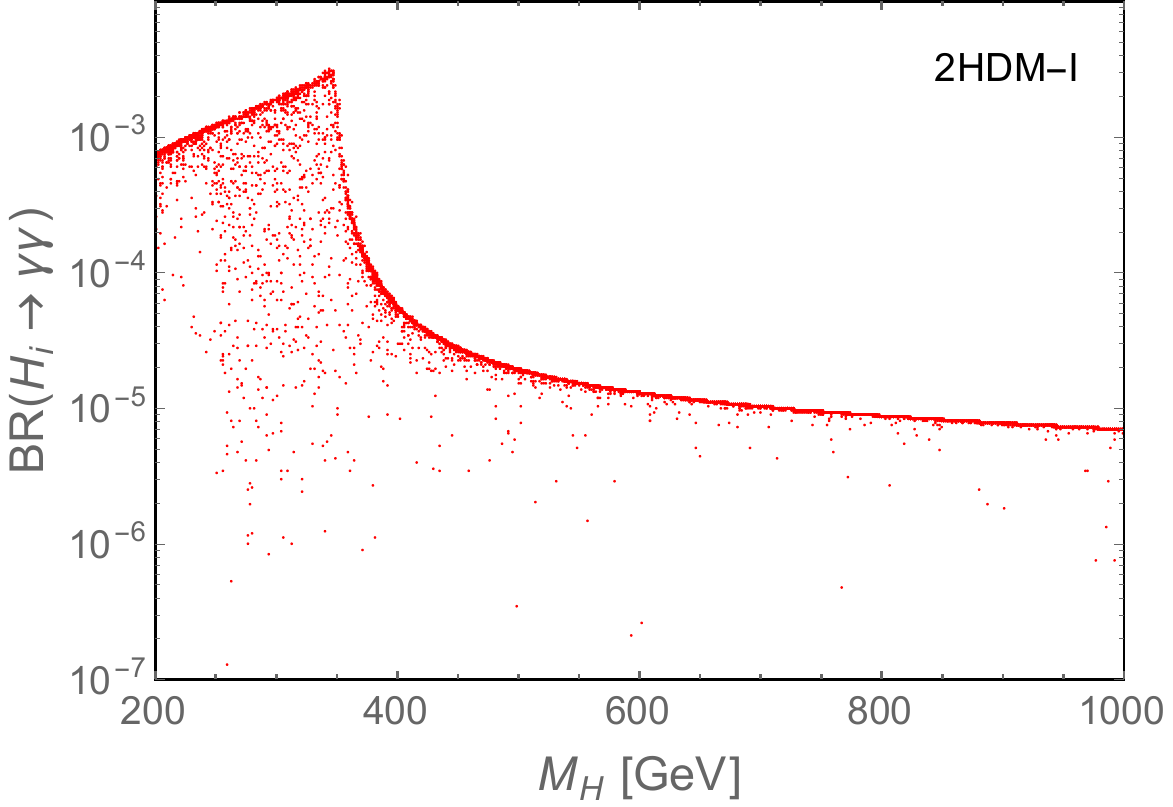} \hspace{-10pt}
  \includegraphics[width=0.4\textwidth]{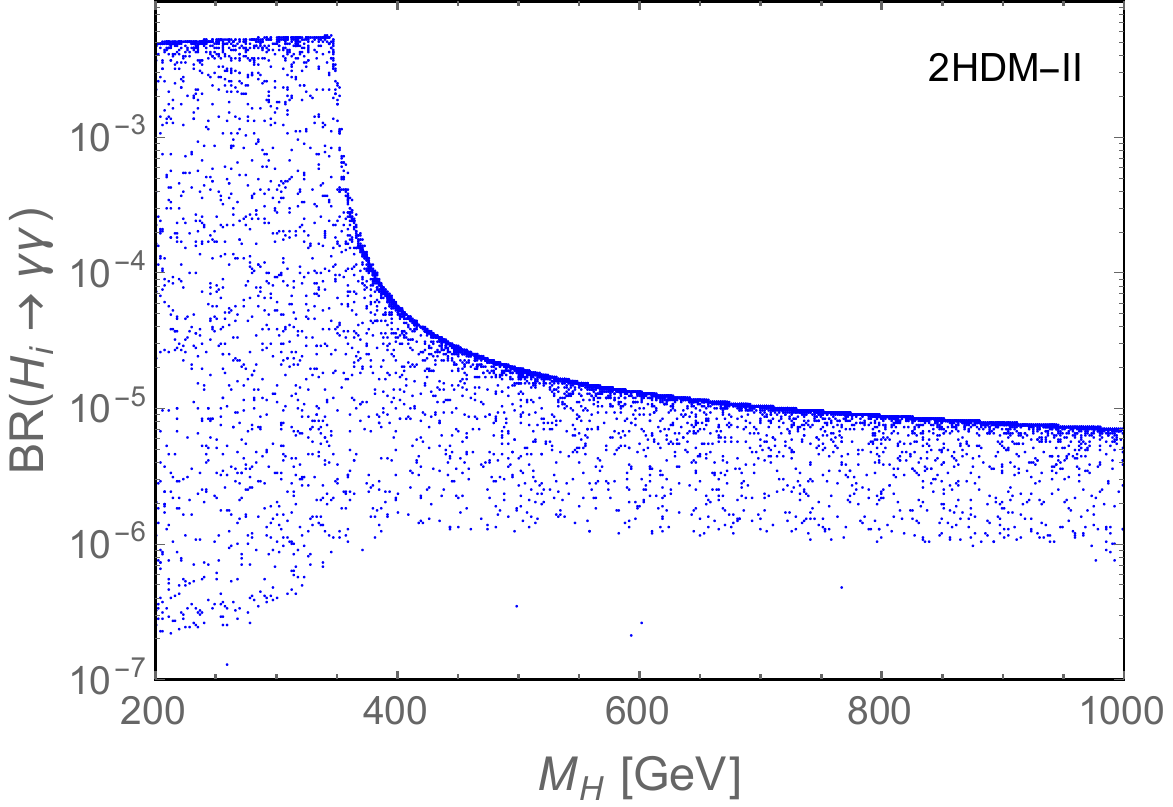}
  \caption{${\rm BR} (H_{2,3} \to \gamma\gamma)$ in CPV 2HDM of type-I (left) and type-II (right), as functions of scalar mass $M_H$, with the mass and mixing parameters varying in the ranges given in Eq.~(\ref{eqn:random}).}
  \label{fig:BR2}
\end{figure*}

Representative examples of the various branching ratios (BRs) of the heavy scalars $H_{2,3}$ in the CPV 2HDM are presented in Fig.~\ref{fig:BR}, as functions of the heavy scalar mass $M_H$, for both the type-I and type-II Yukawa couplings in respectively the left and right panels, with small mass splittings as mentioned $\Delta M_H = 1$ GeV (upper) and 10 GeV (lower), $\tan\beta = 0.5$, and the maximal mixing of the two heavy scalars $\alpha_{c} = \pi/4$. With these mass and mixing parameters fixed, the mixing angle $\alpha_b$ is determined via the relation given in Eq.~(\ref{eqn:relation}), which is considered as a function of the scalar mass $M_H$. In this maximal-mixing case, the decay BRs of the two quasi-degenerate scalars $H_{2,3}$ are almost the same, with small corrections from, e.g., the tiny differences of phase spaces. Here follow more comments on these different decay modes:
\begin{itemize}
  \item As a result of the $\mathcal{O} (1)$ top Yukawa coupling $m_t/v$ in the SM, $H_{2,3}$ decays almost 100\% into top pairs, as long as it is kinematically allowed. As a ``side effect'', the top-loop induced decay $H_{2,3} \to gg$ is generally larger than, or comparable to, other channels besides $t\bar{t}$ in most of the parameter space (the yellow lines in Fig.~\ref{fig:BR}). The decay rates to other lighter fermions, e.g. $b\bar{b}$ and $\tau\bar{\tau}$, depend largely on the SM Yukawa coupling and $\tan\beta$.\footnote{In the CPV 2HDM with quasi-degenerate heavy neutral scalars, the scenarios with large $\tan\beta \sim m_t/m_b$ is excluded by the perturbativity, unitarity and stability constraints, or at least highly disfavored; the favorite regions are around $\tan\beta \sim 1$, see the examples in Figs.~\ref{fig:sensitivity1} and \ref{fig:sensitivity2}.  }
  \item For the quasi-degenerate case of $H_{2,3}$, the mixing angle $\alpha_b$ is generally very small. Even if it is sizable, say $\sim 0.1$, it could be easily excluded by the EDM measurements, cf. Figs.~\ref{fig:sensitivity5} to \ref{fig:sensitivity7}. Therefore, for a small mixing $\alpha_b$ with the SM Higgs, these decay modes into SM $h$, $W$ and $Z$ bosons are in general highly suppressed, if the $t \bar{t}$ channel is open. Resultantly, the constraints from direct searches of $H_{2,3} \to WW/ZZ$, $hh$ and $hZ$ are very limited, effective only when the scalar mass $M_{H} \lesssim 450$ GeV for both $\Delta M_H = 1$ GeV and 10 GeV, unless $\tan\beta$ is to some extent fine-tuned $|\tan\beta - 1| \ll 1$.  See Figs.~\ref{fig:limits1} to \ref{fig:limits5} and Section~\ref{sec:limits} for more details.
  \item In the large $M_H$ limit, the ${\rm BR} (H_{2,3} \to \gamma\gamma)$ is expected to be of order $10^{-5}$, dictated by the couplings and loop factors in Eqs.~(\ref{eqn:Gammaff}) and (\ref{eqn:GammaAA}) (here we have used the fact that the loop functions $|A_{1/2}^H (\tau_t)| \simeq |A_{1/2}^A (\tau_t)|$ when $M_H$ is significantly larger than $2m_t$ but below roughly the TeV range):
      \begin{eqnarray}
      \label{eqn:BRdiphoton}
      {\rm BR} (H_i \to \gamma\gamma) \ &\simeq& \
      \frac{\Gamma (H_i \to \gamma\gamma)}{\Gamma (H_i \to tt)} \nonumber \\
      \ &\simeq& \
      \frac{\alpha_{\rm EM}^2 M_H^2}{54 \pi^2 m_t^2}
      \left| A_{1/2}^{H} (\tau_t) \right|^2 \,,
      \end{eqnarray}
      which is determined predominantly by the heavy scalar mass $M_H$ and has a weak dependence on other parameters in the 2HDM. For the illustration purpose, we present the ${\rm BR} (H_{2,3} \to \gamma\gamma)$ in Fig.~\ref{fig:BR2} in both the type-I and type-II 2HDM, where the mass and mixing parameters vary freely in the ranges below
      \begin{eqnarray}
      \label{eqn:random}
      M_H = M_\pm &\in& [m_h,\, 1 \, {\rm TeV} ] \,, \nonumber \\ 
      \Delta M_H &\in& [1,\, 10] \, {\rm GeV} \,,\nonumber \\ 
      m_{\rm soft} &\in& [100,\, 500] \, {\rm GeV} \,, \nonumber \\ 
      \tan\beta &\in& [0.1,\, 10] \,, \nonumber \\ 
      \alpha_c &\in& [-\pi/2,\, \pi/2] \,,
      \end{eqnarray}
      and $\alpha_b$ can be solved from the relation~(\ref{eqn:relation}). It is transparent in Fig.~\ref{fig:BR2} that in most of regions of interest the branching fraction into diphoton is within a narrow band which is well described by Eq.~(\ref{eqn:BRdiphoton}) (when the theoretical constraints and experimental limits in Section~\ref{sec:limitsall} are taken into consideration, some points with small BRs in Fig.~\ref{fig:BR2} might be excluded).
      Though the BR into diphoton is small, the SM background $gg \to \gamma\gamma$, which arise at one-loop level, is also suppressed compared to other processes. Without severe contamination from the messy QCD processes, $H_{2,3} \to \gamma\gamma$ should be one of the most important channels for direct heavy scalar searches at the LHC,\footnote{The $Z\gamma$ channel will not be explored here since interference effects in the process $gg\rightarrow Z\gamma$ is relatively small, though the BR of which can be relatively larger.} as for the SM Higgs.
\end{itemize}

It is known that for multiple (nearly-)degenerate resonances, the imaginary part of their propagator matrix is not diagonal if they have common decay channels~\cite{Cacciapaglia:2009ic}. This was also discussed for the amplitudes involving the CPV resonances in Ref.~\cite{Pilaftsis:1997dr}. Here we define the $2\times 2$ propagator matrix for $H_{2,3}$ as follows
\begin{widetext}
\begin{eqnarray}
\label{eqn:propagator}
P_{ij} (\hat{s}) &=&
\left( \begin{matrix}
\hat{s} - M_2^2 + i \widehat{\Pi}_{22} (\hat{s}) &
i \widehat{\Pi}_{23} (\hat{s}) \\
i \widehat{\Pi}_{23} (\hat{s}) &
\hat{s} - M_3^2 + i \widehat{\Pi}_{33} (\hat{s})
\end{matrix} \right)^{-1} \nonumber \\
&=& \frac{1}{\det P^{-1}_{ij} (\hat s) }
 \left( \ba{cc}  \hat s - M_3^2 + i \widehat \Pi_{33}(\hat s) & -  i \widehat \Pi_{23}(\hat s)  \\
 -  i \widehat \Pi_{23}(\hat s) &  \hat s - M_2^2 + i \widehat \Pi_{22}(\hat s)  \ea  \right)\,.
\end{eqnarray}
The absorptive parts of the scalar propagator matrix receive contributions from the loops of the SM fermions, vector bosons, associated scalar-vector bosons, and SM Higgs pairs~\cite{Ellis:2004fs}
\begin{eqnarray}
\widehat{\Pi}_{ij} (\hat{s}) =
 \widehat{\Pi}_{ij}^{ff} (\hat{s})
+\widehat{\Pi}_{ij}^{VV} (\hat{s})
+\widehat{\Pi}_{ij}^{hZ} (\hat{s})
+\widehat{\Pi}_{ij}^{hh} (\hat{s})
\end{eqnarray}
where the partial contributions are respectively
\begin{eqnarray}
\widehat \Pi_{ij}^{ff}(\hat s)&=&
K_F(\sqrt{\hat s}) \sum_{f}
\frac{ N_{C}^f \, \hat{s} \, m_f^2}{8\pi v^2} \sqrt{1-4\kappa_f} \nonumber \\
&& \times \Bigg[ \left(1- 2 \kappa_f \right)
(  c_{f,i} c_{f,j}^\ast  +  \tilde c_{f,i}  \tilde c_{f,j}^\ast   )
-2 \kappa_f ( c_{f,i} c_{f,j}^\ast  -  \tilde c_{f,i}  \tilde c_{f,j}^\ast   )  \Bigg] \Theta( \hat s - 4 m_f^2) \,,\\
\widehat \Pi_{ij}^{VV}(\hat s)&=&
\frac{ G_F a_i a_j \delta_V M_i^2 M_j^2}{16\sqrt{2} \pi }
\sqrt{1 - 4\kappa_V } \nonumber \\
&& \times \left[ 1+ 2 \left( \frac{m_V^2}{M_i^2} + \frac{m_V^2}{M_j^2} \right) -
\frac{4m_V^2 (2 \hat s - 3 m_V^2)}{M_i^2 M_j^2} \right]
\Theta(\hat s - 4 m_V^2 )\,,\\
\widehat \Pi_{ij}^{hZ}(\hat s)&=&
\frac{ g_{1i Z} g_{1j Z} M_i^2 M_j^2 }{16\pi\, v^2}
\lambda^{1/2}(1\,,\kappa_h\,, \kappa_Z) \nonumber \\
&& \times \left[ 1 - \frac{m_h^2-m_Z^2}{M_i^2} -\frac{m_h^2-m_Z^2}{M_j^2}
+ \frac{(m_h^2 - m_Z^2 )^2 -4\hat s m_Z^2}{M_i^2 M_j^2} \right]
\Theta(\hat s - (m_h+ m_Z)^2 ) \,, \\
\widehat \Pi_{ij}^{hh}(\hat s)&=&
\frac{S_{ij;\,11} \, \lambda_{11i} \lambda_{11j}}{32\pi}
\sqrt{1 - 4\kappa_h }\, \Theta(\hat s - 4m_h^2 )\,.
\end{eqnarray}
\end{widetext}
where $\kappa_X \equiv m_X^2/\hat s$, $K_f(\sqrt{\hat s})\simeq 1+5.67 {\alpha_s (\sqrt{\hat s})}/{\pi}$ accounting for the high-order corrections, $\lambda(x\,,y\,,z)=x^2 + y^2 + z^2 - 2(xy + yz + xz)$, and $S_{ij;\,11}$ the symmetry factor for identical particles. The total decay widths of $H_{2,3}$ are related to the imaginary parts of the self energies as follows
\begin{eqnarray}
\widehat{\Pi}_{ii} (M_i^2) \simeq M_i \Gamma_i \,,
\end{eqnarray}
excluding the loop decay modes into gluons and photons.
Under the limit of negligible off-diagonal widths, the propagators are reduced to the standard one
\begin{eqnarray}
P_{ii}(\hat s) \to \frac{1}{\hat s - M_i^2 + i \widehat \Pi_{ii} (\hat s) } \,.
\end{eqnarray}

\section{Theoretical and experimental constraints}
\label{sec:limitsall}


To have a self-consistent description, the mass spectrum and scalar potential of the CPV 2HDM should be constrained by the unitarity, perturbativity and vacuum stability requirements, which is summarized in Section~\ref{sec:theoretical}. The current LHC constraints on the heavy scalars are presented in Section~\ref{sec:collider}, including the direct searches in the final state of $WW/ZZ$, $hh$ and $hZ$ and the consistency of differential $t\bar{t}$ data with the SM predictions. When the EW precision tests are considered, the mass splitting $|M_H - M_\pm|$ can not be arbitrarily large, which could imply constraints on $H_{2,3}$ from the charged scalar sector, which is detailed also in Section~\ref{sec:collider}. The EDMs are one of the observables that are most sensitive to the beyond SM CPV, which is collected in Section~\ref{sec:EDM}. All these limits are used to constrain the masses of heavy scalars $H_{2,3}$ in CPV 2HDM and their couplings.

With the parameter setups in Section~\ref{sec:degenerate}, we scan the parameter space by varying $M_H$, $\tan\beta$, $\alpha_b$ (or $\alpha_c$), and present all these limits in the two-dimensional plots of $M_H - \tan\beta$, $M_H - \alpha_c$ and $M_H - \alpha_b$ in Figs.~\ref{fig:limits1} to \ref{fig:limits5}. In the $M_H - \tan\beta$ space, we compare the two scenarios of CP conservation $\alpha_c = 0$ and maximal CP violation $\alpha_c = \pi/4$ (here $\alpha_b$ determined by the relation~\ref{eqn:relation}). Clearer dependence on the CPV angle $\alpha_c$ can be found in the $M_H - \alpha_c$ plots, where we take two benchmark values of $10^{-3}$ and $10^{-2}$ for $\alpha_b$ and $\tan\beta$ is determined by Eq.~(\ref{eqn:relation2}). In the $M_H - \alpha_b$ plots we set $\alpha_c$ to be positively and negatively maximal ($\tan\beta$ is again obtained by the equation Eq.~(\ref{eqn:relation2})), i.e. $\alpha_c = \pm \pi/4$. By comparing the plots in Figs.~\ref{fig:limits4} and \ref{fig:limits5} we can see clearly the implications of changing the sign of $\alpha_c$, in some regions of the parameter space.

\subsection{Unitarity, perturbativity and stability bounds}
\label{sec:theoretical}

The perturbative unitarity constraints are imposed on the model so that it is not very strongly coupled, which are obtained by evaluating the $S$-matrices for the coupled  scalar scattering amplitudes in the CPV 2HDM (see Refs.~\cite{Arhrib:2000is,Kanemura:2015ska} for the CP conserving 2HDM case).
The $S$-matrices for coupled channels with different charge configurations can be packed as follows
\begin{eqnarray}
\label{eqn:a01}
a_0^0 \ &=& \ \frac{1}{16\pi} {\rm diag} (X_{4\times 4}\,, Y_{4\times4}\,, Z_{3\times 3}\,, Z_{3\times3})\,,\\
\label{eqn:a02}
a_0^+ \ &=& \ \frac{1}{16\pi} {\rm diag} ( Y_{4\times 4}\,, Z_{3\times 3}\,, \lambda_3 - \lambda_4 )\,,\\
\label{eqn:a03}
a_0^{++} \ &=& \ \frac{1}{16\pi} Z_{3\times 3} \,,
\end{eqnarray}
with the explicit expressions for the submatrices of $(X_{4\times 4}\,, Y_{4\times4}\,, Z_{3\times 3})$~\cite{Kanemura:2015ska}
\begin{widetext}
\begin{eqnarray}
X_{4\times 4} \ &=& \ \begin{pmatrix}
3\lambda_1 & 2\lambda_3 + \lambda_4 & 0 & 0 \\
2\lambda_3 + \lambda_4 &3\lambda_2 & 0 & 0 \\
0 & 0 & \lambda_3+2\lambda_4+3{\rm Re}\lambda_5  & 3{\rm Im}\lambda_5 \\
0 & 0 &  3{\rm Im}\lambda_5  & \lambda_3+2\lambda_4-3{\rm Re}\lambda_5 \\
\end{pmatrix} \,,
\end{eqnarray}
\end{widetext}
\begin{eqnarray*}
Y_{4\times 4} \ &=& \ \begin{pmatrix}
\lambda_1 & \lambda_4 & 0 & 0 \\
\lambda_4 & \lambda_2 & 0 & 0 \\
0 & 0 & \lambda_3+{\rm Re}\lambda_5 & {\rm Im}\lambda_5 \\
0 & 0 & {\rm Im}\lambda_5 & \lambda_3-{\rm Re}\lambda_5
\end{pmatrix} \,, \\
Z_{3\times 3} \ &=& \ \begin{pmatrix}
\lambda_1 & {\rm Re}\lambda_5 + i{\rm Im}\lambda_5 & 0 \\
{\rm Re}\lambda_5 - i{\rm Im}\lambda_5 & \lambda_2 & 0 \\
0 & 0 & \lambda_3+\lambda_4
\end{pmatrix} \,.
\end{eqnarray*}
The eigenvalues in Eqs.~(\ref{eqn:a01}) to (\ref{eqn:a03}) should be $\in( -\frac12 \,, \frac12)$ under the unitarity constraints. To satisfy the tree-level vacuum stability requirements, we impose the following conditions onto the quartic couplings in the potential~(\ref{eqn:potential}):\footnote{At loop level, these stability conditions might be weakened to some extent, see e.g.~\cite{Staub:2017ktc}.}
\begin{eqnarray}
&&\lambda_{1\,,2}>0\,,\qquad
\lambda_3 > - \sqrt{\lambda_1 \lambda_2}\,, \nonumber \\
&& \lambda_3+\lambda_4 - |\lambda_5 | > - \sqrt{\lambda_1 \lambda_2} \,.
\end{eqnarray}
The perturbativity limits are simply $|\lambda_i| < 4\pi$.


\subsection{Collider constraints}
\label{sec:collider}

\subsubsection{Direct heavy neutral scalar searches}
\label{sec:direct}

The direct searches of heavy neutral scalars have been performed at the LHC, in the decay modes of heavy scalars into the SM particles of $VV = WW$, $ZZ$~\cite{Aad:2015kna, Khachatryan:2015cwa, Aaboud:2016okv}, $hh$~\cite{Aad:2014yja, Khachatryan:2015yea, ATLAS:2016ixk, Khachatryan:2016sey, Aaboud:2016xco} and $hZ$~\cite{Aad:2015wra, Khachatryan:2015tha, TheATLAScollaboration:2016loc}, with the $h$, $W$ and $Z$ bosons decaying further into lighter SM particles.
There have been also searches of heavy CP-even or odd resonance scalars in the diphoton spectra~\cite{Aad:2014ioa, ATLAS:2016eeo, Khachatryan:2016yec}. However, in these searches, the interference effects between the resonance and SM background are not taken into account, and these exclusion limits can {\it not} be na\"ively interpreted on the 2HDM we are considering in which the interference terms are generally much more important than the pure resonances (see the examples in Figs.~\ref{fig:example05} and \ref{fig:example2}). The diphoton searches in Ref.~\cite{Khachatryan:2015qba} are interpreted in terms of the 2HDM~\cite{Craig:2013hca, Dev:2014yca, Dev:2015bta, Dev:2015cba, Dev:2017org}, with a pair of degenerate CP-even and odd scalars $H/A$. However, the scenarios they considered are $M_{H/A} = 200$ GeV and $300$ GeV, which is excluded by the theoretical limits in Section~\ref{sec:theoretical}, even if there is no CPV mixing between the two heavy scalars (cf, e.g. Figs.~\ref{fig:sensitivity1} and \ref{fig:sensitivity2}).
Therefore, we will consider only the direct search limits from the massive final states $h$, $W$ and $Z$ in the discussions below.

\begin{figure*}[!t]
  \centering
  \includegraphics[height=0.3\textwidth]{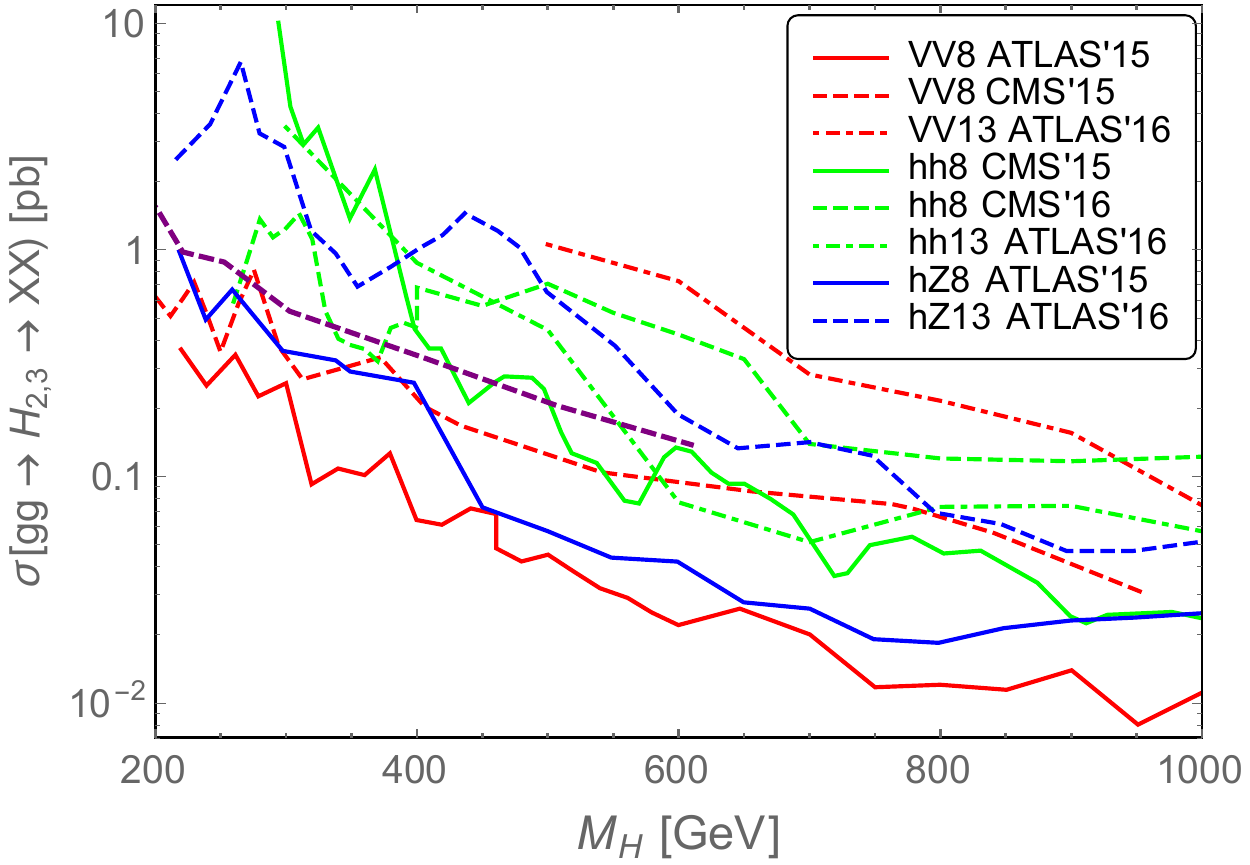}
  \includegraphics[height=0.3\textwidth]{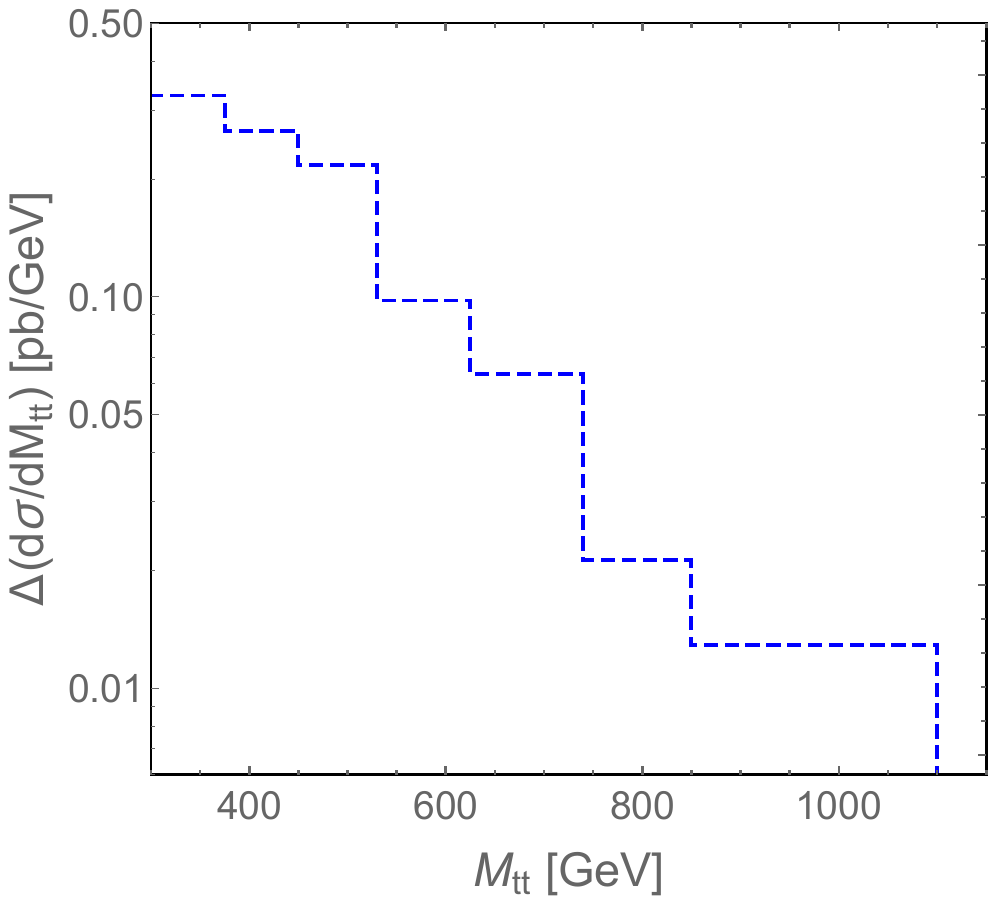}
  \caption{Left: Limits on the cross sections of $gg \to H_{2,3}$ at the LHC, in the subsequent different decay modes: The solid, dashed, and dot-dashed red lines are the limits from the decays $H \to ZZ$ in Ref.~\cite{Aad:2015kna} and $H \to WW/ZZ$ in Ref.~\cite{Khachatryan:2015cwa} and \cite{Aaboud:2016okv}; the solid, dashed and dot-dashed green lines from $H \to hh$~\cite{Khachatryan:2015yea}, \cite{Khachatryan:2016sey} and \cite{ATLAS:2016ixk}; the solid and dashed blue lines from $H \to hZ$~\cite{Aad:2015wra} and \cite{TheATLAScollaboration:2016loc}. In this plot we show also the direct charged scalar search limits from $pp \to H^\pm X$~\cite{Khachatryan:2015qxa} (dashed purple). Right: The 95\% CL uncertainties of the differential cross section ${\rm d} \sigma / {\rm d}M_{tt}$~\cite{Khachatryan:2016mnb}, which is used to constrain the (CPV) couplings of $H_{2,3}$ to the top quark. See text for more details.}
  \label{fig:LHC}
\end{figure*}

To constrain the CPV 2HDM, we collect all the current most stringent direct search limits in these different decay channels in Fig.~\ref{fig:LHC} at both $\sqrt{s} = 8$ TeV and 13 TeV. The degenerate heavy scalars are produced predominantly from gluon fusion, as in the most general 2HDM scenarios. In the left panel of Fig.~\ref{fig:LHC}, the red, green and blue lines stand respectively for the limits in the final states of $WW/ZZ$, $hh$ and $hZ$. We do not show the limits beyond $1$ TeV, as in CPV 2HDM with quasi-degenerate $H_{2,3}$ the masses range $M_H \gtrsim 1$ TeV is excluded, or at least highly disfavored, by the stringent theoretical bounds in Section~\ref{sec:theoretical} on the quartic couplings $\lambda_i$ (cf. the limits in Figs.~\ref{fig:sensitivity1} to \ref{fig:sensitivity7}).

To impose the current LHC constraints on the cross sections
\begin{eqnarray}
&&\sigma (pp\to H_{2,3} \to XX)  \nonumber \\
&& = \sum_{i=2,3} \sigma (gg\to H_i) \times {\rm Br} (H_i \to XX) \,,
\end{eqnarray}
we consider for simplicity the leading order production of heavy scalars from gluon fusion by rescaling the production rate for a SM-like Higgs
\begin{eqnarray}
\frac{\sigma (gg \to H_i)}{\sigma (gg \to h_{\rm SM})} =
\frac{
\left| \sum_q c_{q,i} A_{1/2}^{H} (\tau_q) \right|^2 +
\left| \sum_q \tilde{c}_{q,i} A_{1/2}^{A} (\tau_q) \right|^2  }{
\left| \sum_q A_{1/2}^{H} (\tau_q) \right|^2} \nonumber \\
\end{eqnarray}
with $\tau_q = M_{H_i}^2 / 4m_q^2$, and then evaluate the leading order ${\rm BR} (H_i \to XX)$ from the partial decay widths in Eqs.~(\ref{eqn:Gammaff}) to (\ref{eqn:Gammagg}). As mentioned in Section~\ref{sec:degenerate}, we scan the parameter space of CPV 2HDM, by changing the parameters $M_H$, $\tan\beta$, $\alpha_b$ (or $\alpha_c$), with the constraint in Eq.~(\ref{eqn:relation}) taken into consideration and $\Delta M_H = 1$ GeV or 10 GeV. All the 95\% CL limits from direct LHC searches in the $WW/ZZ$, $hh$ and $hZ$ channels are presented, respectively, as the red, green and blue shaded regions in Figs.~\ref{fig:limits1} to \ref{fig:limits5}, as functions of the heavy scalar mass $M_H$ and $\tan\beta$, $\alpha_c$ or $\alpha_b$. See Section~\ref{sec:limits} for the details.

\subsubsection{Differential $t\bar{t}$ cross section}

The heavy neutral scalars $H_{2,3}$ of 2HDM couple to the SM fermions, even in the CP conservation limit of $\alpha_{b,c}=0$. As aforementioned and exemplified in Fig.~\ref{fig:BR}, $H_{2,3}$ decay predominantly into the top quark pairs. There have been dedicated searches of the (pseudo)scalars $H/A \to t\bar{t}$ in 2HDM performed by ATLAS, with the interference terms taken into consideration. However, only two specific scalar masses are considered: $M_{H/A} = 500$ and 750 GeV~\cite{ATLAS:2016pyq}. To constrain the CPV 2HDM in a more general sense, we resort to the differential cross section measurements with respect to the invariant mass of the two top jets ${\rm d} \sigma / {\rm d} M_{t \bar{t}}$~\cite{Khachatryan:2016mnb}. The 2HDM processes $gg \to H_{2,3} \to t\bar{t}$ arise at one-loop level through the top quark mediated $H_i gg$ loop, and interfere with the tree-level SM background $gg \to t\bar{t}$. The invariant mass $M_{t\bar t}$ could likely be distorted, depending on $\tan\beta$, the scalar mass $M_{H}$, the mass splitting $\Delta M_H$ and the mixing parameters. The consistency of experimental data and theoretical predictions imposes stringent constraints on the couplings of $H_{2,3}$ to the top quark, which is largely complementary to the direct searches of $H_{2,3}$ in the $h$, $W$ and $Z$ bosonic final states.

The parton-level analytical expressions for $H_{2,3} \to t \bar{t}$ in CPV 2HDM are to some extent similar to that for diphoton channel in Section~\ref{sec:diphoton}, with the amplitudes $H_{i} \gamma\gamma$ amplitudes replaced by those for $H_{i} t \bar{t}$. As in the diphoton case, the resonance signal $gg \to H_{2,3} \to t\bar{t}$ interferes with the tree-level SM background $gg \to t\bar{t}$. The explicit formulas for the resonance and interference terms can be found, e.g., in Ref.~\cite{Carena:2016npr}; for the sake of completeness, we collect the differential cross sections ${\rm d} \sigma / {\rm d} M_{t \bar{t}}$ in Appendix~\ref{sec:ttbar}. The 95\% CL experimental uncertainties $\Delta ({\rm d} \sigma / {\rm d} M_{t \bar{t}})$ at $\sqrt{s} = 13$ TeV are presented in the right panel of Fig.~\ref{fig:LHC}, which is dominated by the systematic and statistical errors of the experimental data~\cite{Khachatryan:2016mnb}. To constrain the beyond SM CP conserving and violating couplings, in particular those to the top quark, we evaluate the differential cross sections ${\rm d} \sigma / {\rm d} M_{t \bar{t}}$ in the CPV 2HDM, as functions of the scalar masses and mixing parameters, and compare them to the experimental limits given in Fig.~\ref{fig:LHC} by requiring that the integrated cross sections in these seven bins from $M_{tt} = 300$ GeV to 1100 GeV are {\it all} smaller than the experimental uncertainties. The excluded regions in the parameter space of $M_H$, $\tan\beta$ and $\alpha_{b,c}$ are presented in Figs.~\ref{fig:limits1} to \ref{fig:limits5} as the pink lines, and in Figs.~\ref{fig:sensitivity1} to \ref{fig:sensitivity7} as the shaded pink regions.

\subsubsection{Limits from the charged scalar sector}
\label{sec:chargedscalar}

The scalar mass spectrum of CPV 2HDM and the mixing angles are subject to the EW precision tests. In particular, with the two neutral scalar $H_{2,3}$ almost degenerate, the mass splitting $|M_{H} - M_{H^\pm}|$ of heavy neutral and charged scalars can not be arbitrarily large, which is tightly constrained by the oblique parameters. Therefore, all the mass limits on the charged scalar $H^\pm$ can be ``transferred'' to the neutral scalars of 2HDM, no matter where these limits are from. These limits from the charged scalar sector can be, in some sense, considered as ``indirect'' limits on the neutral scalars in the framework of 2HDM, and might be dramatically changed when the scalar sector is altered, e.g. more scalar singlet(s) and/or multiplet(s) are introduced. When ``transferred'' from the charged scalar sector to the neutral scalar sector, the mass limits would be weakened by the magnitude of ${\cal O} (100 \, {\rm GeV})$, which is dictated by the $S$ and $T$ parameters, and ultimately determined by the mass and mixing parameters in CPV 2HDM. In the case of $\beta-\alpha=\pi/2$ with $\alpha_{b,c}\neq 0$, the expressions for $S$ and $T$ can be significantly simplified and are collected in Appendix~\ref{sec:oblique}~\cite{Chen:2015gaa}. As the oblique parameter $T$ is much more sensitive to the mass splitting $|M_{H} - M_{\pm}|$ than $S$, in the numerical calculations we will consider for simplicity only the constraints from the current global EW fit of $T$~\cite{Baak:2014ora}:
\begin{eqnarray}
T = 0.09 \pm 0.13 \,.
\end{eqnarray}

\begin{figure}[!t]
  \centering
  \includegraphics[width=0.4\textwidth]{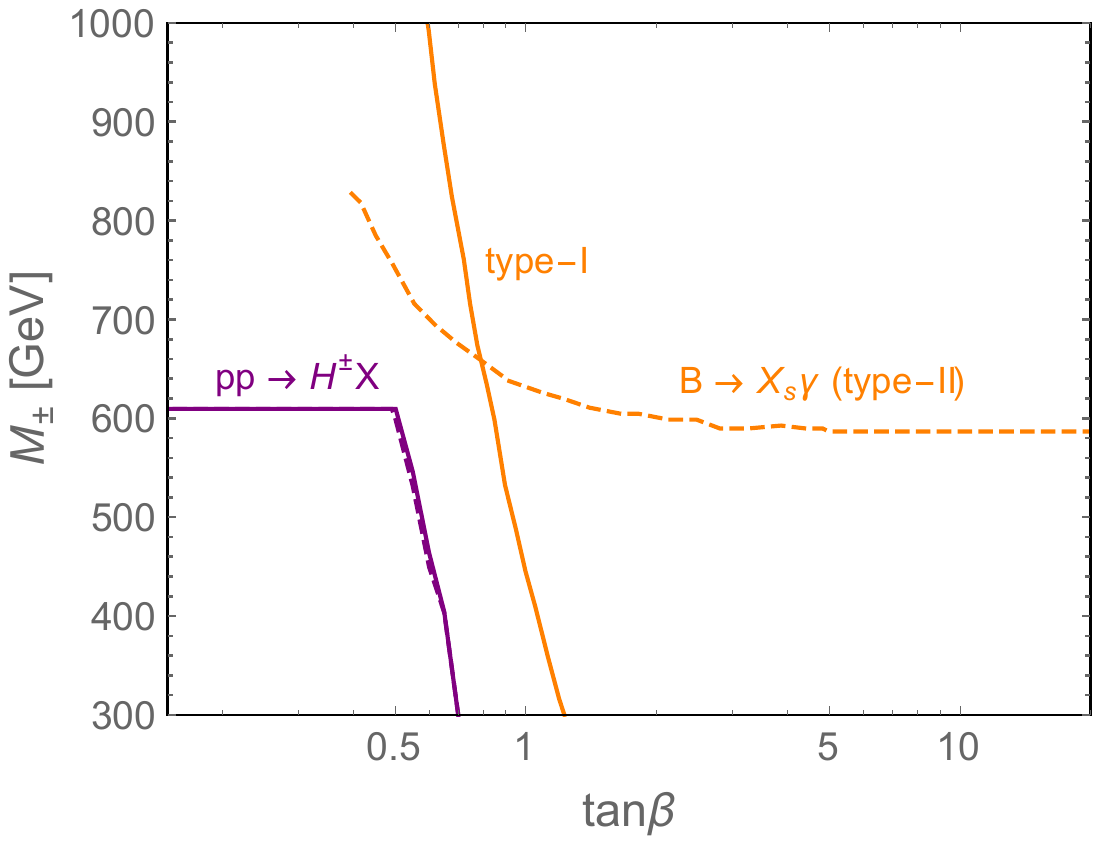}
  \caption{Lower limits on the mass of charged scalar $H^\pm$ in 2HDM, as functions of $\tan\beta$, from the direct searches at the LHC~\cite{Khachatryan:2015qxa} and the observations of rare decay $B \to X_s \gamma$~\cite{Misiak:2017bgg}. The solid and dashed lines are respectively for the type-I and type-II models. These limits could be used to constrain the heavy quasi-degenerate neutral scalars $H_{2,3}$ when combined with the experimental limits on oblique parameter $T$~\cite{Baak:2014ora}.}
  \label{fig:Hpm}
\end{figure}

On the experimental side, charged scalars have been searched at the LHC in associate production with a top quark (and a bottom quark), i.e. $pp \to H^\pm X$, with the subsequent decay of $H^\pm \to tb, \, \tau\nu$~\cite{Khachatryan:2015qxa,
Aad:2015typ, Aaboud:2016dig}. In the 2HDM, the charged scalar $H^\pm$ decays mostly into the top-bottom quarks, with the coupling strength depending on $\tan\beta$ and whether it is of type-I or type-II. The current most stringent limits on the cross section $\sigma(pp \to H^\pm)$ is from Ref.~\cite{Khachatryan:2015qxa} and shown in the left panel Fig.~\ref{fig:LHC} as the purple line, and the lower limit on $M_{\pm}$ is presented in Fig.~\ref{fig:Hpm}, for both the type-I and type-II models, as function of $\tan\beta$. In obtaining the $M_\pm$ mass limits, we follow the leading order parton-level cross section $\sigma (bg \to H^- t)$ in Ref.~\cite{Borzumati:1999th}, multiply a factor of 1.5 to account for the subleading processes~\cite{Djouadi:2005gj}, with the Yukawa couplings given in Eq.~(\ref{eq:2HDM_Yuk2}).

With couplings to the SM fermions, the charged scalar $H^\pm$ in 2HDM contribute significantly to some rare flavor-changing decay processes which is highly suppressed in the SM. With $\sim 10^9$ $B$ mesons collected at Belle~\cite{Belle:2016ufb}, the partial width of the radiative decay $B \to X_s \gamma$ is precisely measured, imposing severe constraints on the charged scalar $H^\pm$ in 2HDM~\cite{Misiak:2015xwa,Misiak:2017bgg}. The extra contributions of $H^\pm$ to the rare $B$ decays depend on the Yukawa couplings, i.e. whether they are type-I or type-II, and also on $\tan\beta$, as shown in Fig.~\ref{fig:Hpm}. There are also limits on the charged scalar $H^\pm$ from other flavor observations such as $\Delta m_B$ and $\epsilon_K$, but these are expected to be weaker and are not considered here~\cite{Branco:2011iw}.
In the leptonic sector, there are also limits on the charged scalar $H^\pm$ from the anomalous magnetic moment of muon $(g-2)_\mu$, see e.g.~\cite{Chen:2015gaa}, which however is much weaker, and will be neglected in this work. To apply the direct search and $B$ decay limits on $H^\pm$ in Fig.~\ref{fig:Hpm} to the neutral scalars $H_{2,3}$, we adopt the formula for $\Delta T$ given in Eq.~(\ref{eqn:oblique2}), with the limits presented in Figs.~\ref{fig:limits1} to \ref{fig:limits5}.

\subsubsection{Constraining CPV 2HDM}
\label{sec:limits}

\begin{figure*}[!t]
  \centering
  \includegraphics[width=0.4\textwidth]{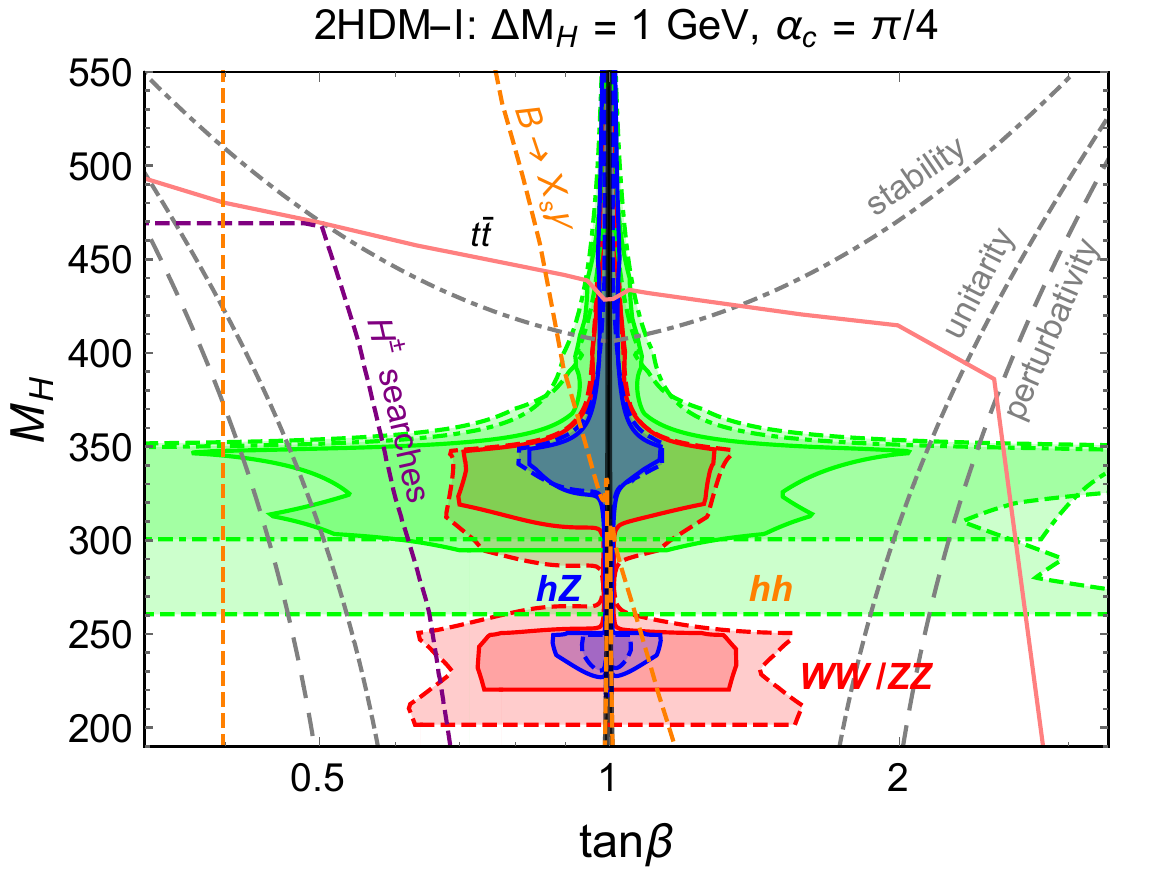} \hspace{-7.5pt}
  \includegraphics[width=0.4\textwidth]{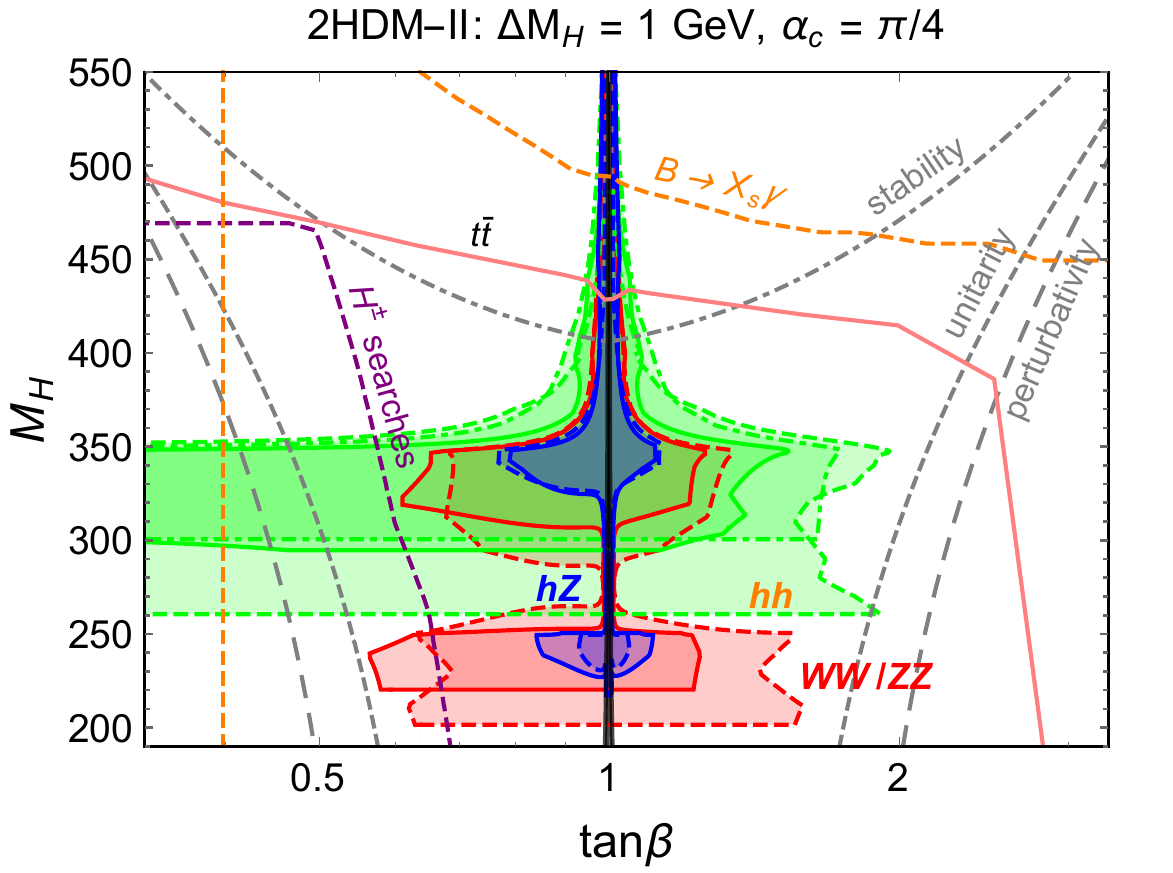} \\
  \includegraphics[width=0.4\textwidth]{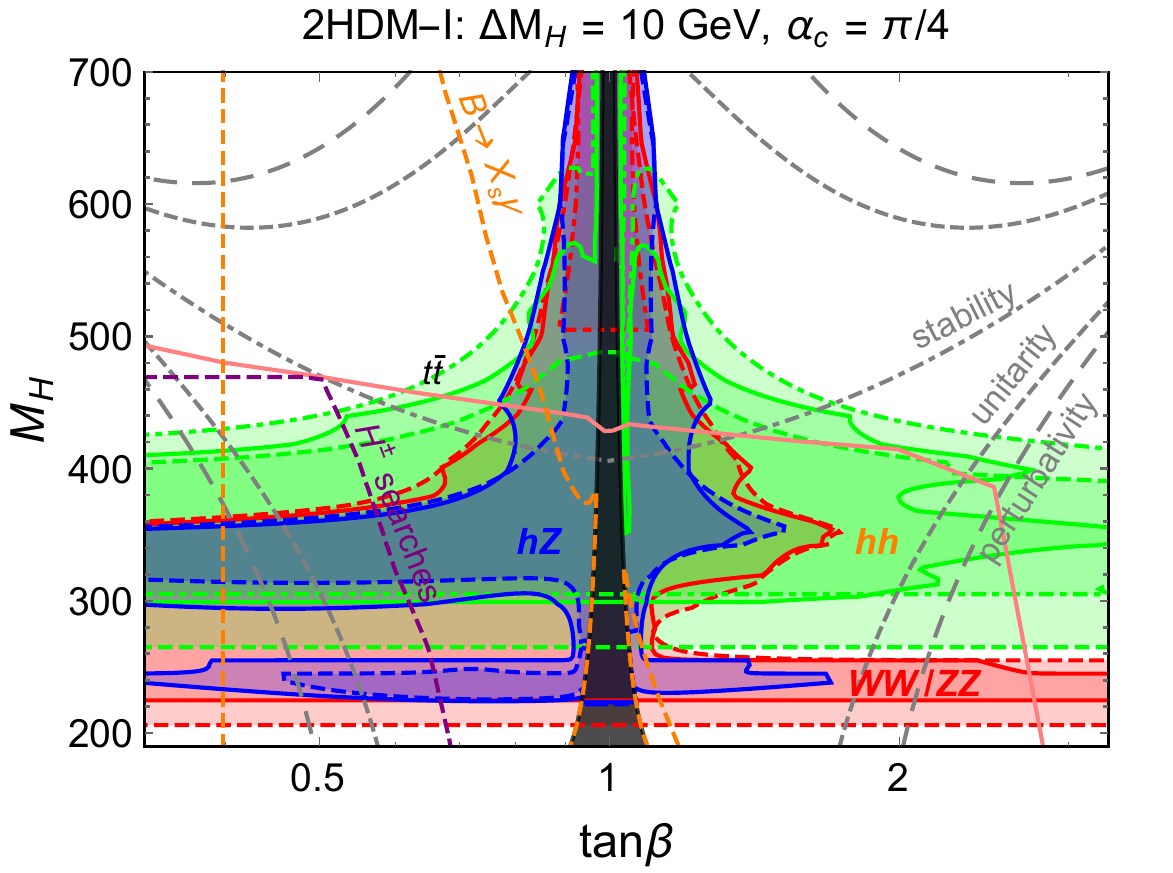} \hspace{-7.5pt}
  \includegraphics[width=0.4\textwidth]{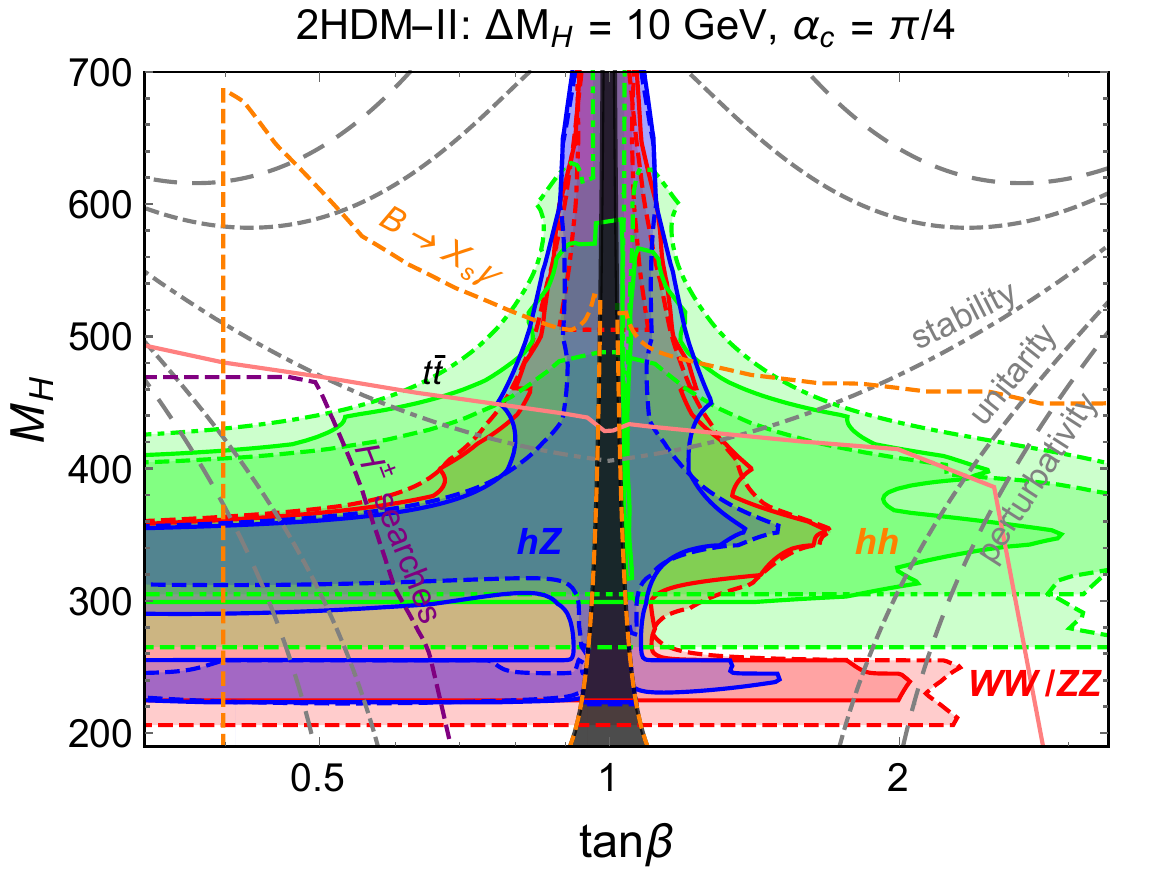}
  \caption{Experimental limits on the CPV 2HDM of type-I (left) and type-II (right) with maximal CP violation $\alpha_c = \pi/4$ and two quasi-degenerate scalars $H_{2,3}$ with a small mass splitting of $\Delta M_H = 1$ GeV (upper) or $10$ GeV (lower). The red, green and blue shaded regions are respectively from the direct searches of heavy neutral scalars in the final states of $WW/ZZ$, $hh$ and $hZ$ collected in Fig.~\ref{fig:LHC}~\cite{Aad:2015kna, Khachatryan:2015cwa, Aaboud:2016okv, Khachatryan:2015yea, Khachatryan:2016sey, ATLAS:2016ixk, Aad:2015wra, TheATLAScollaboration:2016loc}, with the same line legends as in that figure, e.g. the solid red line represents the limits from Ref.~\cite{Aad:2015kna}. In this figure we also show the limits from the uncertainties of differential cross section ${\rm d}\sigma / {\rm d} M_{tt}$ at the parton-level~\cite{Khachatryan:2016mnb} (solid pink), the limits from the direct searches of charged scalars at the LHC~\cite{Khachatryan:2015qxa} (dashed purple) and precise measurements of $B \to X_s \gamma$~\cite{Misiak:2017bgg} (dashed orange). The short-dashed, long-dashed and dot-dashed gray lines are respectively form the limits of theoretical arguments of unitarity, perturbativity and stability of the scalar potential. All the regions below the unshaded lines (and the regions above the upper short and dashed gray lines in the two lower panels) are excluded. See text for more details.}
  \label{fig:limits1}
\end{figure*}

\begin{figure*}[!t]
  \centering
  \includegraphics[width=0.4\textwidth]{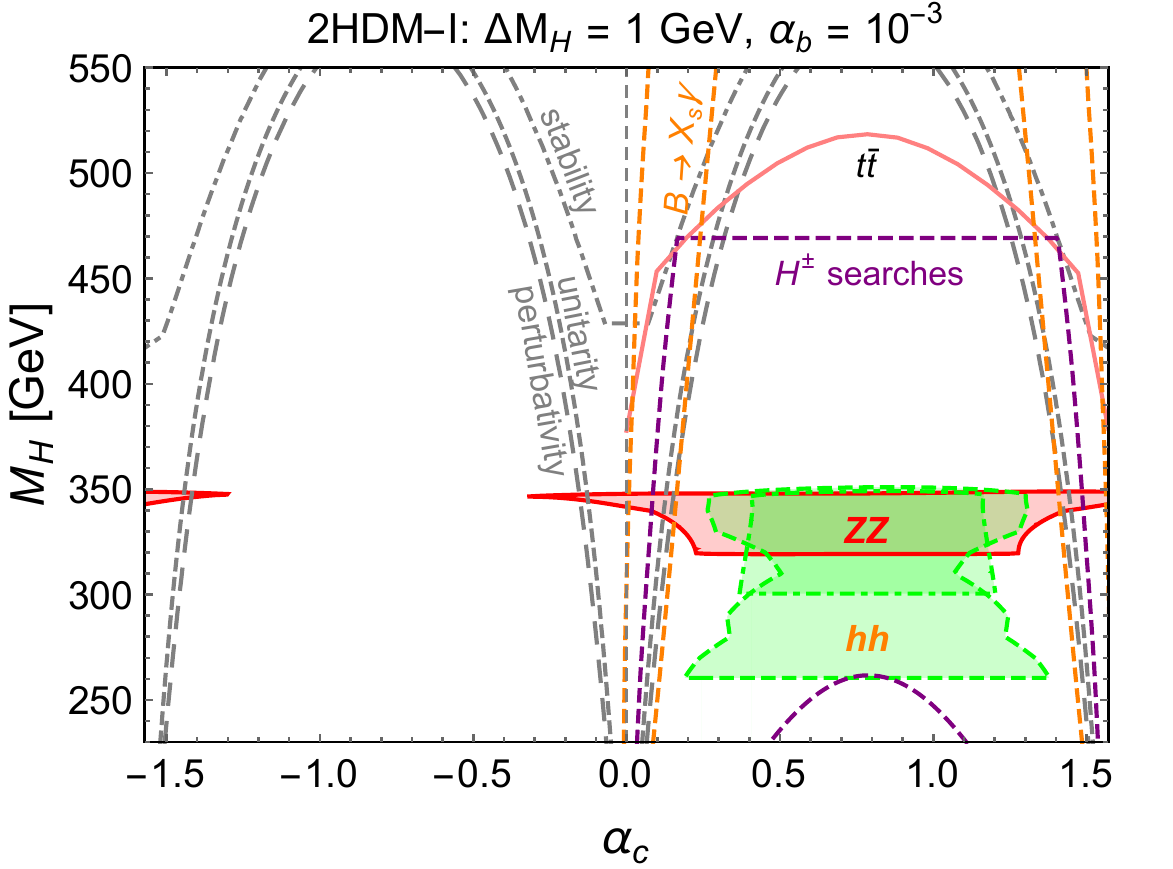} \hspace{-7.5pt}
  \includegraphics[width=0.4\textwidth]{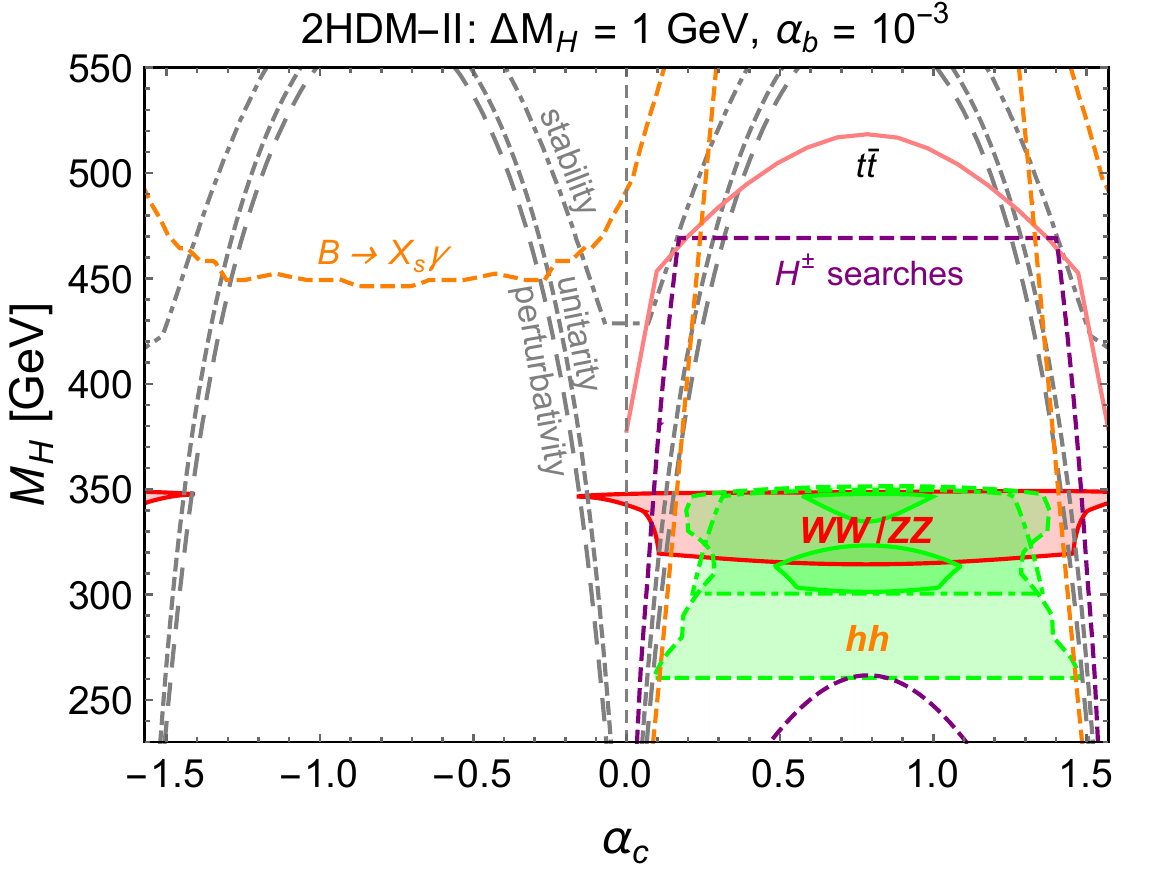}
  \caption{The same as in Fig.~\ref{fig:limits1} in the $M_H - \alpha_c$ plane, with $\alpha_b = 10^{-3}$ and $\Delta M_H = 1$ GeV. The scenarios with larger splitting $\Delta M_H = 10$ GeV is excluded by the theoretical arguments of perturbativity, unitarity and stability.}
  \label{fig:limits2}
\end{figure*}

\begin{figure*}[!t]
  \centering
  \includegraphics[width=0.4\textwidth]{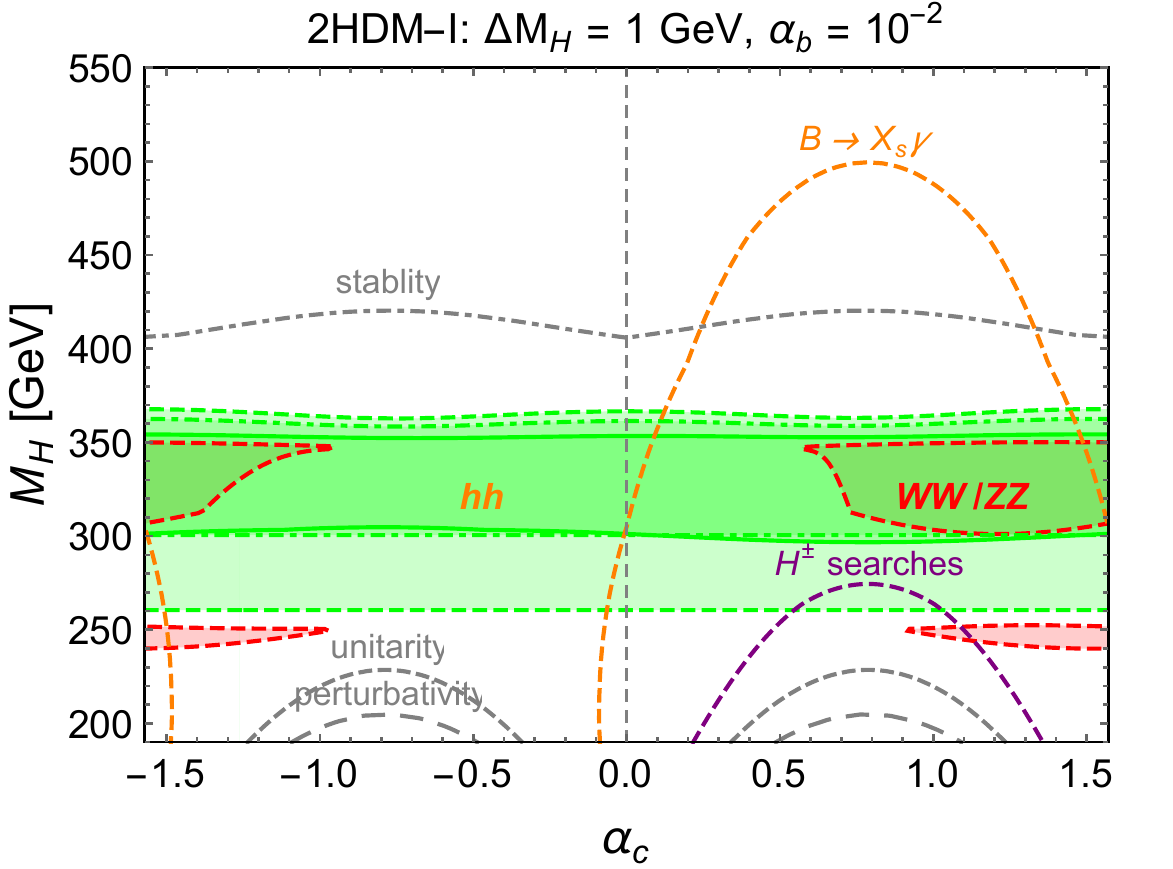} \hspace{-7.5pt}
  \includegraphics[width=0.4\textwidth]{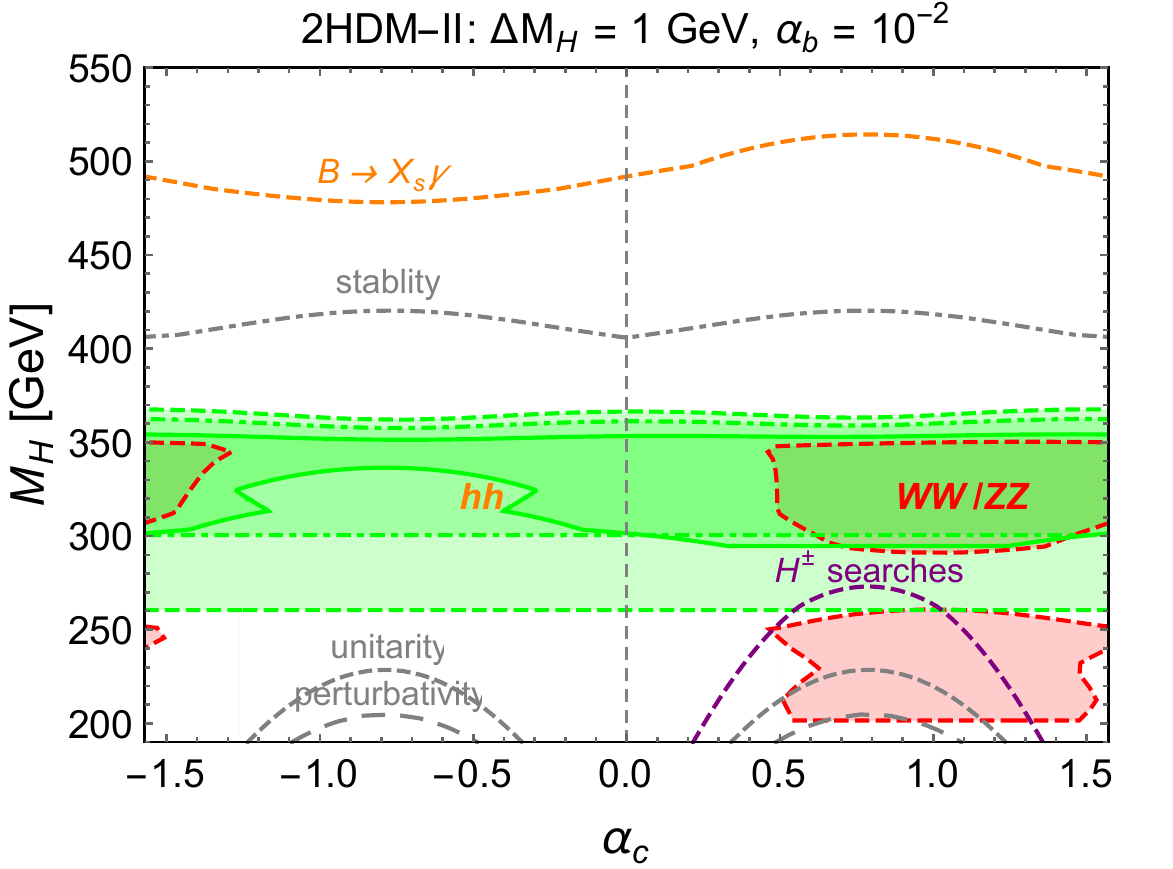} \\
  \includegraphics[width=0.4\textwidth]{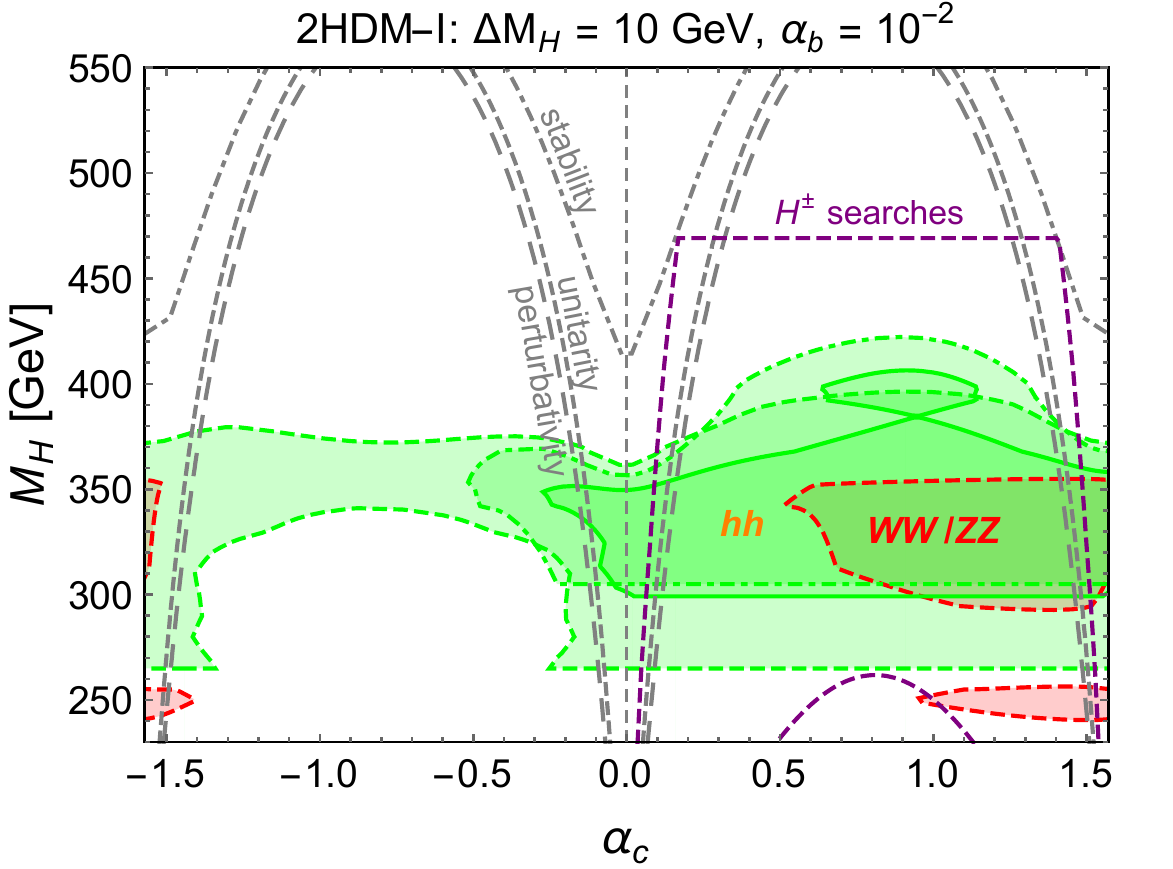} \hspace{-7.5pt}
  \includegraphics[width=0.4\textwidth]{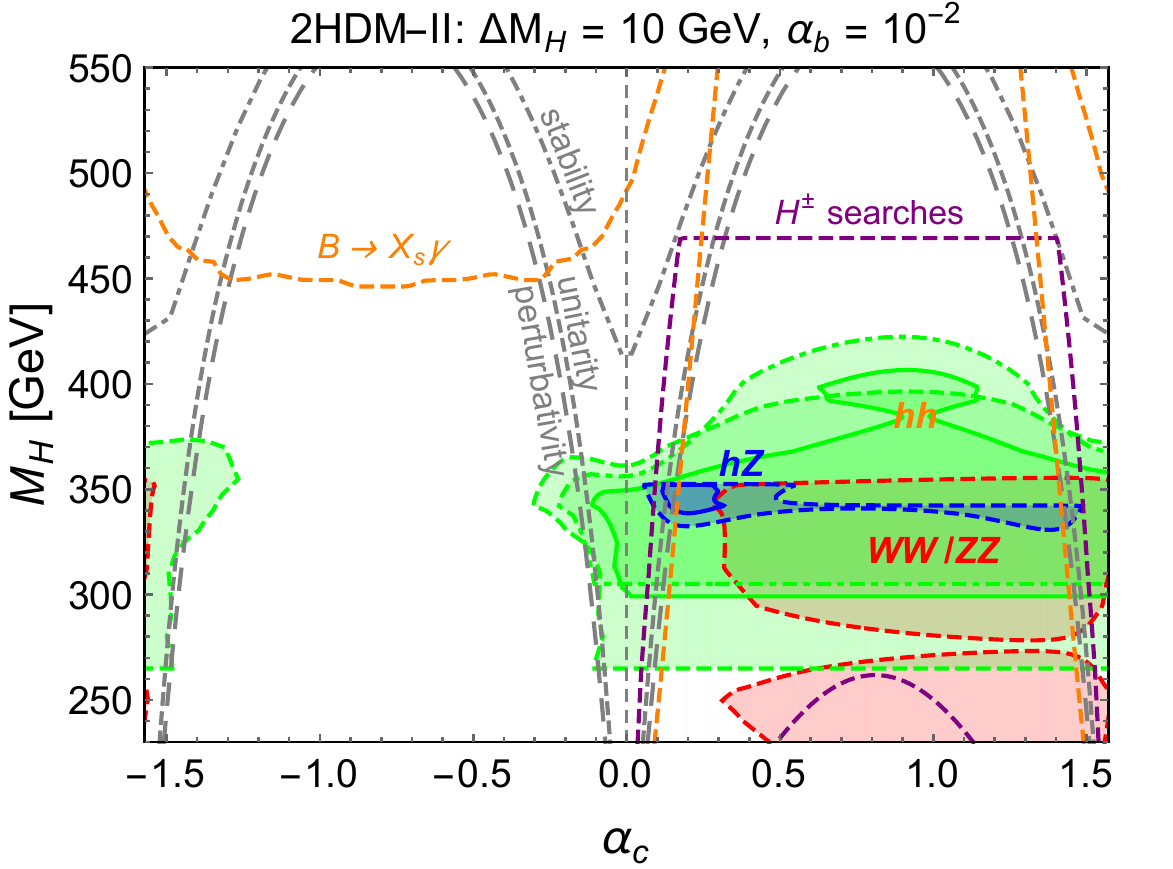}
  \caption{The same as in Fig.~\ref{fig:limits2} with $\alpha_b = 10^{-2}$, $\Delta M_H = 1$ GeV (upper) and 10 GeV (lower).}
  \label{fig:limits3}
\end{figure*}

\begin{figure*}[!t]
  \centering
  \includegraphics[width=0.4\textwidth]{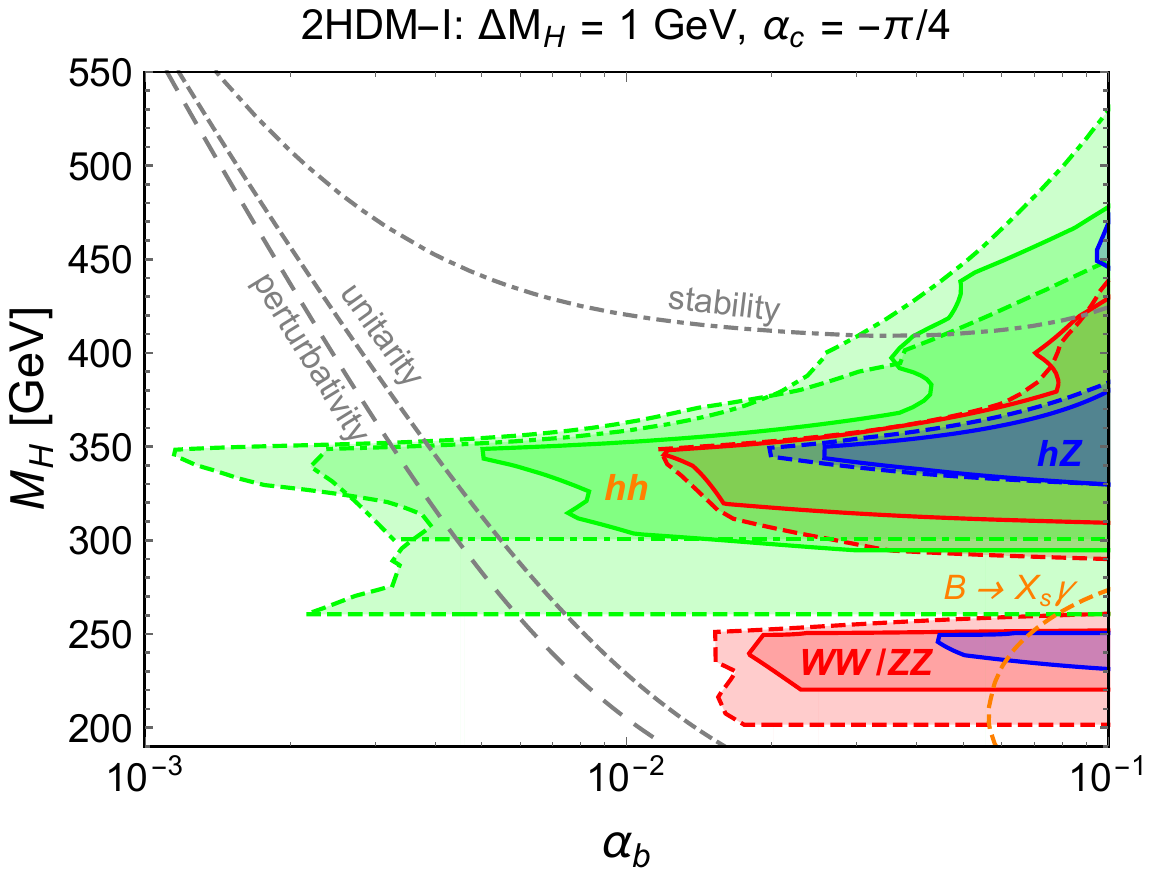} \hspace{-7.5pt}
  \includegraphics[width=0.4\textwidth]{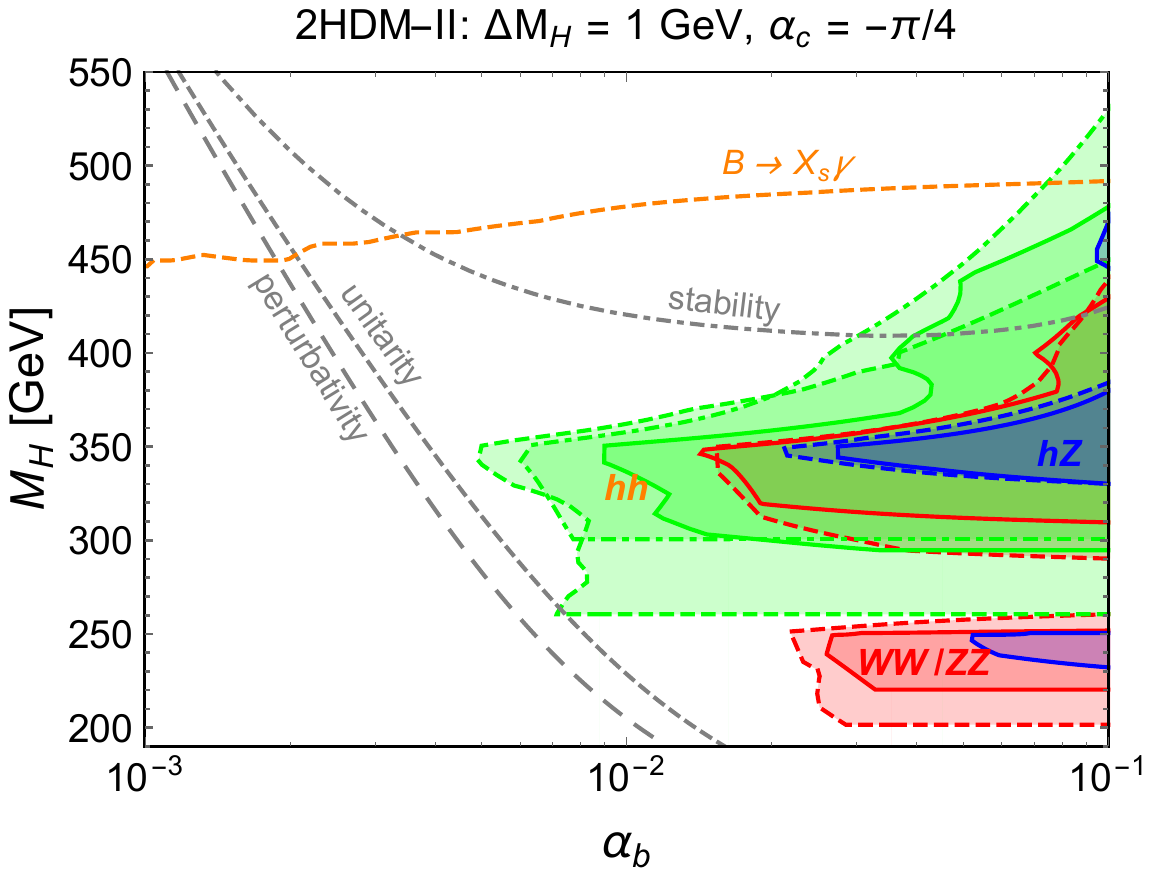} \\
  \includegraphics[width=0.4\textwidth]{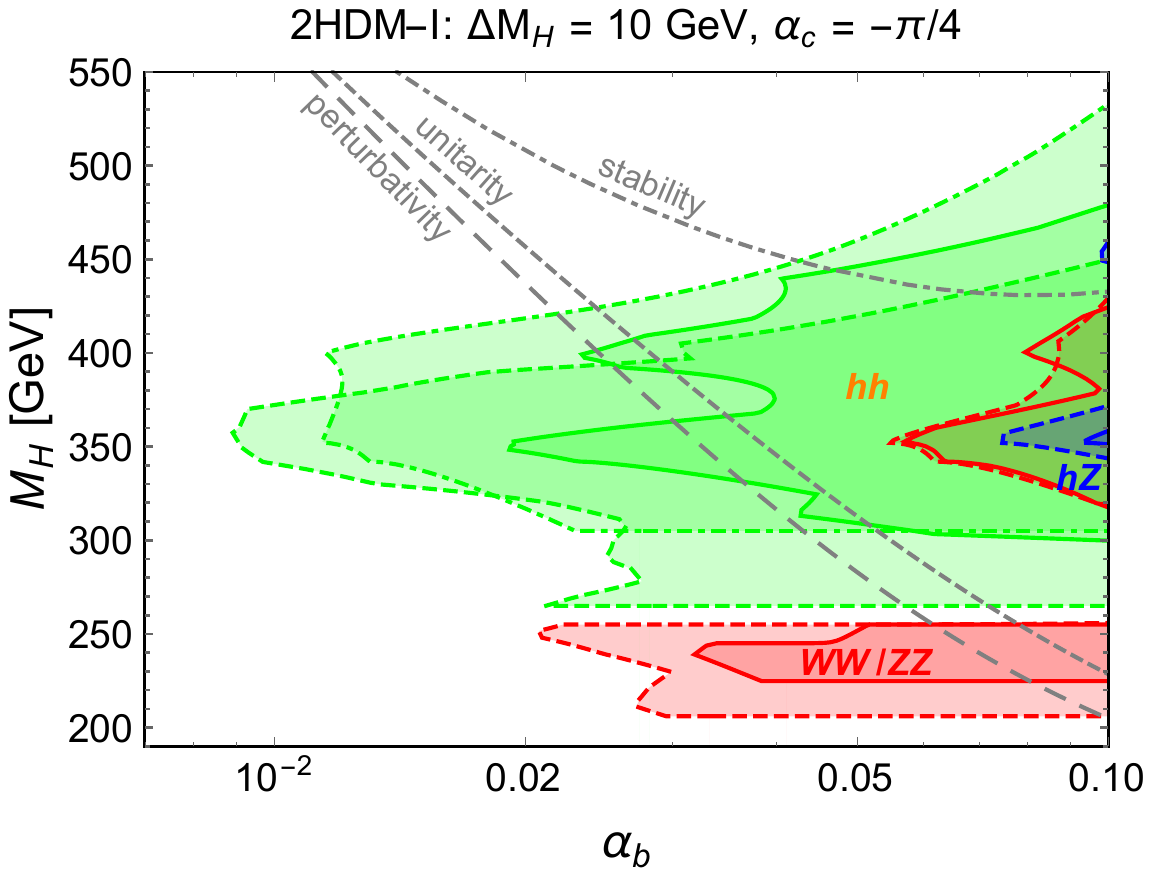} \hspace{-7.5pt}
  \includegraphics[width=0.4\textwidth]{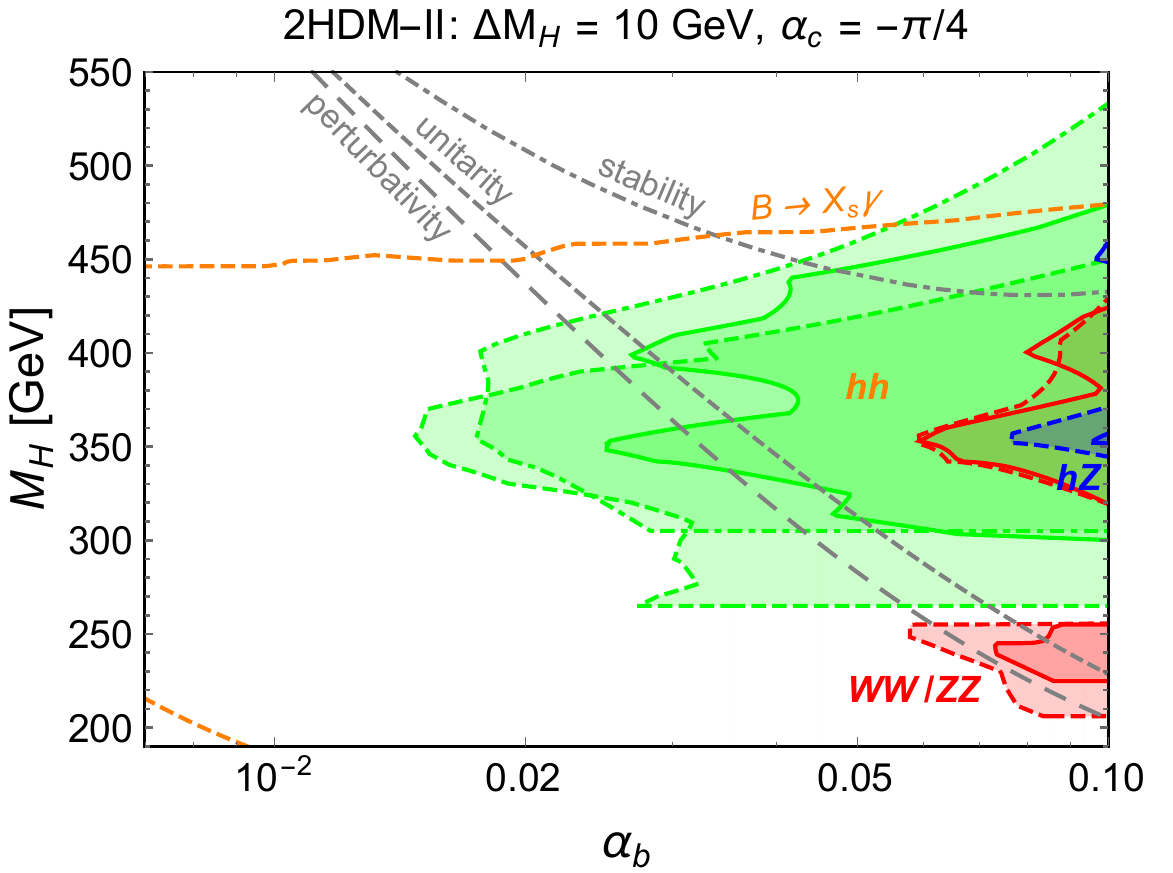}
  \caption{The same as in Fig.~\ref{fig:limits1} in the $M_H - \alpha_b$ plane, with $\alpha_c = -\pi/4$, $\Delta M_H = 1$ GeV (upper) and 10 GeV (lower). }
  \label{fig:limits4}
\end{figure*}

\begin{figure*}[!t]
  \centering
  \includegraphics[width=0.4\textwidth]{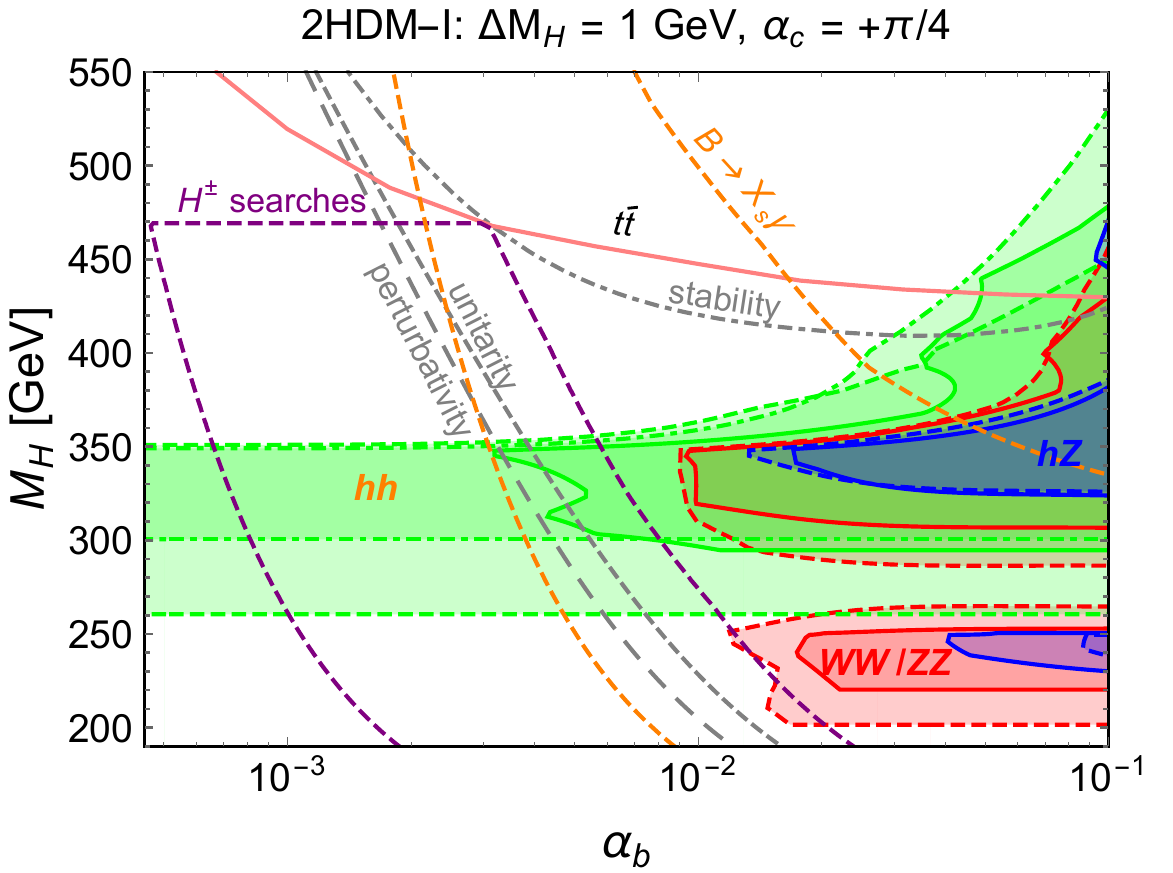} \hspace{-7.5pt}
  \includegraphics[width=0.4\textwidth]{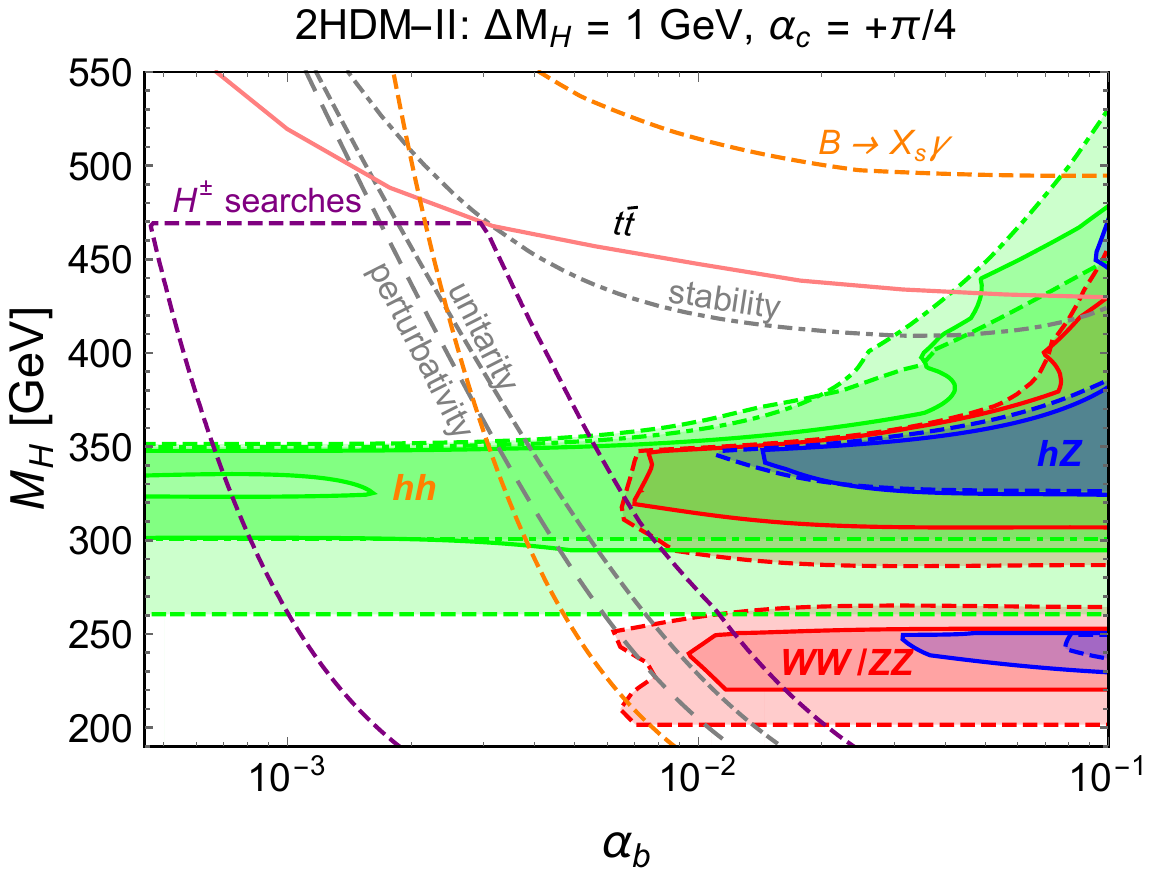} \\
  \includegraphics[width=0.4\textwidth]{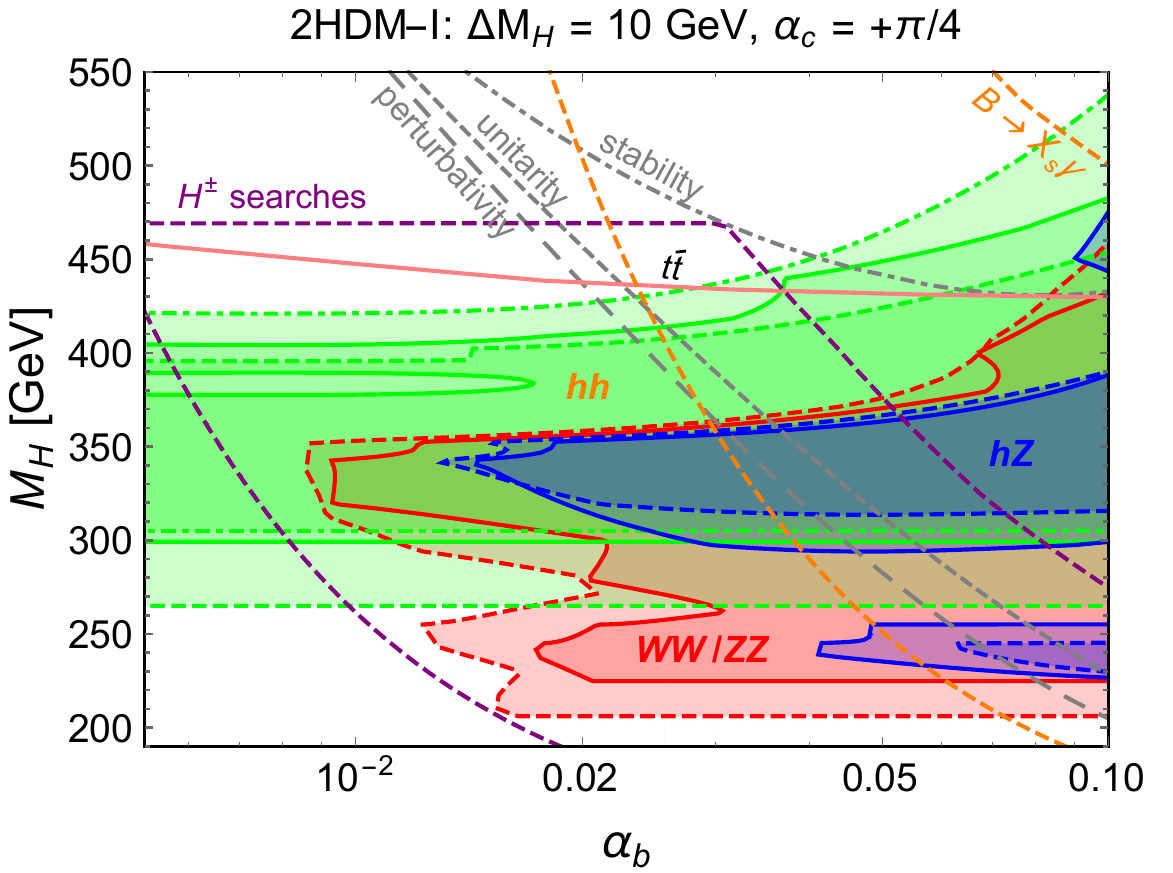} \hspace{-7.5pt}
  \includegraphics[width=0.4\textwidth]{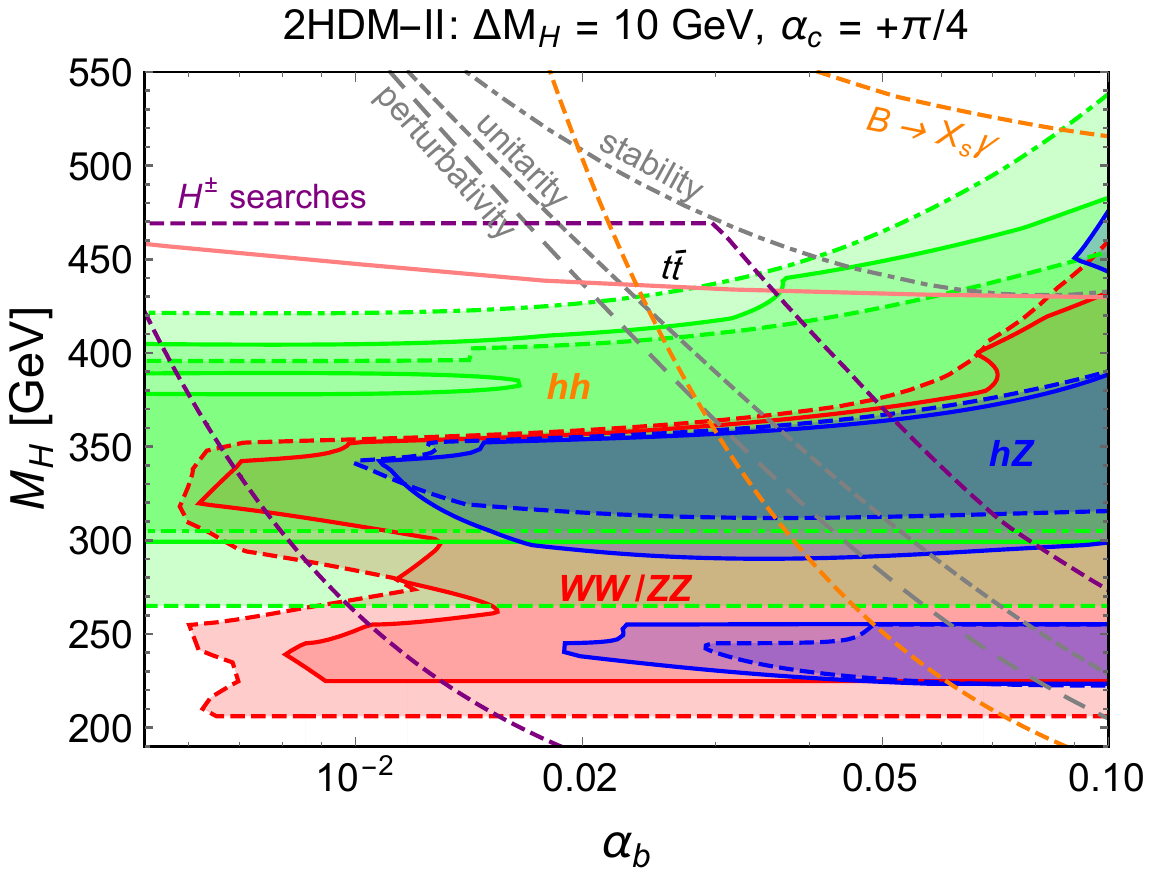}
  \caption{The same as in Fig.~\ref{fig:limits4} in the $M_H - \alpha_b$ plane, with $\alpha_c = + \pi/4$.}
  \label{fig:limits5}
\end{figure*}

All the direct search limits of $H_{2,3}$ in the final states of $WW/ZZ$, $hh$ and $hZ$, the constraints from the differential $t\bar{t}$ cross sections, and the limits from the charged scalar $H^\pm$ (direct search of $H^\pm$ at the LHC and the constraints from $B \to X_s \gamma$) are collected in Figs.~\ref{fig:limits1} to \ref{fig:limits5}, in the two-dimensional planes of $M_H - \tan\beta$, $M_H - \alpha_c$ and $M_H - \alpha_b$. In these plots, the legends are the same: the shaded regions are all excluded by the direct searches of heavy neutral scalars, with the red, green and blue colors stand respectively for the limits in the final states of $WW/ZZ$, $hh$ and $hZ$, using the same line legends (solid, dashed or dot-dashed) as in Fig.~\ref{fig:LHC}. The limits from $t\bar{t}$ data are depicted in pink, while the constraints from direct $H^\pm$ searches and $B \to X_s \gamma$ in dashed purple and orange. All the experimental limits are at the 95\% CL. The theoretical limits from perturbativity, unitarity and stability are labelled as the gray lines. All the regions below these colorful and gray lines are excluded.  The electron and mercury EDM limits are more relevant to heavier $H_{2,3}$, and not shown in these plots but presented in Figs.~\ref{fig:sensitivity1} to \ref{fig:sensitivity7}.

As mentioned and exemplified in Section~\ref{sec:decay}, the BRs of $H_{2,3}$ into the SM $WW/ZZ$, $hh$ and $hZ$ bosons are generally very small when the top quark channel is kinematically allowed, and these direct search data could exclude some regions where the scalars are not too heavy, i.e. $M_H \lesssim 450$ GeV, in general less constraining than the ``indirect'' limits from $t\bar{t}$, $H^\pm$ direct searches, rare $B$ decay and EDM data. The direct searches limits from $WW/ZZ$, $hh$ and $hZ$ are collectively depicted in yellow in Figs.~\ref{fig:sensitivity2}, \ref{fig:sensitivity4}, \ref{fig:sensitivity5} and \ref{fig:sensitivity6}. The readers who are more interested in the diphoton prospects at the LHC and the EDM limits can skip all the following details in this subsection.

We first demonstrate the important collider limits on the heavy neutral scalars $H_{2,3}$ in CPV 2HDM in the $M_H - \tan\beta$ plane. One should note that in the CP conserving limit of $\alpha_{b,c} = 0$, the decay modes $H_{2,3} \to WW/ZZ,\, hh,\, hZ$ are all highly suppressed,
and we do not have any limits on $H_{2,3}$ from the direct searches at the LHC. However, the limits from differential $t\bar{t}$ data are still there, as the scalar $H_{2,3}$ both couple to the SM fermions, no matter how the mixing changes. In addition, the oblique parameter $T$ does not vanish even in the limit of $\alpha_{b,c} = 0$ (cf.~Eq.~(\ref{eqn:oblique2})), which render limits on the neutral scalars $H_{2,3}$ from the $H^\pm$ searches and $B \to X_s \gamma$ data. These limits from the $t\bar{t}$ data, $B$ decay data and the direct searches of $H^\pm$ in the CP conserving limit of $\alpha_{b,c} = 0$ can be found in Fig.~\ref{fig:sensitivity1} where we also show the diphoton prospects.

The collider limits on $H_{2,3}$ with the maximal $\alpha_c = \pi/4$ in the $M_H - \tan\beta$ space are presented in Fig.~\ref{fig:limits1}, for both the type-I and type-II Yukawa couplings.
The unphysical regions are painted in black, within which we can not find real solutions for the mixing angles $\alpha_{b,c}$ in Eq.~(\ref{eqn:relation}). As mentioned in Section~\ref{sec:direct}, the scalars $H_{2,3}$ are produced predominantly from gluon fusion $gg \to H_{2,3}$. When the heavy scalar masses $M_H \simeq 2 m_t \simeq 350$ GeV,  we have a resonance-like effect for the direct search limits, due to the enhanced top loop amplitude in the production process, therefore excluding broader regions. Comparing the upper and lower panels in Fig.~\ref{fig:limits1} with respectively $\Delta M_H = 1$ GeV and 10 GeV, a larger mass splitting $\Delta M_H$ pushes the mixing $\alpha_b$ and the ${\rm BR} (H_{2,3} \to WW/ZZ,\, hh,\, hZ)$ larger (cf. the example given in Fig.~\ref{fig:BR}), then broader regions are excluded in the lower plots for all these bosonic decay modes. Note that in Fig.~\ref{fig:limits1}, the direct search data could exclude larger values of $M_H$ when
\begin{eqnarray}
|\tan\beta-1|\ll 1 \,,
\end{eqnarray}
as in the limit of $\tan\beta \to 1$, the CPV angle $\alpha_b$ is largely enhanced by $\tan2\beta$ in Eq.~(\ref{eqn:relation}), when other mass and mixing parameters fixed.

The collider limits projected into the $M_{H} - \alpha_c$ plane are collected in Figs.~\ref{fig:limits2} and \ref{fig:limits3}, with respectively the benchmarks values of $\alpha_b = 10^{-3}$ and $10^{-2}$. Note that with $\alpha_b = 10^{-3}$ and $\Delta M_H = 10$ GeV, the whole region in the $M_H - \alpha_c$ plane is excluded by the perturbativity, unitarity and stability limits, thus we have only the plots with a smaller splitting $\Delta M_H = 1$ GeV in Fig.~\ref{fig:limits2}. For fixed values of $\alpha_b$, a positive $\alpha_c >0$ leads to a solution $\tan\beta < 1$ via Eq.~(\ref{eqn:relation2}), and the limits from differential $t\bar{t}$ data are more stringent than the case with a negative $\alpha_c < 0$ for which $\tan\beta >1$ (see also the $t\bar{t}$ limit in Fig.~\ref{fig:limits1}). With a larger $\alpha_b = 10^{-2}$, for positive $\alpha_c$ the $\tan\beta$ in Fig.~\ref{fig:limits3} is larger and the couplings of $H_{2,3}$ to the top quark get smaller, then the $t\bar{t}$ limits are much weaker. Therefore the $t\bar{t}$ limits are not shown in Fig.~\ref{fig:limits3}. With the same reason, the $H^\pm$ direct search limits in Fig.~\ref{fig:limits3} are much weaker than those in Fig.~\ref{fig:limits2}, as the direct search data in Fig.~\ref{fig:Hpm} is effective only when $\tan\beta \lesssim 0.5$. We can also see the dependence of the theoretical limits on the mass and mixing parameters, in particular by comparing the upper and lower plots with different $\Delta M_H$ in Fig.~\ref{fig:limits3}. Anyway, the direct neutral scalar limits are well below the theoretical limits and $B \to X_s \gamma$ constraints in the two-dimension space of $M_H$ and $\alpha_c$, as long as the mass splitting is small.

In Figs.~\ref{fig:limits4} and \ref{fig:limits5}, we present the collider limits in the two-dimensional space of $M_H$ and $\alpha_b$, with respectively $\alpha_c = -\pi/4$ and $+ \pi/4$, and $\tan\beta$ determined via the relation in Eq.~(\ref{eqn:relation2}). In these plots we can clearly see the dependence of the $WW/ZZ$, $hh$ and $hZ$ limits on the mixing angle $\alpha_b$. The cross sections $\sigma (gg \to H_{2,3} \to WW/ZZ,\,hh,\,hZ)$ are, roughly, proportional to the mixing of the SM Higgs $h$ with the heavy scalars, therefore a large $\alpha_b$ excludes broader range of heavy scalar mass $M_H$. However, a large $\alpha_b$, say $\sim 0.1$, is excluded or highly disfavored by the EDM measurements; see Figs.~\ref{fig:sensitivity5} to \ref{fig:sensitivity7}. With $\alpha_c<0$ and $>0$, $\tan\beta$ is greater and smaller than one, respectively, in Figs.~\ref{fig:limits4} and \ref{fig:limits5}, therefore the limits from $t\bar{t}$ and $H^\pm$ direct searches at the LHC are much more stringent in the latter case, as just mentioned. As a direct consequence of $\tan\beta < 1$ and larger couplings of $H_{2,3}$ to the top quark, the $WW/ZZ$, $hh$ and $hZ$ data exclude larger regions in Fig.~\ref{fig:limits5} than in Fig.~\ref{fig:limits4}. As in Figs.~\ref{fig:limits1} to \ref{fig:limits3}, all the direct search limits in Figs.~\ref{fig:limits4} and \ref{fig:limits5} are below $M_H \lesssim 450$ in almost the whole parameter space, and less important than other limits, e.g. from EDM, when we are focusing on the diphoton searches in the degenerate limit.


\subsection{EDM constraints}
\label{sec:EDM}

With the  EDM of a fermion denoted by $d^E_f$ and the chromo-EDM (CEDM) of a quark by $d^C_q$, the relevant (C)EDM interaction Lagrangian is given by
\begin{eqnarray}
 \label{CEDM}
{\cal L}_{\rm
  (C)EDM}\ &=&
  -\frac{i}{2}\,d^E_f\,F^{\mu\nu}\,\bar{f}\,\sigma_{\mu\nu}\gamma_5\,f \nonumber \\
&& - \frac{i}{2}\,d_q^C\,G^{a\,\mu\nu}\,\bar{q}\,\sigma_{\mu\nu}\gamma_5 T^a q\,,
\end{eqnarray}
with $F^{\mu\nu}$ and $G^{a\,\mu\nu}$ the electromagnetic and strong  field strengths,  and $T^a=\lambda^a/2$ the generators of  the  SU(3)$_C$ group. The  gluonic  dimension-six Weinberg operator is described by the interaction Lagrangian:
\begin{equation}
{\cal L}_{\rm Weinberg}\ =\
\frac{1}{6}\,d^{\,G}\,f_{abc}\,\epsilon^{\mu\nu\lambda\sigma}G^a_{\rho\mu}\,
G_{\lambda\sigma},
{G^c}_{\nu}^{~~\rho}\; ,
\end{equation}
  In  the CPV 2HDM, the
Weinberg operator $d^{\,G}$ is the neutral Higgs
contribution~\cite{Weinberg:1989dx,Dicus:1989va}
\begin{equation}
  \label{dG}
d^{\,G}\ =\: (d^{\,G})^{H} \,.
\end{equation}

In the CPV 2HDM, the CP-odd electron-nucleon interactions $C_S$ coming from the CP-odd four-fermion interactions
\begin{eqnarray}
\label{CS}
{\cal L}_{C_S}\ =\ C_S^{4f} \,\bar{e}i\gamma_5e\,\bar{N}N\,
\end{eqnarray}
with
\begin{eqnarray}
(C_S)^{4f}&=&\frac{(29 {\rm MeV}) g_d g_e c_d \tilde{c}_e}{m_d M_H^2} +\frac{ (220 {\rm MeV}) g_s g_e c_s \tilde{c_e}\kappa}{m_s M_H^2} \,, \nonumber
\end{eqnarray}
with $g_f = m_f / v$ and $\kappa\approx 0.5\pm0.25$~\cite{Ellis:2008zy}.
then, it is appropriate to define an effective electron EDM entering paramagnetic system as~\cite{Chupp:2014gka},
\begin{eqnarray}
d^{\rm eff}_{\rm para}\approx d_e + \frac{\alpha_{C_S^{4f}}}{\alpha_{d_e}}C_S^{4f} \,.
\end{eqnarray}
the coefficients $\alpha_i$ are provided by atomic calculations~\cite{Dzuba,Jung:2013mg}. As the contributions to $C_S$ are mediated by the scalars and hence highly suppressed, the ACME results can be safely interpreted as an upper limit for the electron EDM $d_e^E$. Therefore, we will impose on 2HDM the latest eEDM constraint from the ACME collaboration of~\cite{Baron:2013eja},
\begin{eqnarray}
|d_e| < 8.7\times 10^{-29}\,e\cdot {\rm cm}\,.
\end{eqnarray}

The mercury EDM receives a dominant contribution from the nuclear Schiff moment $S$, which is generated by long-range, pion-exchange time-reversal-violating and parity-violating (TVPV) nucleon-nucleon interactions,
\begin{eqnarray}
\label{eq:piNN1}
\mathcal{L}_{\pi NN}^\mathrm{TVPV} = {\bar N}\left[ \bar g_{\pi}^{(0)} {\vec\tau}\cdot{\vec \pi} +  \bar g_{\pi}^{(1)} \pi^0 +\bar g_{\pi}^{(2)}(2\tau_3\pi^0-{\vec\tau}\cdot{\vec \pi})\right]N . \nonumber \\
\end{eqnarray}
In a general context, the isoscalar and isovector couplings $\bar g_{\pi}^{(0)},~\bar g_{\pi}^{(1)}$ dominate over the isotensor coupling $\bar g_{\pi}^{(2)}$, then the mercury EDM is approximately given by~\cite{Engel:2013lsa},\footnote{It should be kept in mind that the calculations of mercury EDMs are subjected to the uncertainties of hadronic
matrix elements, for a recent review, see Ref.~\cite{Yamanaka}. The upper limits on the nuclear Schiff moment $S$ has been estimated in Ref.~\cite{Singh}.}
\begin{eqnarray}
d_{\rm Hg} = \kappa_S S \approx \kappa_S \frac{2 m_N g_A}{F_\pi} \left(a_0 \bar g_{\pi}^{(0)} + a_1 \bar g_{\pi}^{(1)}\right) \ ,
\end{eqnarray}
where $g_A \approx 1.26$, $F_\pi=186\, {\rm MeV}$, the nuclear matrix elements $a_0=0.01\,e\,{\rm fm}^3$, $a_1=\pm0.02\,e\,{\rm fm}^3$~\cite{oai:arXiv.org:hep-ph/0203202}, and
\begin{eqnarray}
&&\bar g_{\pi}^{(0)} = \tilde \eta_{(0)} ({\tilde\delta}_u + {\tilde\delta}_d) + \gamma^{\tilde{G}}_{(0)} C_{\tilde{G}} \,, \\
&&\bar g_{\pi}^{(1)} = \tilde \eta_{(1)} ({\tilde\delta}_u - {\tilde\delta}_d) + \gamma^{\tilde{G}}_{(1)} C_{\tilde{G}} \,.
\end{eqnarray}
To perform the numerical calculations, we use the following hadronic matrix elements {\cite{Engel:2013lsa}}
\begin{eqnarray}
&& \tilde \eta_{(0)}=-2\times10^{-7} \,, \quad
\tilde \eta_{(1)}=-4\times10^{-7} \,, \nonumber \\
&&\gamma^{\tilde{G}}_{(0)}\approx \gamma^{\tilde{G}}_{(1)} = 2\times 10^{-6} \,,
\end{eqnarray}
and assume a new atomic sensitivity coefficient $\kappa_S=-2.8\times10^{-4}\,{\rm fm}^{-2}$ \cite{oai:arXiv.org:hep-ph/0203202}. Throughout our calculations, we will impose the latest constraint on the mercury EDM~\cite{Graner:2016ses} 
\begin{eqnarray}
\label{eq:HgEDM2016}
| d_{\rm Hg}  |&<& 7.4\times 10^{-30} \, e\cdot {\rm cm}\,,
\end{eqnarray}
which could constrain tightly the 2HDM parameter space and is largely complementary to the ACME result.

To calculate the mercury EDM, we need to incorporate the effect of renormalization group running of the Wilson coefficients from the new physics scale down to the hadronic scale.
The Wilson coefficients of effective operators related to the electron EDM, CEDM and Weinberg three gluon operators are, respectively,
\begin{eqnarray}
\delta_f \equiv -\frac{\Lambda^2 d_f^E}{2 e Q_q m_q}, \quad
\tilde \delta_q \equiv -\frac{\Lambda^2 d_q^C}{2 m_q}, \quad
C_{\tilde G} = \frac{\Lambda^2 d^G}{3 g_s} \,,
\end{eqnarray}
with $m_q$ and $Q_q$ respectively the quark masses and charges, and $\Lambda$ representing the CPV 2HDM scale which is chosen to be $v=246$ GeV. These effective coefficients can be generated from the following effective Lagrangian
\begin{eqnarray}
{\cal L}&=&i\sum_f\, \frac{\delta_f}{\Lambda^2} m_{f} e F^{\mu\nu} \bar f \sigma_{\mu\nu} \gamma_5 f  \nonumber \\
&& + i \sum_q\, \frac{\tilde{\delta}_q}{\Lambda^2} m_{q} g_s G^{a\mu\nu} \bar q \sigma_{\mu\nu} \gamma_5 T^a q \nonumber\\
&&+ \frac{C_{\tilde{G}}}{2 \Lambda^2} g_s f^{abc} \epsilon^{\mu\nu\lambda\sigma}G^a_{\rho\mu}\,
G_{\lambda\sigma}.
{G^c}_{\nu}^{~~\rho} \,.
\end{eqnarray}

Details of the EDM evaluations in the CPV 2HDM are summarized in Appendix~\ref{sec:EDMapp}. All the separate contributions to electron and mercury EDMs are proportional to the CP violating coefficients, e.g. the $\tilde{c}_f$ couplings in Table~\ref{tab:couplings}, thus in a large region of the parameter space of 2HDM, these CPV couplings are tightly constrained, as shown in Figs.~\ref{fig:sensitivity1} to \ref{fig:sensitivity7}.


\section{The $\gamma\gamma$ channel at hadron colliders}
\label{sec:diphoton}

The diphoton process $gg \to H_{2,3} \to \gamma\gamma$ in (CPV) 2HDM at hadron colliders is analogous to that in the SM $gg \to h \to \gamma\gamma$ , where the production of scalar(s) is from gluon fusion mediated predominately by the SM top quark, and the scalar(s) decays radiatively into two photons through the SM fermion and $W$ loops (with subleading contribution from the $H^\pm$ loop). If extra heavy vector-like fermions or heavy charged vector bosons are introduced, the production rate and BR into diphoton might be dramatically enhanced~\cite{Djouadi:2016ack, Dev:2016vle, Dev:2017dui}. Within the well-motivated framework of 2HDM without any more beyond SM particles, the diphoton signal at hadron colliders is unambiguously determined by the Yukawa and gauge couplings in Table~\ref{tab:couplings} and the mass and quartic couplings in the scalar potential~(\ref{eqn:potential}).

The diphoton signal at the LHC from a single heavy scalar decay has recently be studied in the Ref.~\cite{Jung:2015sna,Djouadi:2016ack,Jung:2015gta}, which applies also to the 2HDM with the two heavy scalars $H/A$ significantly separated apart,
in which case the interference of the two heavy resonances is in general negligible. The scenarios with (quasi-)degenerate heavy scalars $H_{2,3}$ are a straightforward generalization of the single-resonance case, with much richer phenomenologies linked to CPV in 2HDM, as stated above. As detailed below, the degenerate resonances can be searched at the LHC in the diphoton channel as well as other decay modes such as $t\bar{t}$. Furthermore, a large mixing $\alpha_c$ of the quasi-degenerate scalars $H_{2,3}$ could enhance significantly the cross section at the resonance peak, compared to a quasi-degenerate case without any CPV mixing ($\alpha_c = 0$), by roughly a factor of 50\% or even more in a large region of the parameter space. Therefore, the CPV in the scalar sector could also be directly probed at a high energy collider, by simply examining the cross section at the resonance peak. Searches of $H_{2,3} \to \gamma\gamma$ are not only largely complementary to other channels such as the final states of $hh$ and $t\bar{t}$, but also to other probes of CPV beyond the SM like EDM experiments. In particular, the CPV in the scalar sector of 2HDM might be small enough to evade the EDM limits but that is still probable at the high energy colliders. Throughout this paper we will consider only the sensitivities at the $\sqrt{s} = 14$ TeV HL-LHC with an integrated luminosity of 3000 fb$^{-1}$. At a future 100 TeV collider like FCC-hh~\cite{Arkani-Hamed:2015vfh, Golling:2016gvc, Contino:2016spe} or SPPC~\cite{CEPC-SPPCStudyGroup:2015csa}, with a larger production cross section, the significance could be largely improved.

\subsection{The differential cross sections}

At hadron colliders, for the diphoton events we have both the tree-level backgrounds from $q\bar{q} \to \gamma\gamma$ and the one-loop level process $gg \to \gamma\gamma$. The quark parton processes do not interfere with the diphoton signal from the heavy scalars, but are comparable to or even larger than the gluon-initialized backgrounds, both of which are included in calculation of the signal sensitivities below. The parton-level differential cross section for the $q\bar{q}$ backgrounds is
\begin{eqnarray}
\frac{{\rm d}}{{\rm d} z} \hat\sigma (q\bar{q} \to \gamma\gamma) =
\sum_{q} \frac{\pi \alpha_{\rm EM}^2 Q_q^4}{3\hat{s}}
\left( \frac{\hat{t}}{\hat{u}} + \frac{\hat{u}}{\hat{t}} \right) \,,
\end{eqnarray}
where $z = \cos\theta$ the scattering angle, and we have summed up all the initial quark flavors. The parton-level cross section for $gg \to \gamma\gamma$ sums up the SM background and heavy scalar resonance contributions,
\begin{eqnarray}\label{eq:total_xsec}
\frac{{\rm d } }{{\rm d}z} \hat \sigma^{\rm tot} (gg\to \gamma\gamma) &=&
k_F \frac{\alpha_{\rm EM}^2 \alpha_s^2 (\sqrt{\hat s}) }{64\pi\, \hat s} \nonumber \\
&& \times \sum_{ \{\lambda \} } \Big|   \mM_{ \{  \lambda \}}^{ \rm bkg }  + \mM_{ \{  \lambda \}}^{\rm res} \Big|^2\,,
\end{eqnarray}
with $k_F \simeq 2$ is the $k$-factor for high order QCD corrections~\cite{Djouadi:2016ack}. The pure signal cross section is obtained by subtracting the SM background in Eq.~\eqref{eq:total_xsec}:
\begin{eqnarray}
\frac{{\rm d} \hat \sigma^{\rm sig}}{{\rm d}z} &=&\frac{{\rm d} \hat \sigma^{\rm res}}{{\rm d}z} +\frac{{\rm d} \hat \sigma^{\rm int}}{{\rm d}z} \,,
\end{eqnarray}
with
\begin{eqnarray}
\label{eqn:sigmahat}
\frac{{\rm d} \hat{\sigma}^{\rm res}}{{\rm d}z} &=& k_F \,
\frac{\alpha_{\rm EM}^2 \alpha_s^2 (\sqrt{\hat s})}{64\pi \hat s}
 \sum_{\{ \lambda \}} \left| {\cal M}^{{\rm res}}_{\{ \lambda \}} \right|^2 \,, \\
\frac{{\rm d} \hat{\sigma}^{\rm int}}{{\rm d}z} &=&  - k_F
\frac{\alpha_{\rm EM}^2 \alpha_s^2 (\sqrt{\hat s})}{64\pi \hat s}
 \sum_{\{ \lambda \}}
 {\cal M}^{{\rm res}}_{\{ \lambda \}} {\cal M}^{{\rm bkg}\,\ast}_{\{ \lambda \}} + {\rm c.c.} \,,
\end{eqnarray}
and the minus sign in the interference terms are from the additional fermion loops, and one only needs to include the helicity configurations of $\{  \lambda \} = (\pm \pm \pm \pm)\,, (\pm \pm \mp \mp)$. Explicitly, the reduced helicity amplitudes for the continuous SM background $gg\to \gamma\gamma$ are
\begin{eqnarray}
&& {\cal M}^{\rm bkg}_{\pm\pm\pm\pm} = {\cal M}_1 \,, \\
&& {\cal M}^{\rm bkg}_{\pm\pm\mp\mp} = {\cal M}_2 \,,
\end{eqnarray}
with, in the massless quark limit~\cite{Dicus:1987fk},
\begin{eqnarray}
{\rm Re}\, {\cal M}_{1} &=&
\left( \sum_q Q_q^2 \right) \left\{ 1 + \frac{\hat t - \hat u}{\hat s} \log \Big|  \frac{\hat t }{\hat u} \Big| \right. \nonumber \\
&& \left. + \frac{\hat t^2 + \hat u^2 }{2 \hat s^2} \Big[ \log^2\Big| \frac{\hat t}{\hat u}  \Big|+ \pi^2 \theta(\frac{\hat t }{\hat u } )    \Big] \right\} \,,\\
{\rm Im}\, {\cal M}_{1}^{} &=&
-\left( \sum_q Q_q^2 \right) \pi \Big[ \theta(\hat t) - \theta (\hat u)  \Big] \nonumber \\
&& \times \left( \frac{\hat t - \hat u }{\hat s} + \frac{\hat t^2 + \hat u^2 }{\hat s^2} \log \Big| \frac{\hat t}{\hat u }  \Big|   \right)\,, \\
{\cal M}_{2} &=& - \left( \sum_q Q_q^2 \right) \,.
\end{eqnarray}
Since the resonance masses we probe below are typically heavier than the $t\bar t$ threshold, i.e., $\hat s \sim M_H \gtrsim 2 m_t$, we sum over all the six flavors of quarks, which leads to $\sum_q Q_q^2 = 5/3$.

In the simpler case with a single (CPV) Higgs boson $H$, the corresponding resonance helicity amplitudes can be written in the rude form of
\begin{eqnarray}
{\cal M}^{\rm res}_{\{ \lambda_1 \lambda_2 \}} \sim
{\cal M}^{}_{\{ \lambda_1 \}} (gg \to H) \; P_H \;
{\cal M}^{}_{\{ \lambda_2 \}} (H \to \gamma\gamma) \nonumber \\
\end{eqnarray}
with $P_H$ the standard propagator for a single heavy scalar $H$. For a pair of quasi-degenerate CPV scalars $H_i$, the corresponding resonance helicity amplitudes can be generalized by including the $2\times2$ propagator matrix $P_{jk}$ in Eq.~(\ref{eqn:propagator}):
\begin{eqnarray}
&& {\cal M}^{{\rm res}}_{\pm\pm\pm\pm} =
\frac{ G_F \hat s^2}{128 \pi^2} \sum_{j,k}
\left( c_{g,j} \pm i \tilde{c}_{g,j} \right) P_{jk}
\left( c_{\gamma,k} \pm i \tilde{c}_{\gamma,k} \right) \,, \nonumber \\ && \\
&& {\cal M}^{{\rm res}}_{\pm\pm\mp\mp} =
\frac{ G_F \hat s^2}{128 \pi^2} \sum_{j,k}
\left( c_{g,j} \pm i \tilde{c}_{g,j} \right) P_{jk}
\left( c_{\gamma,k} \mp i \tilde{c}_{\gamma,k} \right) \,, \nonumber \\
\end{eqnarray}
with $j$, $k = $2, 3. When moduli squared and summed up, we have altogether 64 terms. With some of the them cancelled out,
\begin{eqnarray}
&& \sum_{\{ \lambda \}} \left| {\cal M}^{{\rm res}}_{\{ \lambda \}} \right|^2 = 4
\left( \frac{ G_F \hat s^2}{128 \pi^2} \right)^2
\sum_{jkmn}
\left( c_{g,j} c_{g,m}^\ast + \tilde{c}_{g,j} \tilde{c}_{g,m}^\ast \right)  \nonumber \\
&& \qquad\qquad \times
P_{jk} P_{mn}^\ast
\left( c_{\gamma,k} c_{\gamma,n}^\ast + \tilde{c}_{\gamma,k} \tilde{c}_{\gamma,n}^\ast \right) \,,
\end{eqnarray}
where the four indices $j,\,k,\,m,\,n$ all run from 2 to 3. Here with the summation we have included both the diagonal and off-diagonal terms; for the latter case the indices $j\neq k$ and $m\neq n$ stand for the interferences of the two nearly-degenerate heavy scalars. The CP-even and odd contributions to the effective coupling of $H_i gg$ (with $i=$2,3) are respectively
\begin{eqnarray}
c_{g,i} &=& \sum_q c_{q,i} A_{1/2}^H (\tau_q ) \,, \\
\tilde{c}_{g,i} &=& \sum_q \tilde{c}_{q,i} A_{1/2}^A (\tau_q )
\end{eqnarray}
where $\tau_X = \hat s / 4 m_X^2$. For the $H_i \gamma\gamma$ couplings, 
\begin{eqnarray}
\label{eqn:gammaloop}
c_{\gamma,i} &=&
- \sum_{j,=1,2} \frac{{\cal R}_{ij} \tilde\lambda_{j+-} v}{2M^2_{H^\pm}} A_{0} (\tau_{H^\pm}^{}) \nonumber \\
&&+ \sum_f c_{f,i} N_C^f Q_f^2 A_{1/2}^{H} (\tau_f^{}) 
  + a_i A_{1}^{H} (\tau_W^{}) \\
\label{eqn:gammaloop2}
\tilde{c}_{\gamma,i} &=&
- \frac{{\cal R}_{i3} \tilde\lambda_{3+-} v}{2M^2_{H^\pm}} A_{0} (\tau_{H^\pm}^{}) \nonumber\\ 
&&+ \sum_f \tilde{c}_{f,i} N_C^f Q_f^2 A_{1/2}^{A} (\tau_f^{}) \,,
\end{eqnarray}
with the trilinear scalar coupling given in Eqs.~(\ref{eqn:lambda+-1}) to (\ref{eqn:lambda+-3}). The prefactor ${\cal R}_{ij} \tilde\lambda_{j+-} v / 2M_{H^\pm}^2$ for the $H^\pm$ loop is intrinsically a function of the quartic couplings, which turns out to be small as long as the couplings $\lambda_i$ in the scalar potential are perturbative. Furthermore, the charged scalar term in Eq.~(\ref{eqn:gammaloop2}) is generally also suppressed by the CP-violating coupling Im$\lambda_5$, which makes the scalar loop contribution even smaller for the CP-odd contributions.

The generalization above is also valid for the interference terms. Summing up the helicities, we have
\begin{widetext}
\begin{eqnarray}
&& \sum_{\{ \lambda \} = \pm\pm\pm\pm}
 {\cal M}^{{\rm res}}_{\{ \lambda \}} {\cal M}^{{\rm bkg}\,\ast}_{\{ \lambda \}} + {\rm c.c.}
\propto 2 \sum_{ij}
\left[ c_{g,i} P_{ij} c_{\gamma,j} -
\tilde c_{g,i} P_{ij} \tilde c_{\gamma,j} \right] {\cal M}_1^{{\rm bkg}\, \ast} +{\rm c.c.} \,, \\
&& \sum_{\{ \lambda \} = \pm\pm\mp\mp}
 {\cal M}^{{\rm res}}_{\{ \lambda \}} {\cal M}^{{\rm bkg}\,\ast}_{\{ \lambda \}} + {\rm c.c.}
\propto 2 \sum_{ij}
\left[ c_{g,i} P_{ij} c_{\gamma,j} +
\tilde c_{g,i} P_{ij} \tilde c_{\gamma,j} \right] {\cal M}_2^{{\rm bkg}\, \ast} +{\rm c.c.} \,,
\end{eqnarray}
then the interfering amplitude square:
\begin{eqnarray}
\sum_{\{ \lambda \}}
 {\cal M}^{{\rm res}}_{\{ \lambda \}} {\cal M}^{{\rm bkg}\,\ast}_{\{ \lambda \}} + {\rm c.c.} &=& \frac{ G_F \hat s^2}{128 \pi^2} \left[
2 \left( \sum_{ij}
c_{g,i} P_{ij} c_{\gamma,j} \right)
\left( {\cal M}_1^{{\rm bkg}\, \ast} +
{\cal M}_2^{{\rm bkg}\, \ast} \right) \right. \nonumber \\
&& \left. +2 \left( \sum_{ij}
\tilde c_{g,i} P_{ij} \tilde c_{\gamma,j} \right)
\left( {\cal M}_2^{{\rm bkg}\, \ast} -
{\cal M}_1^{{\rm bkg}\, \ast} \right)
+{\rm c.c.} \right] \,.
\end{eqnarray}
Splitting the interference terms into the real and imaginary parts, they reads
\begin{eqnarray}
\label{eqn:Re}
&& 4 \left[
\sum_{ij} \left( c_{g,i} c_{\gamma,j} \right)^{\rm Re} P_{ij}^{\rm Re}
\left( {\cal M}_1^{{\rm Re}} + {\cal M}_2^{{\rm Re}} \right) +
\sum_{ij} \left( c_{g,i} c_{\gamma,j} \right)^{\rm Im} P_{ij}^{\rm Re}
\left( {\cal M}_1^{{\rm Im}} + {\cal M}_2^{{\rm Im}} \right)
\right. \nonumber \\
&& \left. +
\sum_{ij} \left( \tilde c_{g,i} \tilde c_{\gamma,j} \right)^{\rm Re} P_{ij}^{\rm Re}
\left( {\cal M}_2^{{\rm Re}} - {\cal M}_1^{{\rm Re}} \right) +
\sum_{ij} \left( \tilde c_{g,i} \tilde c_{\gamma,j} \right)^{\rm Re} P_{ij}^{\rm Im}
\left( {\cal M}_2^{{\rm Im}} - {\cal M}_1^{{\rm Im}} \right)
\right] \,, \\
\label{eqn:Im}
&& 4 \left[
\sum_{ij} \left( c_{g,i} c_{\gamma,j} \right)^{\rm Re} P_{ij}^{\rm Re}
\left( {\cal M}_1^{{\rm Im}} + {\cal M}_2^{{\rm Im}} \right) -
\sum_{ij} \left( c_{g,i} c_{\gamma,j} \right)^{\rm Im} P_{ij}^{\rm Im}
\left( {\cal M}_1^{{\rm Re}} + {\cal M}_2^{{\rm Re}} \right)
\right. \nonumber \\
&& \left. +
\sum_{ij} \left( \tilde c_{g,i} \tilde c_{\gamma,j} \right)^{\rm Re}
P_{ij}^{\rm Im}
\left( {\cal M}_2^{{\rm Re}} - {\cal M}_1^{{\rm Re}} \right) -
\sum_{ij} \left( \tilde c_{g,i} \tilde c_{\gamma,j} \right)^{\rm Im}
P_{ij}^{\rm Im}
\left( {\cal M}_2^{{\rm Re}} - {\cal M}_1^{{\rm Re}} \right)
\right] \,,
\end{eqnarray}
\end{widetext}
where for the SM background ${\cal M}_2^{{\rm Im}} = 0$, while the imaginary parts of the loop functions come from the loops with $\hat s > 4 M^2_{\rm loop}$. When $H_2$ decouples from $H_3$ in the propagator matrix, i.e. $P_{23} = 0$, the imaginary part of the propagator $P_{ii}^{\rm Im} \to M_i^2 \Gamma_{H_i}^2$ in the limit of $\hat s \to M_i^2$.

It is straightforward to obtain the differential cross sections with respect to the diphoton invariant mass $M_{\gamma\gamma}$, by integrating over the scattering angle $z$ and convoluting with the gluon distribution luminosity ${\cal L}_{gg}$ in proton~\cite{Ball:2014uwa}:
\begin{eqnarray}
\label{eqn:diffcs1}
\frac{{\rm d} \sigma^{\rm res}}{{\rm d} M_{\gamma\gamma}} &=&
\frac{2}{M_{\gamma\gamma}} \hat{\sigma}^{\rm res} (\hat{s} = M_{\gamma\gamma}^2) {\cal L}_{gg} \,, \\
\label{eqn:diffcs2}
\frac{{\rm d} \sigma^{\rm int}}{{\rm d} M_{\gamma\gamma}} &=&
\frac{2}{M_{\gamma\gamma}} \hat{\sigma}^{\rm int} (\hat{s} = M_{\gamma\gamma}^2) {\cal L}_{gg} \,.
\end{eqnarray}

\subsection{Diphoton signal at hadron colliders}

\begin{figure*}[!t]
  \centering
  \includegraphics[height=0.32\textwidth]{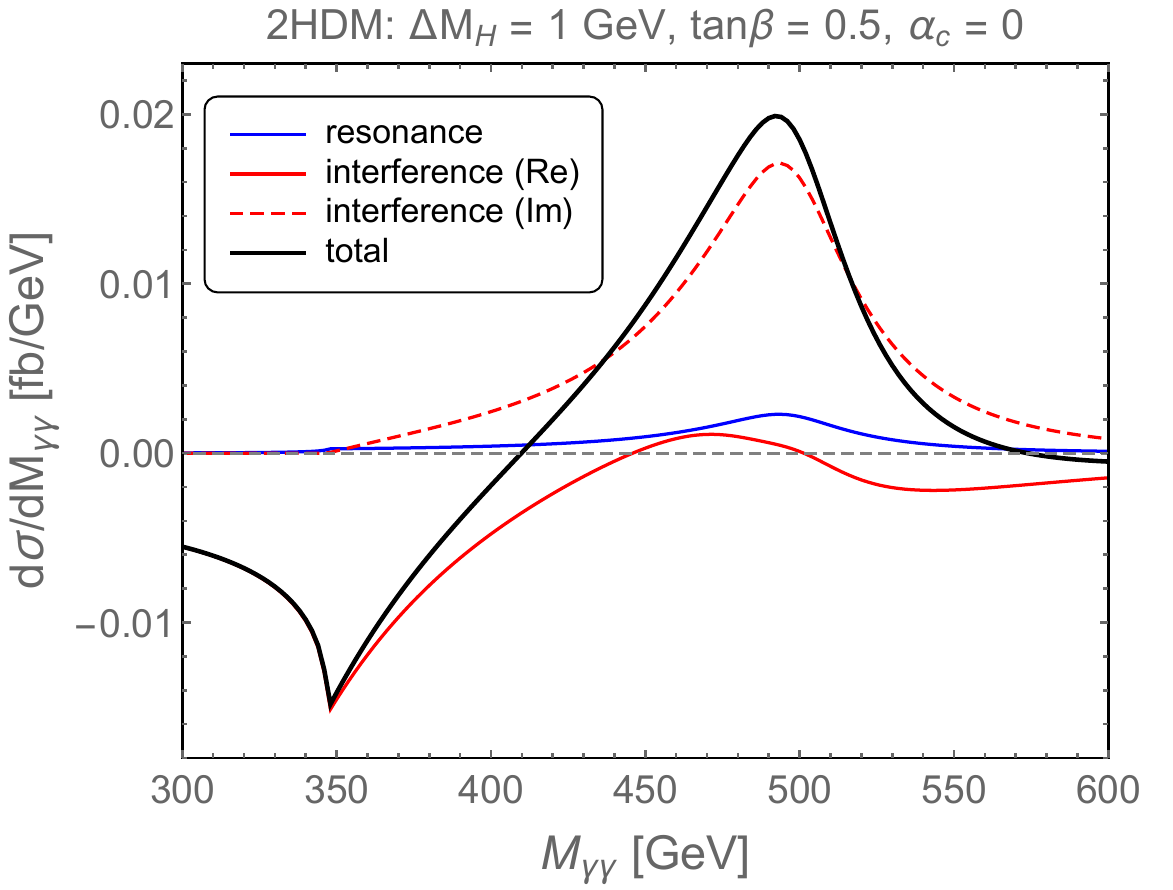} \hspace{-7.5pt}
  \includegraphics[height=0.32\textwidth]{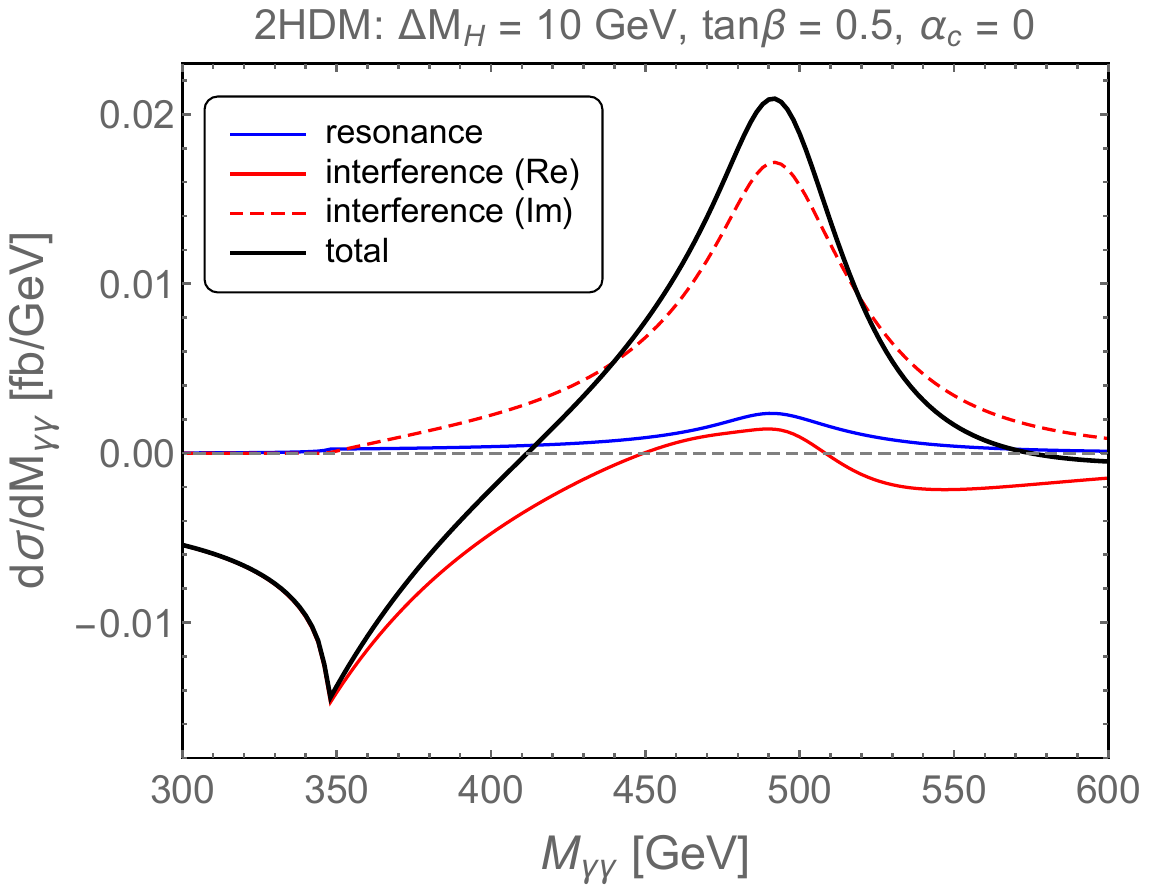} 
  \caption{Examples of the diphoton spectra ${\rm d} \sigma / d M_{\gamma\gamma}$ in the CPV 2HDM, with the heavy scalar mass $M_H = 500$ GeV with a small splitting $\Delta M_H = 1$ GeV (left) or 10 GeV (right), $\tan\beta = 0.5$ and $\alpha_c = 0$. In these plots we show both the pure resonance (blue) and real and imaginary interference (solid and dashed red) contributions, as well as the total spectra (black). For the Yukawa couplings of type-I and type-II, these spectra are almost the same.}
  \label{fig:example05}
\end{figure*}

\begin{figure*}[!t]
  \centering
  \includegraphics[height=0.32\textwidth]{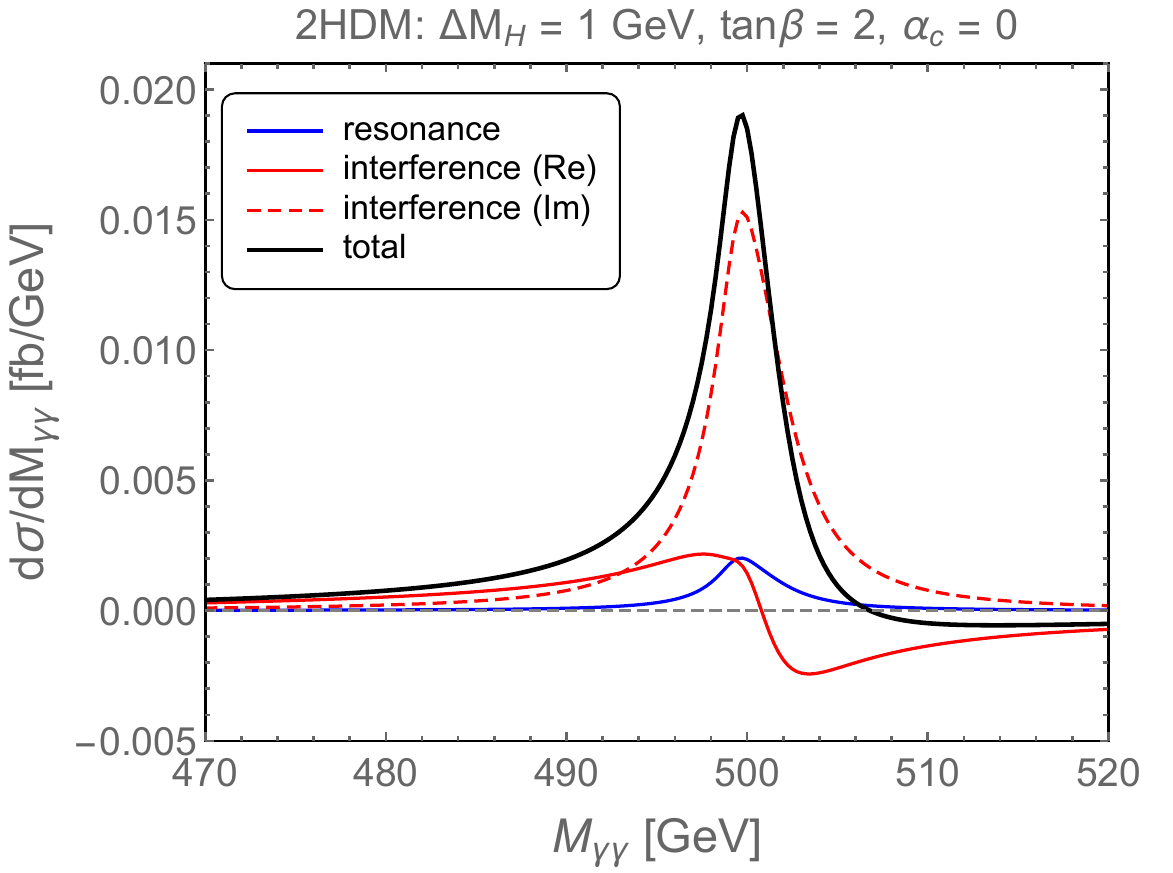} \hspace{-7.5pt}
  \includegraphics[height=0.32\textwidth]{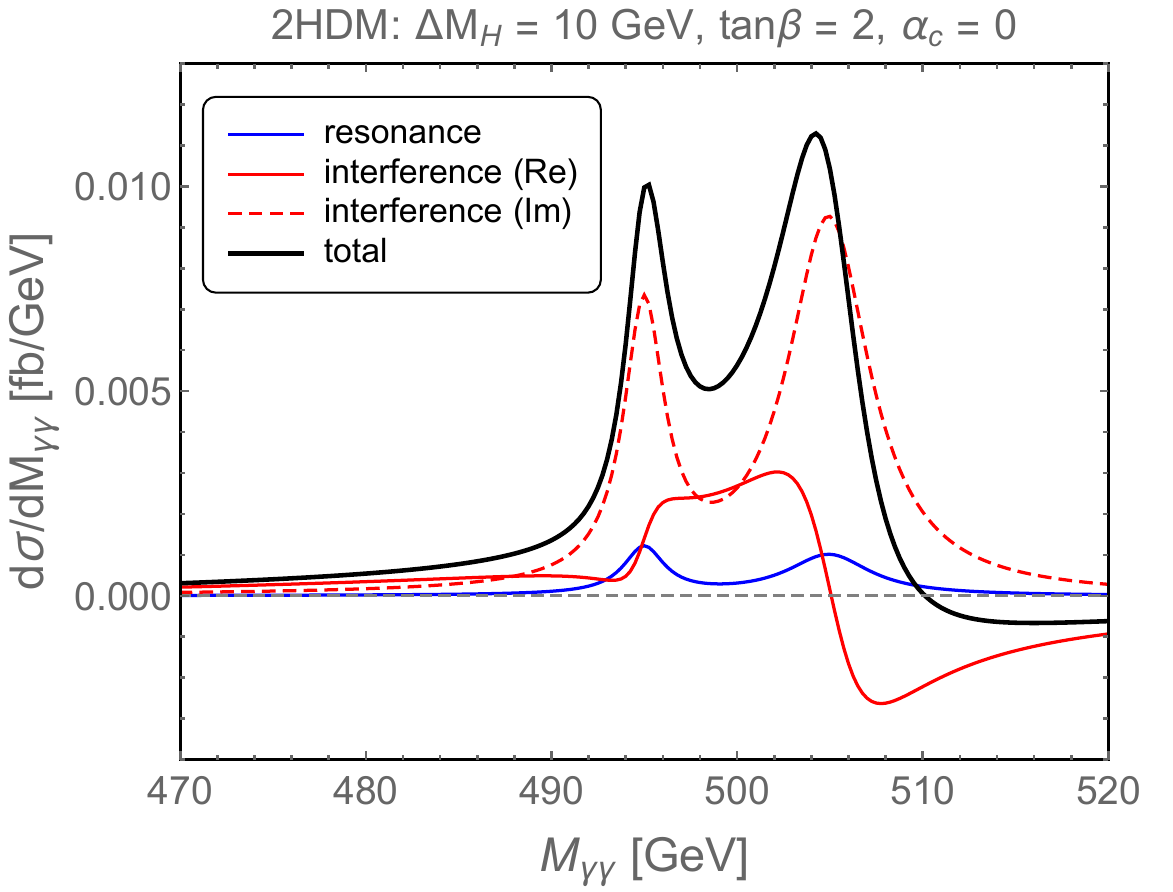} 
  \caption{The same as in Fig.~\ref{fig:example05}, with $\tan\beta = 2$.}
  \label{fig:example2}
\end{figure*}

With the parton-level differential cross sections for both the resonance and interference terms given in Eqs.~(\ref{eqn:diffcs1}) and (\ref{eqn:diffcs2}), we are ready to predict the diphoton signals at the LHC from the decay of heavy scalars in the CPV 2HDM. To demonstrate the most important features in the diphoton signals, two representative examples are presented in Figs.~\ref{fig:example05} and \ref{fig:example2} with respectively $\tan\beta = 0.5$ and $2$ in both the type-I and type-II 2HDM. Other parameters are set as follows: the heavy scalar mass $M_H = 500$ GeV with a splitting $\Delta M_H = 1$ GeV or 10 GeV and the vanishing CPV in the heavy scalar sector $\alpha_c = 0$ for which $\alpha_b = 0$. In the two figures we show both the separate contributions from the pure degenerate resonances and the real and imaginary interference terms in Eqs.~(\ref{eqn:Re}) and (\ref{eqn:Im}). For the sake of concreteness, we set $\sqrt{s} = 14$ TeV, and integrate over the scattering angle $z = \cos\theta$ from 0 to $z_{\rm max} = 0.5$. For the scalar mediators $H_{2,3}$ the signal process $gg \to H_i \to \gamma\gamma$ does not depend on $z = \cos\theta$, while the SM background $q\bar{q} \to \gamma\gamma$ and $gg \to \gamma\gamma$ both peak in the forward direction. Without optimising the kinematics we adopt a na\"ive cut on the angle $|\cos\theta| < 0.5$. A dedicated study would improve to some extent the projected sensitivities in Section~\ref{sec:prospects}.

In both the two benchmark scenarios in Figs.~\ref{fig:example05} and \ref{fig:example2}, the diphoton signal above the continuous SM background is always dominated by the interference terms, as expected: The resonance signal $gg \to H_i \to \gamma\gamma$ arises at two-loop level, and is much smaller than the background-resonance interfering which is comparatively enhanced by the one-loop background process $gg \to \gamma\gamma$. When the invariance mass of the two photons are close to the heavy scalar mass, i.e. $M_{\gamma\gamma} \simeq M_H$, the real interference effects are destructive, induced from the extra fermion loops in the heavy scalar mediated diagrams.\footnote{It is also possible that the real interference effects are constructive, as long as the contributions of $W^\pm$ and $H^\pm$ loop to the $H_i \gamma\gamma$ couplings are grater than the SM fermion loops. However, in the CPV 2HDM with two nearly-degenerate heavy scalars, the constructive scenarios are highly disfavored by the couplings of the heavy scalars.}  However, at the resonance $M_{\gamma\gamma} \simeq M_H$, the differential diphoton cross section is always dominated by the imaginary parts, as $(M_{\gamma\gamma} - M_H)^2 \lesssim M_H \Gamma_H$ (neglecting the heavy scalar mixing effects, i.e. the off-diagonal elements of the propagator $P_{ij}$), with the heavy scalar decay width $\Gamma_H$ largely enhanced by the ${\cal O} (1)$ top quark Yukawa coupling in the SM. One could note in the couplings in Table~\ref{tab:couplings} that the couplings of heavy scalars to top quark are roughly $\propto (\tan\beta)^{-1}$, then the total width $\Gamma_H$ has, roughly, a second power dependence on $\tan\beta$, i.e. $\Gamma_H \propto (\tan\beta)^{-2}$. Therefore the resonance in Fig.~\ref{fig:example2} in much narrower than that in Fig.~\ref{fig:example05}, roughly by a factor of $(2/0.5)^{-2} = 1/16$. When the mass splitting is larger, e.g. $\Delta M_H = 10$ GeV, as a result of the narrow width for the model with $\tan\beta = 2$, the width $\Gamma_H < \Delta M_H$ and the heavy scalars are significantly separated apart, as seen in Fig.~\ref{fig:example2}. In contrast the diphoton spectra for $\tan\beta = 0.5$ does not change too much.

There is apparently a dip in the vicinity of $M_{\gamma\gamma} \simeq 2 m_{t} \simeq 350$ GeV in the real interference contributions of Fig.~\ref{fig:example05}, which is due to the opening of the $H_{2,3} \to t\bar{t}$ decay mode and sharp increase of the decay width $\Gamma_H$ (neglecting here again the heavy scalar mixing elements in the propagator). In other words, opening of the top decay mode could diminish significantly the propagator $P_{ij}$, which, however, depend largely on the value of $\tan\beta$. As for the resonance widths in Figs.~\ref{fig:example05} and \ref{fig:example2}, the depth of the dip is roughly proportional to
\begin{eqnarray}
P_{ij}^{-1}(M_{tt}) \sim \Gamma_{H} \propto (\tan\beta)^{-2} \,.
\end{eqnarray}
A smaller $\tan\beta$ could thus induce effectively a more significant dip at $M_{\gamma\gamma} \simeq 2 m_t$ (In Fig.~\ref{fig:example2} we do not show explicitly the dip at $2m_t$, which is much smaller than those in Fig.~\ref{fig:example05}, as expected).

Combing both the effects of $\tan\beta$ on the resonance width and the dip at $M_{\gamma\gamma} \simeq 2 m_t$, a smaller $\tan\beta$ could make the double-scalar resonance broader and the dip deeper, then the $\gamma\gamma$ spectrum is expected to be more severely distorted, even without any CPV in the scalar sector of 2HDM, i.e. $\alpha_{b,c} = 0$. That is also the reason why the differential $t\bar{t}$ cross sections in the right panel of Fig.~\ref{fig:LHC} exclude larger regions when $\tan\beta$ is small (cf. the pink regions in Figs.~\ref{fig:sensitivity1} and \ref{fig:sensitivity2}): the significant dip and broad resonance in Fig.~\ref{fig:example05} could easily be excluded by the uncertainties of $t\bar{t}$ data.

\begin{figure*}[!t]
  \centering
  \includegraphics[height=0.32\textwidth]{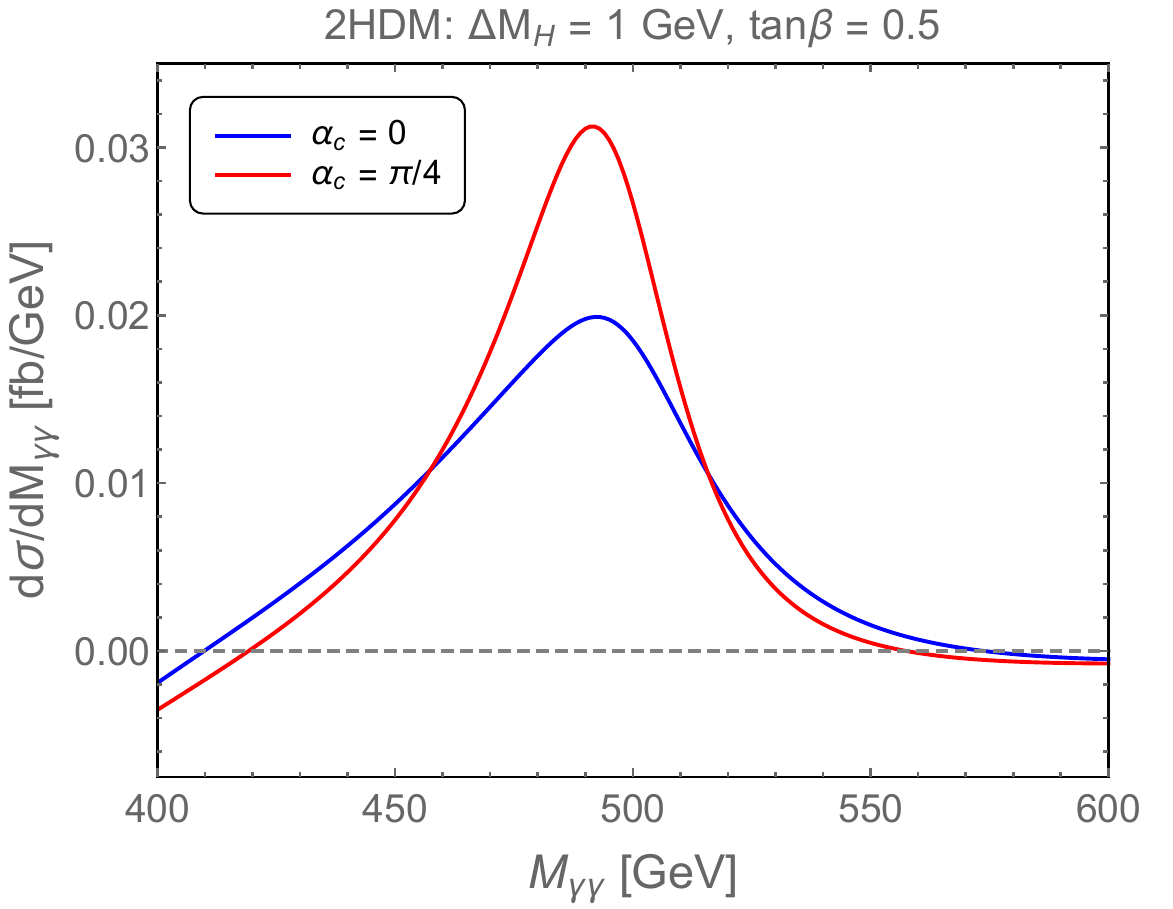} \hspace{-7.5pt}
  \includegraphics[height=0.32\textwidth]{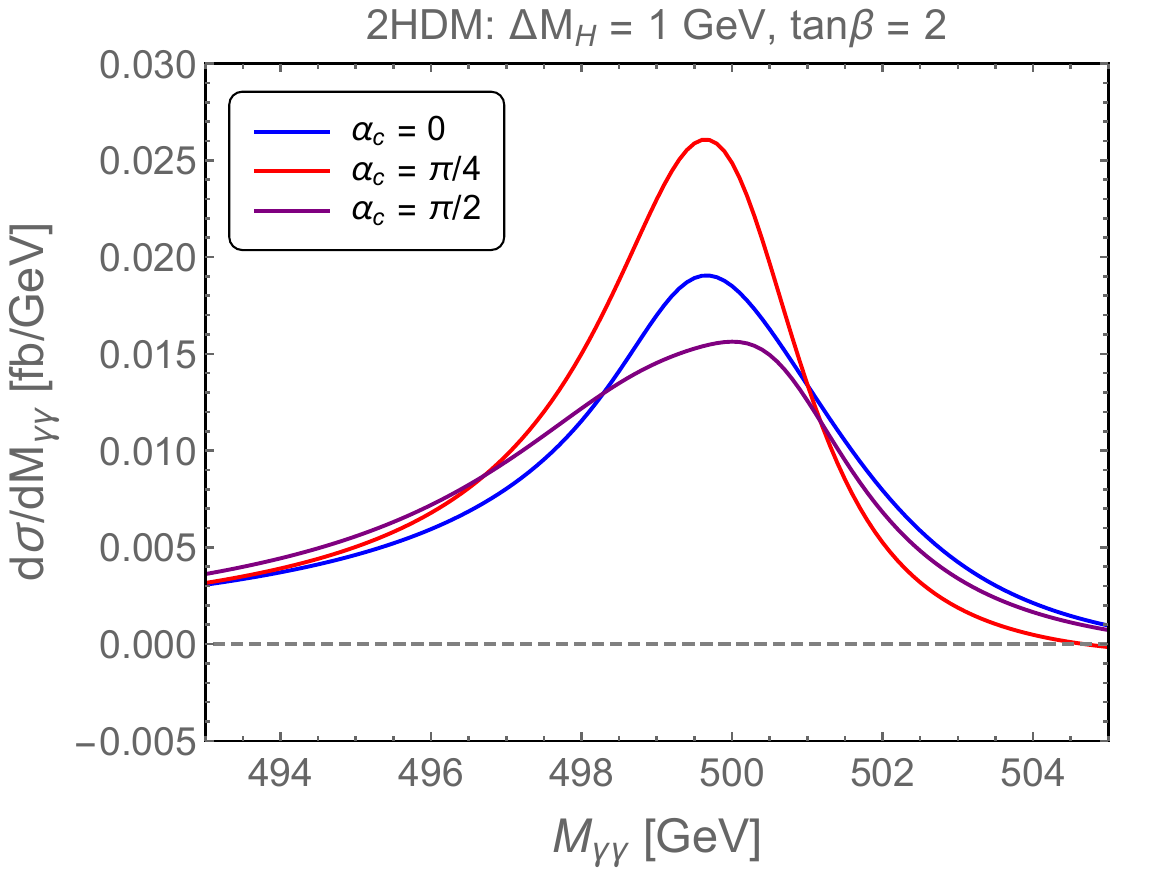} 
  \caption{Total diphoton spectra in vicinity of the resonance peak, in the examples given in Figs.~\ref{fig:example05} and \ref{fig:example2} with mass splitting $\Delta M_H = 1$ GeV, $\alpha_c = 0$ and $\pi/4$ (and $\pi/2$ for $\tan\beta = 2$).}
  \label{fig:example_CP}
\end{figure*}

With a maximal mixing $\alpha_c = \pi/4$ of the two nearly-degenerate scalars, the differential diphoton cross section could be significantly enhanced at the resonance $M_{\gamma\gamma} \simeq M_H \cong M_{2,3}$, as clearly shown in Fig.~\ref{fig:example_CP}. When $M_{\gamma\gamma}$ is far away from the resonance $M_H$, the CPV effects would be highly suppressed. The CPV effect on the diphoton spectrum in the presence of CPV 2HDM could be directly tested in the high-energy collisions at the LHC, and is largely complementary to other current limits and future probes of CPV, e.g. those from the EDM measurements. It depends on some parameters in the scalar sector of 2HDM:
\begin{itemize}
  \item The CP effect is more significant when the scalars are lighter, and vanishes in the limit of ${M_{\gamma\gamma}} \simeq M_H \to \infty$, as the production cross section diminishes when the scalars are heavier, and the diphoton spectrum is further suppressed by the mass $M_{\gamma\gamma} \simeq M_H$ in the dominator of Eqs.~(\ref{eqn:diffcs1}) and (\ref{eqn:diffcs2}), as shown in Fig.~\ref{fig:sensitivity_CP}.
  \item The CPV in the scalar sector depends also on $\tan\beta$, as most of the couplings of $H_{2,3}$ to the SM particles involve as functions of $\tan\beta$, in particular the couplings to the top quark. When $\tan\beta = 0.5$, both the two value of $\alpha_c = 0$ and $\pi/2$ generate almost the same spectrum; in contrast, when $\tan\beta = 2$, some of the subleading terms in the couplings of $H_{2,3}$ to the top quark and other SM particles becomes important, thus in Fig.~\ref{fig:example_CP} the spectra with $\alpha_c = \pi/2$ differ slightly from those with $\alpha = 0$.
  \item  As the mixing angle $\alpha_b$ connecting the SM Higgs to $M_3$ is typically very small in the 2HDM scenarios with quasi-degenerate heavy scalars, the impact of CPV on the diphoton spectra in Fig.~\ref{fig:example_CP} is mainly from the CPV mixing $\alpha_c$ of the two heavy scalars $H_{2,3}$. In principle, the effects from $\alpha_b$ could also be significant when $\alpha_b$ is large, however, it is excluded or tightly constrained by the EDM limits, see the plots in Figs.~\ref{fig:sensitivity5} to \ref{fig:sensitivity7}.
\end{itemize}

\subsection{Prospects at the LHC}
\label{sec:prospects}

To calculate the expected numbers of signal events from $gg \to H_{2,3} \to \gamma\gamma$, we sum up the resonance and interference terms given in Eqs.~(\ref{eqn:diffcs1}) and (\ref{eqn:diffcs2}), and compare them with the SM background $q\bar{q} \to \gamma\gamma$ and $gg \to \gamma\gamma$. In specific, we integrate the differential cross sections ${\rm d} \sigma / {\rm d} M_{\gamma\gamma}$ for both the SM backgrounds and 2HDM signals with a universal bin width of 10 GeV:
\begin{eqnarray}
\Delta \sigma_{\gamma\gamma} (M_0) = \int_{M_0 - 5 \, {\rm GeV}}^{M_0 + 5 \, {\rm GeV}}
{\rm d} M_{\gamma\gamma} \;
\frac{{\rm d} \sigma_{}}{{\rm d} M_{\gamma\gamma}} \,,
\end{eqnarray}
with the list of cross sections as functions of the diphoton invariant mass $M_{\gamma\gamma} = M_0$. To suppress the SM background we have set an upper bound on the angle $z < 0.5$ as in the previous subsection. We estimate how many background and signal events $N_{\gamma\gamma}$ could be expected in each of the diphoton bins at $\sqrt{s} = 14$ TeV and with the total luminosity of 3000 fb$^{-1}$. By counting simply the numbers of events in the diphoton spectra, we obtain the 95\% CL sensitivities via the standard $\chi^2$-method:
\begin{eqnarray}
\label{eqn:chi2}
\chi^2 = \sum_{\rm bins} \left( \frac{N_{\gamma\gamma}^{\rm signal} (M_0)}
{\sqrt{N_{\gamma\gamma}^{\rm bkg}(M_0)}} \right)^2 \,,
\end{eqnarray}
where all the sensitivities in the bins are summed up. To take into account the detector effects we assume an efficiency of 95\% for photon identification. The uncertainty in the photon energy scale at high transverse momentum is typically $\lesssim 2\%$, depending on the rapidity of photon~\cite{Aaboud:2016tru}. We have checked the smearing effect on the photon spectra and sensitivities, and found that it is very small and completely negligible.

\begin{figure*}[!t]
  \centering
  \includegraphics[width=0.4\textwidth]{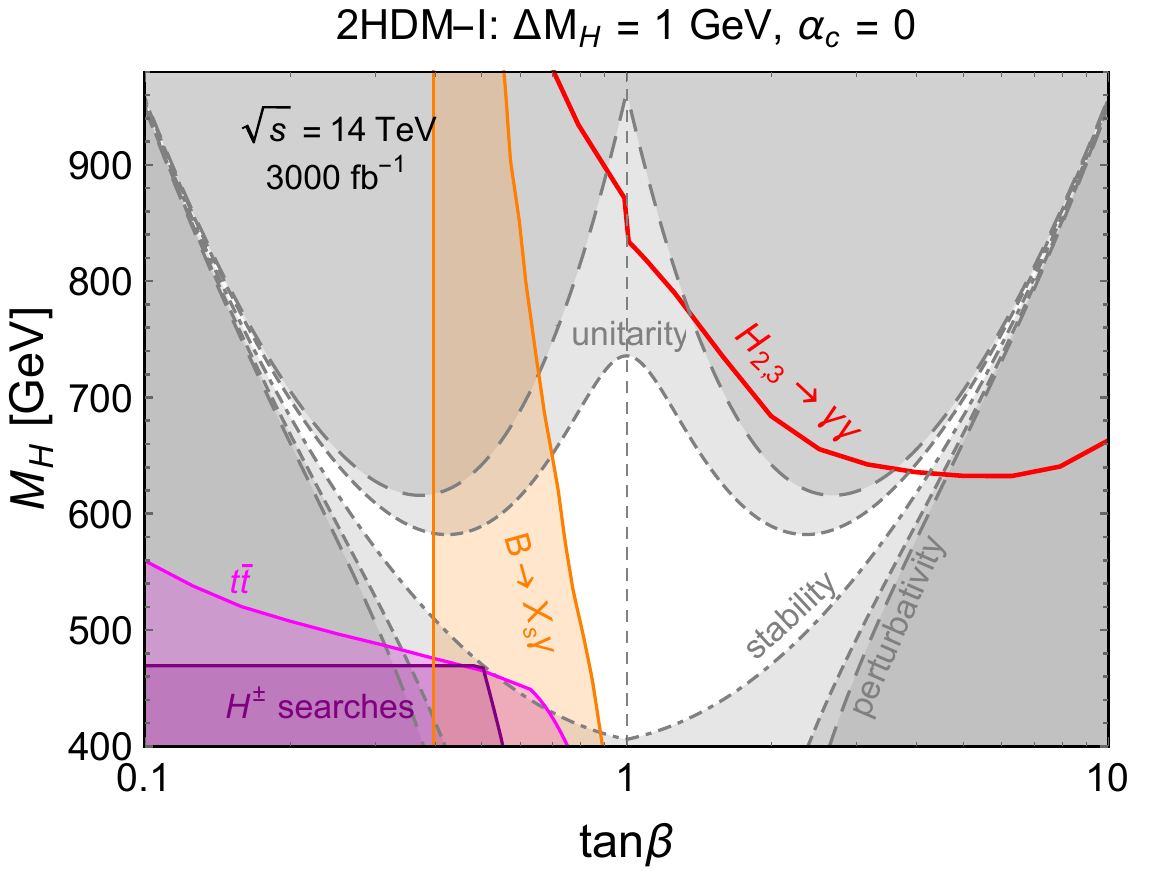} \hspace{-7.5pt}
  \includegraphics[width=0.4\textwidth]{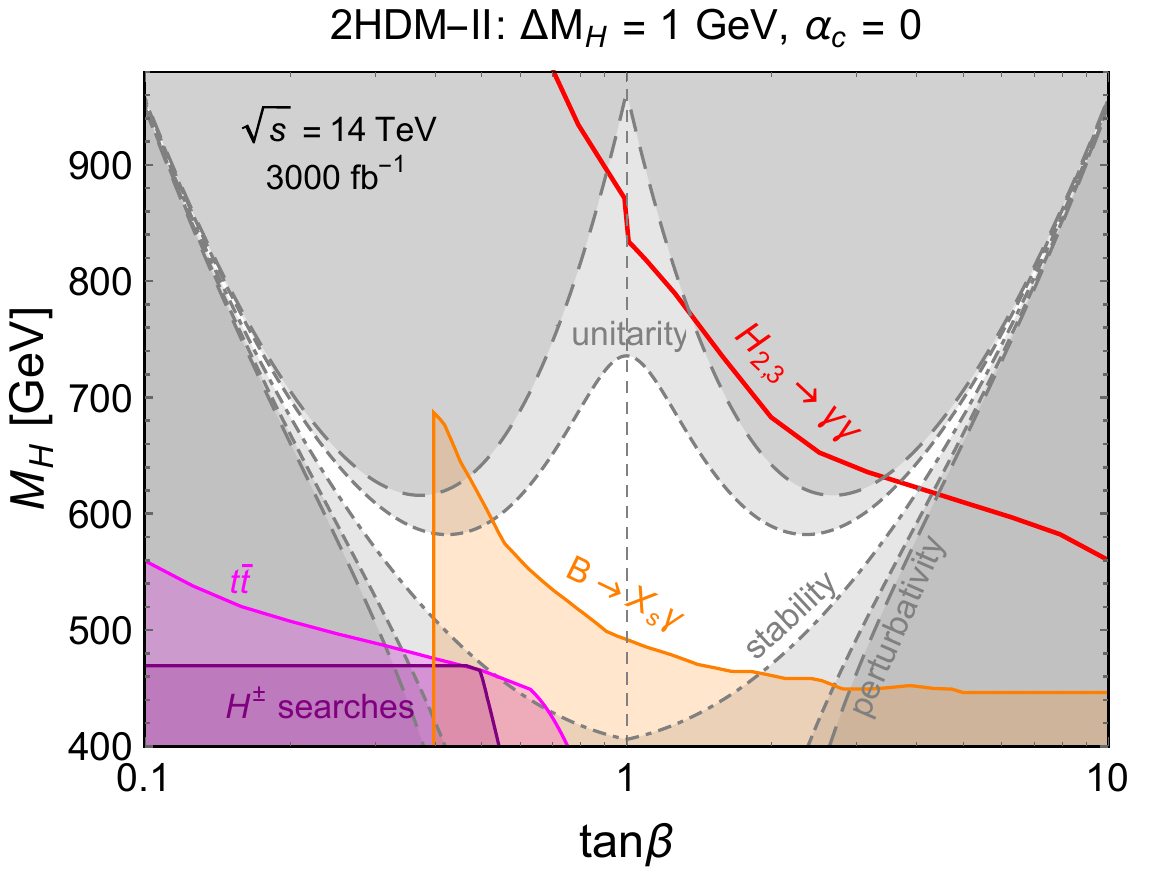}
  \caption{Diphoton prospects of quasi-degenerate scalars $H_{2,3}$ in CPV 2HDM in the parameter space of $M_H$ and $\tan\beta$, with type-I (left) and type-II (right) Yukawa couplings, with a small mass splitting $\Delta M_H = 1$ GeV and $\alpha_c = 0$. The regions below the solid red lines are probable at the 95\% CL at the HL-LHC with $\sqrt{s} = 14$ TeV and an integrated luminosity of 3000 fb$^{-1}$, by searches of $gg \to H_{2,3} \to \gamma\gamma$, with the SM background $gg \to \gamma\gamma$ and the interference of background and resonance taken into consideration (Note that the red lines go beyond 1 TeV when $\tan\beta \lesssim 1$, not shown in these plots; the small $\tan\beta$ regions are all covered implicitly). The shaded regions are excluded, respectively, by differential $t\bar{t}$ data~\cite{Khachatryan:2016mnb} (pink) in the right panel of Fig.~\ref{fig:LHC}, direct $H^\pm$ searches~\cite{Khachatryan:2015qxa} (purple) and $B \to X_s \gamma$~\cite{Misiak:2017bgg} (orange) in Fig.~\ref{fig:Hpm}. The gray regions are excluded by the theoretical arguments of perutrbativity, stability and unitarity. The limits and prospects in the two plots with a larger $\Delta M_H = 10$ GeV are almost the same.}
  \label{fig:sensitivity1}
\end{figure*}

\begin{figure*}[!t]
  \centering
  \includegraphics[width=0.4\textwidth]{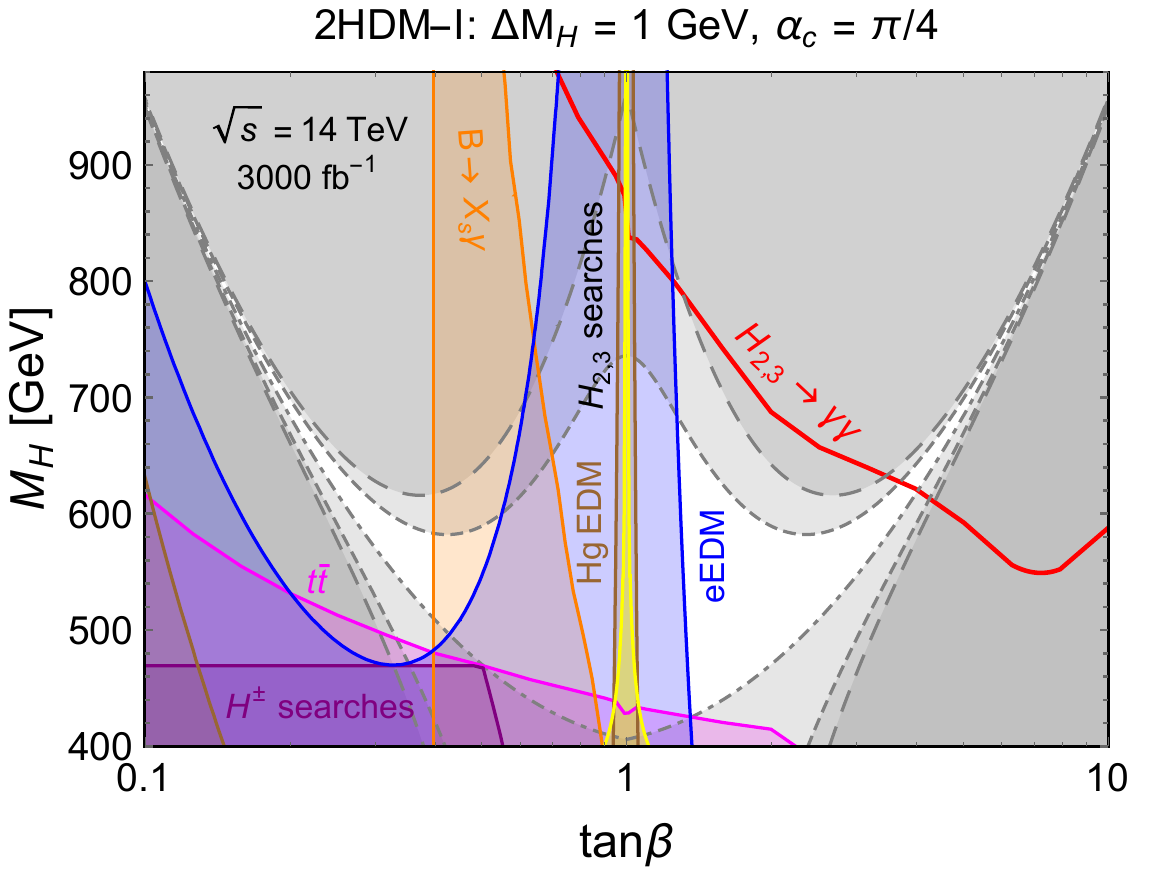} \hspace{-7.5pt}
  \includegraphics[width=0.4\textwidth]{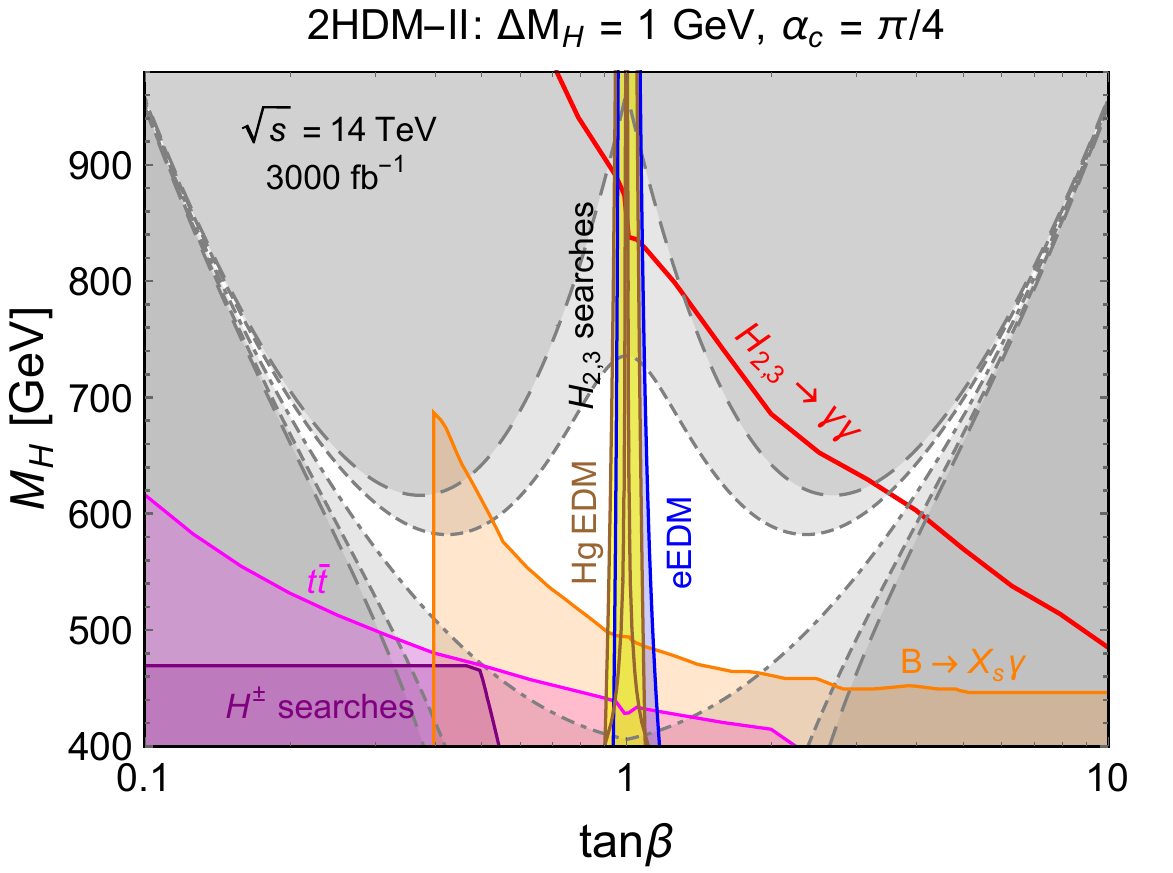}
  \includegraphics[width=0.4\textwidth]{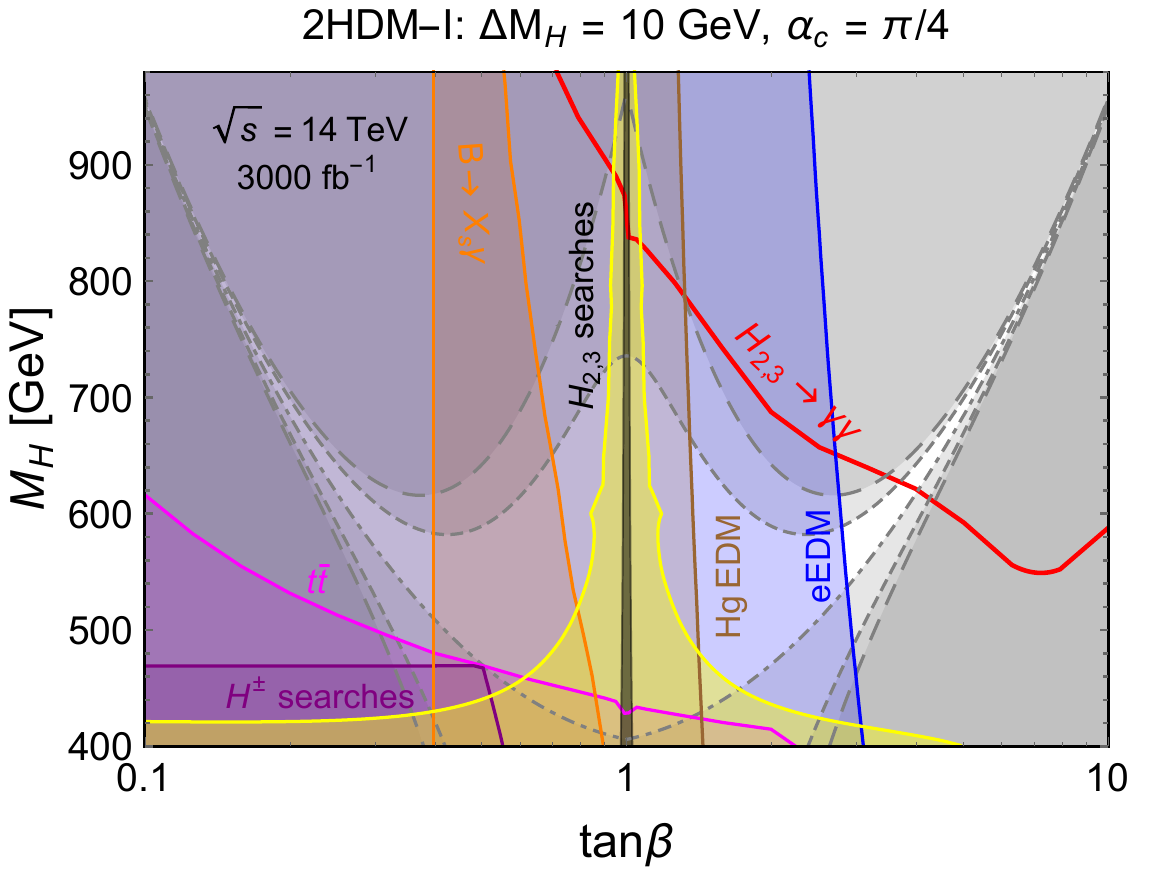} \hspace{-7.5pt}
  \includegraphics[width=0.4\textwidth]{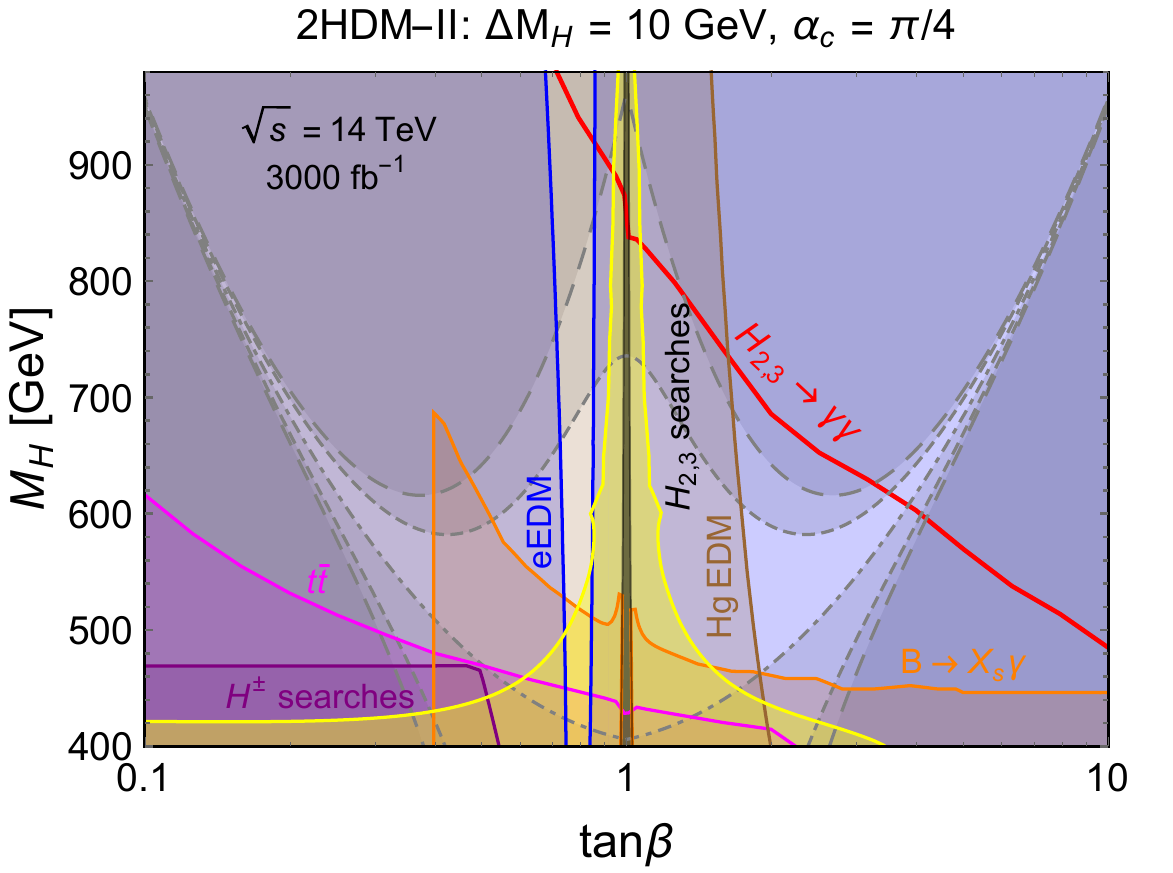}
  \caption{The same as in Fig.~\ref{fig:sensitivity1}, with $\alpha_c = \pi/4$, $\Delta M_H  =1$ GeV (upper) and 10 GeV (lower). More limits are shown in these plots: the combined direct searches of $H_{2,3} \to WW/ZZ$, $hh$ and $hZ$ at the LHC~\cite{Aad:2015kna, Khachatryan:2015cwa, Aaboud:2016okv, Khachatryan:2015yea, Khachatryan:2016sey, ATLAS:2016ixk, Aad:2015wra, TheATLAScollaboration:2016loc} (yellow), the EDM measurements of electron~\cite{Baron:2013eja} (blue) and mercury~\cite{Graner:2016ses} (brown). Within the dark bands, we can not find any physical solution for the relation~(\ref{eqn:relation}).}
  \label{fig:sensitivity2}
\end{figure*}

\begin{figure*}[!t]
  \centering
  \includegraphics[width=0.4\textwidth]{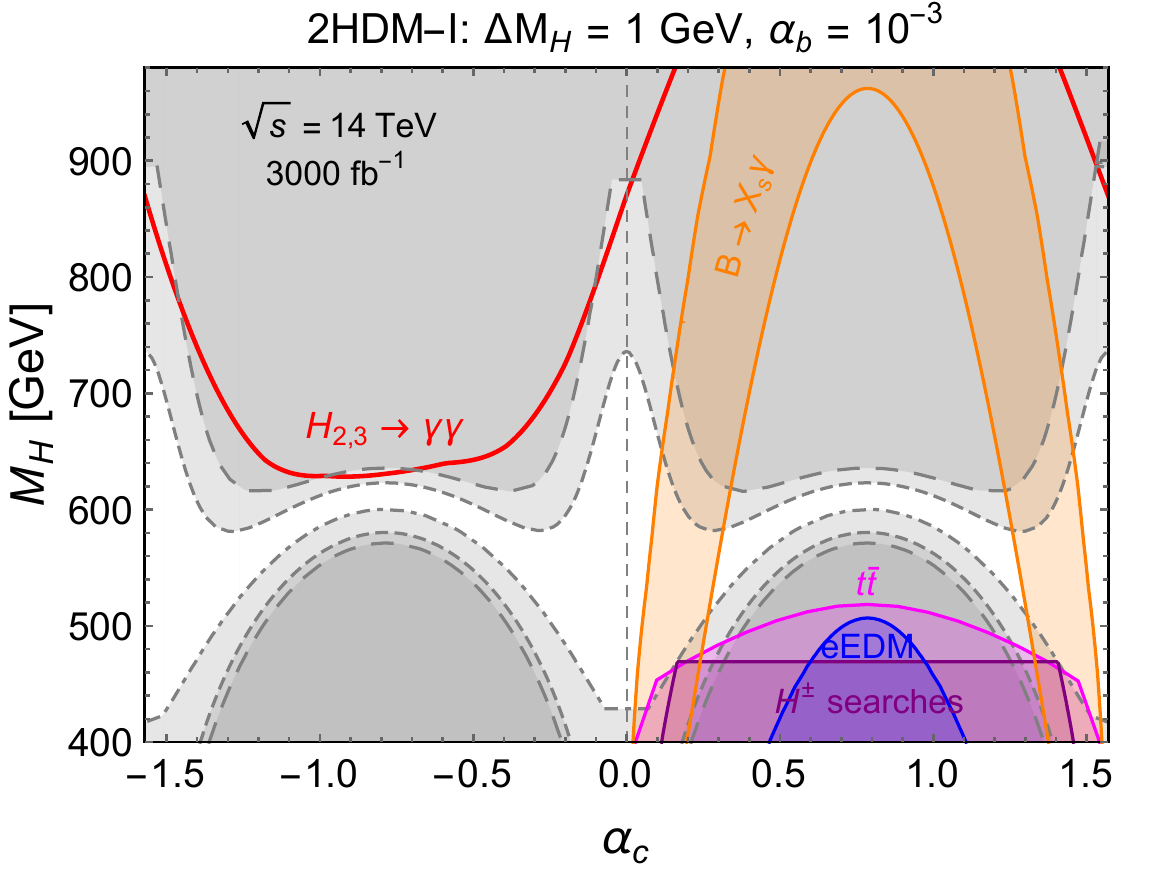} \hspace{-7.5pt}
  \includegraphics[width=0.4\textwidth]{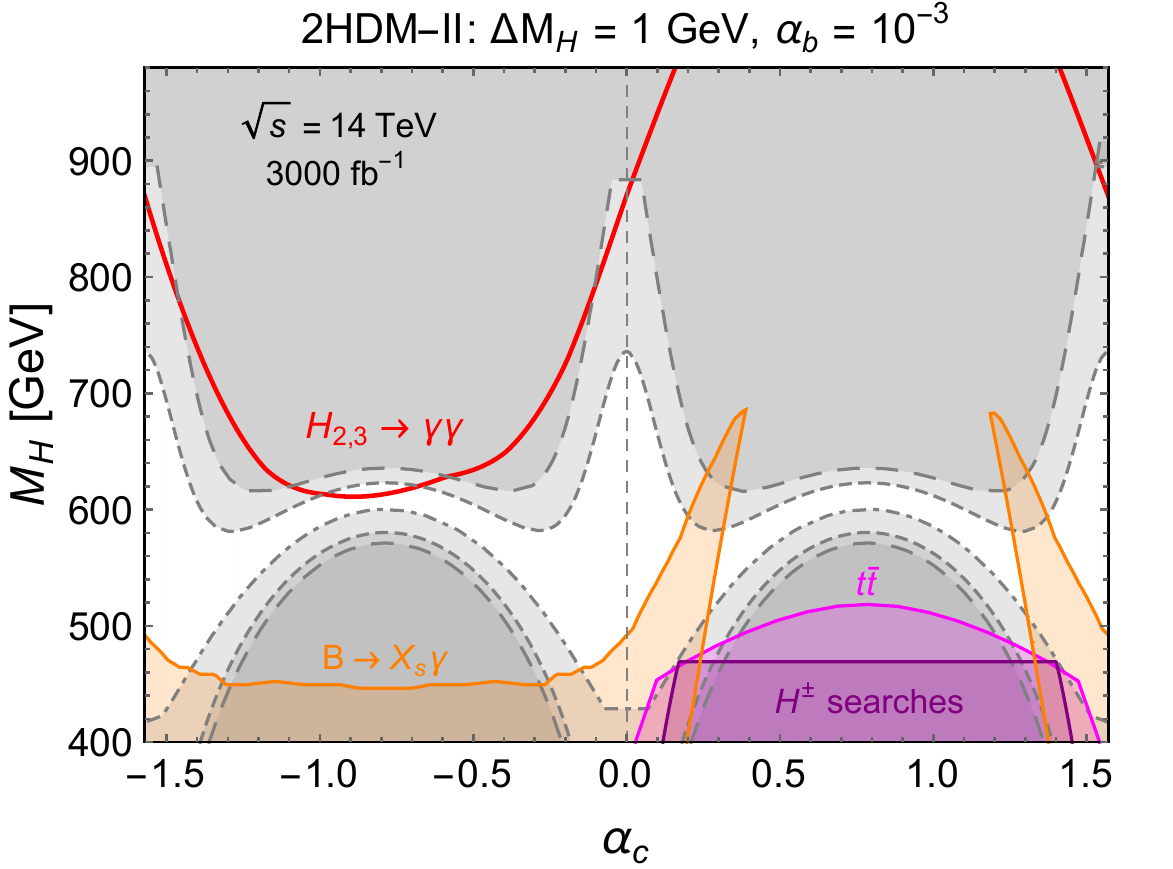}
  \caption{The same as in Figs.~\ref{fig:sensitivity1} and \ref{fig:sensitivity2} in the $M_H - \alpha_c$ plane, with $\alpha_b = 10^{-3}$ and $\Delta M_H  =1$ GeV. The scenarios with larger splitting $\Delta M_H = 10$ GeV is excluded by the theoretical arguments of perturbativity, unitarity and stability. }
  \label{fig:sensitivity3}
\end{figure*}

\begin{figure*}[!t]
  \centering
  \includegraphics[width=0.4\textwidth]{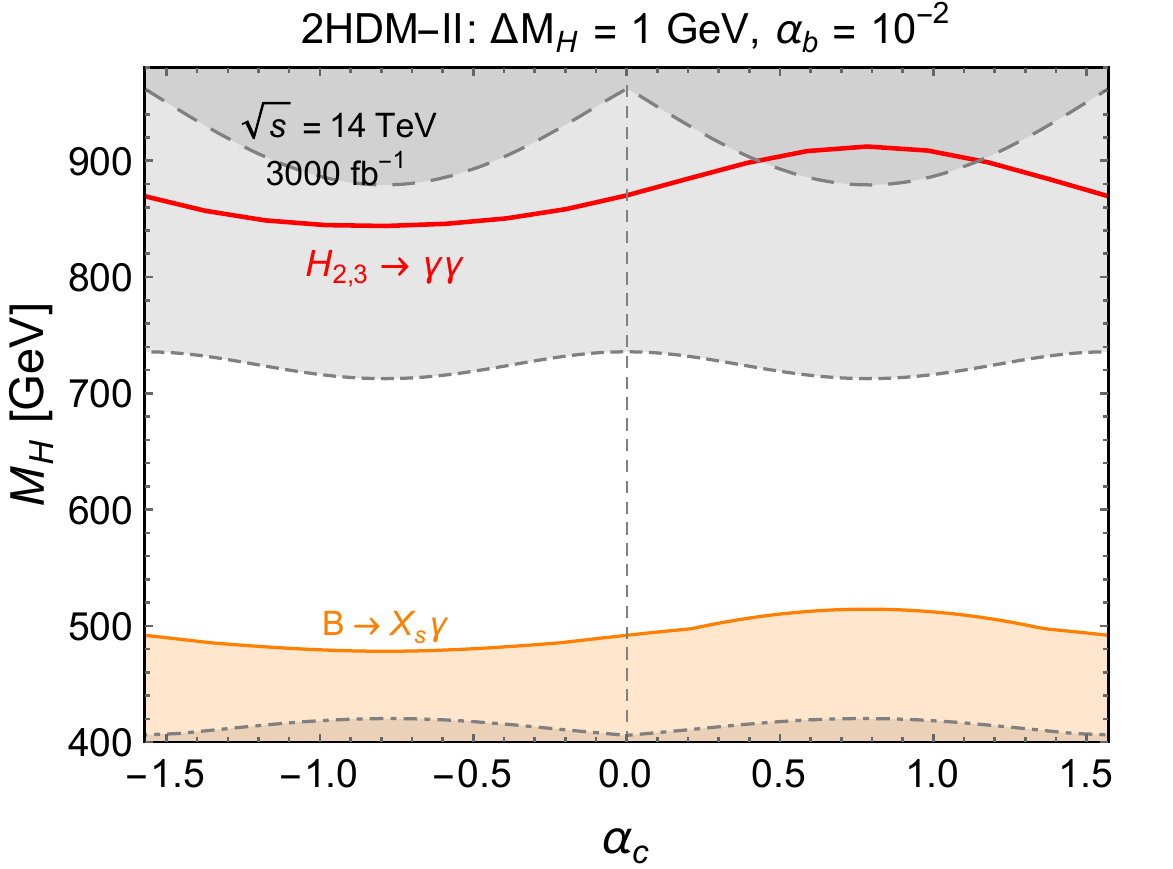} \\
  \includegraphics[width=0.4\textwidth]{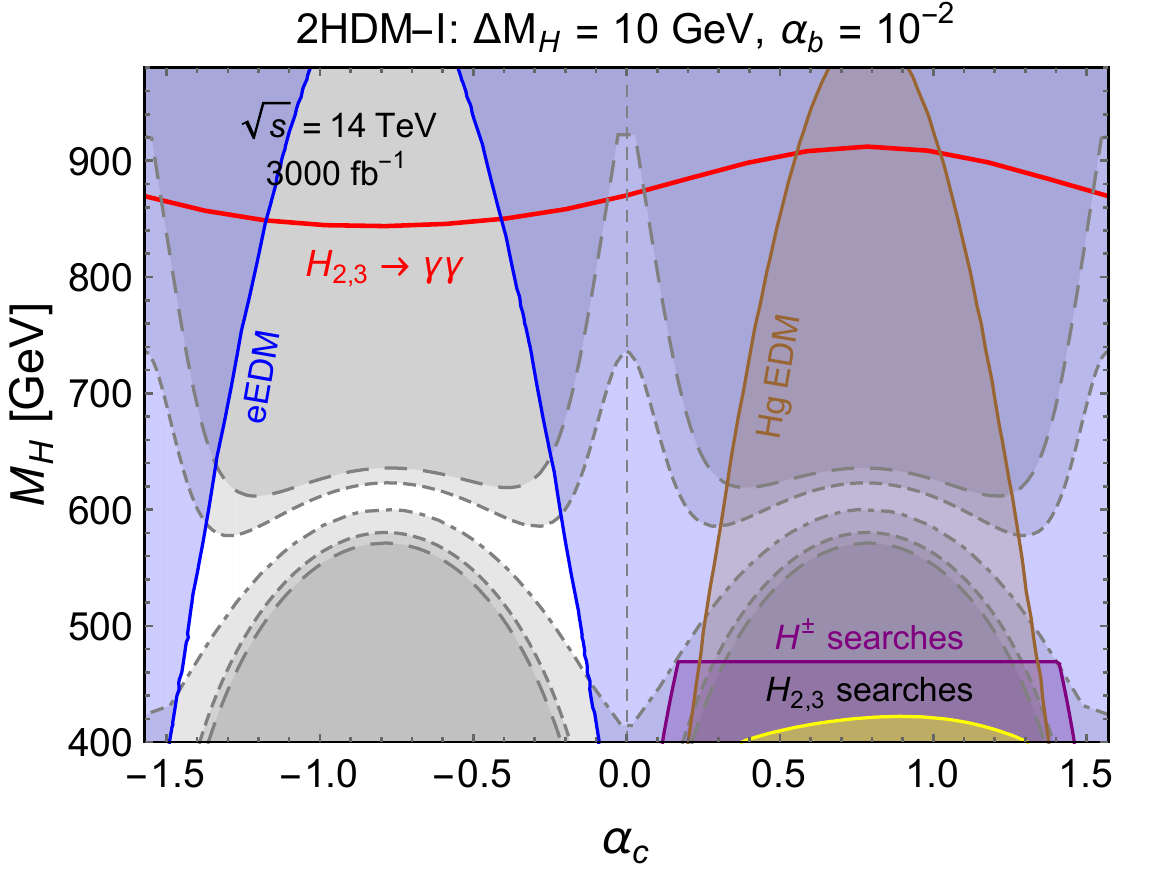} \hspace{-7.5pt}
  \includegraphics[width=0.4\textwidth]{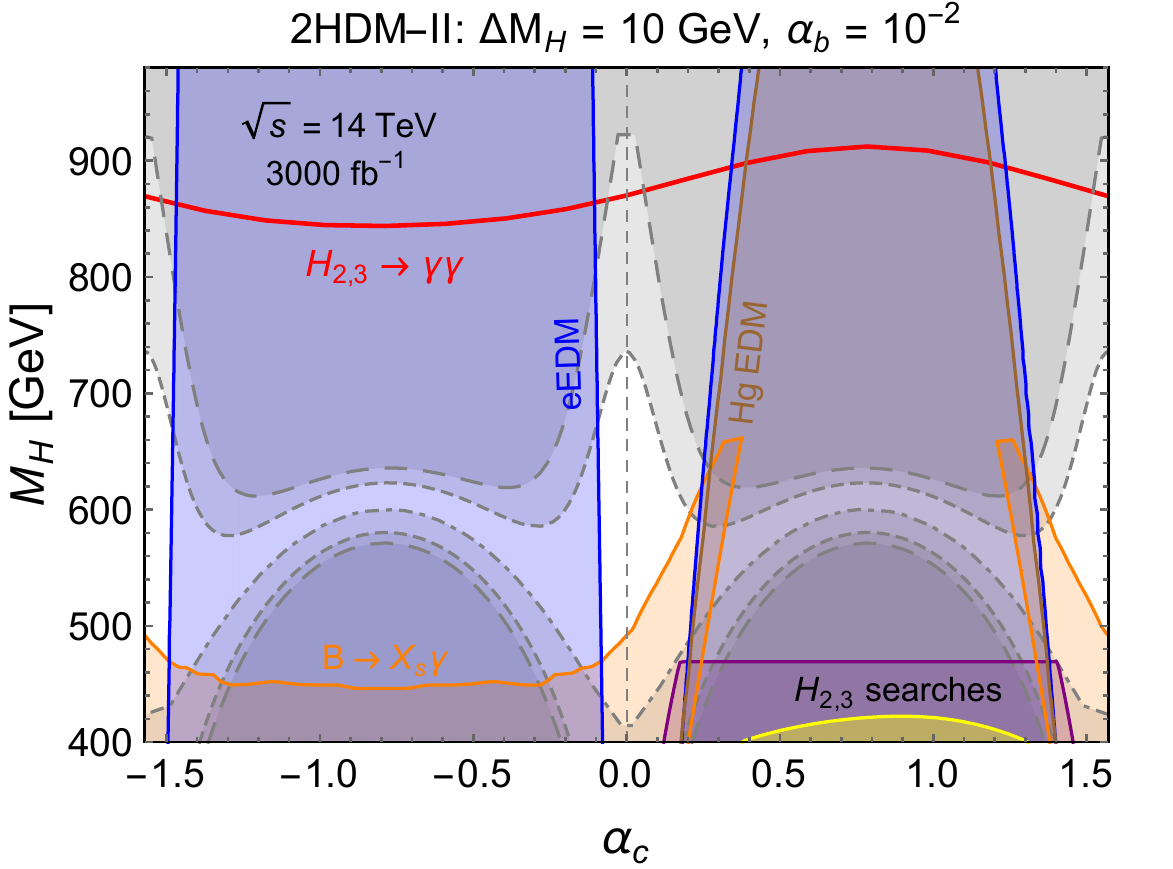}
  \caption{The same as in Figs.~\ref{fig:sensitivity1} and \ref{fig:sensitivity2} in the $M_H - \alpha_c$ plane, with $\alpha_b = 10^{-2}$, $\Delta M_H  =1$ GeV (upper) and 10 GeV (lower). The type-I 2HDM with $\Delta M_H = 1$ GeV is excluded by the electron EDM measurements~\cite{Baron:2013eja}.}
  \label{fig:sensitivity4}
\end{figure*}

\begin{figure*}[!t]
  \centering
  \includegraphics[width=0.4\textwidth]{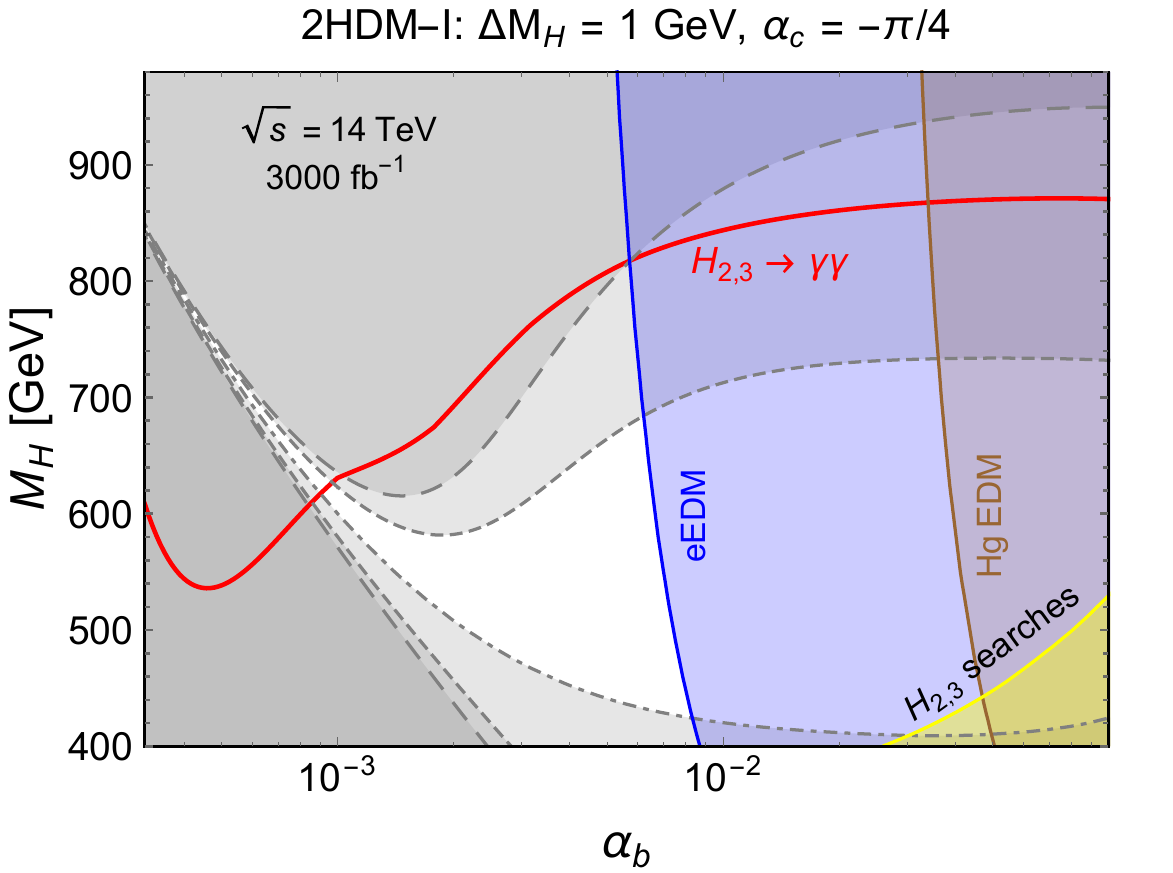} \hspace{-7.5pt}
  \includegraphics[width=0.4\textwidth]{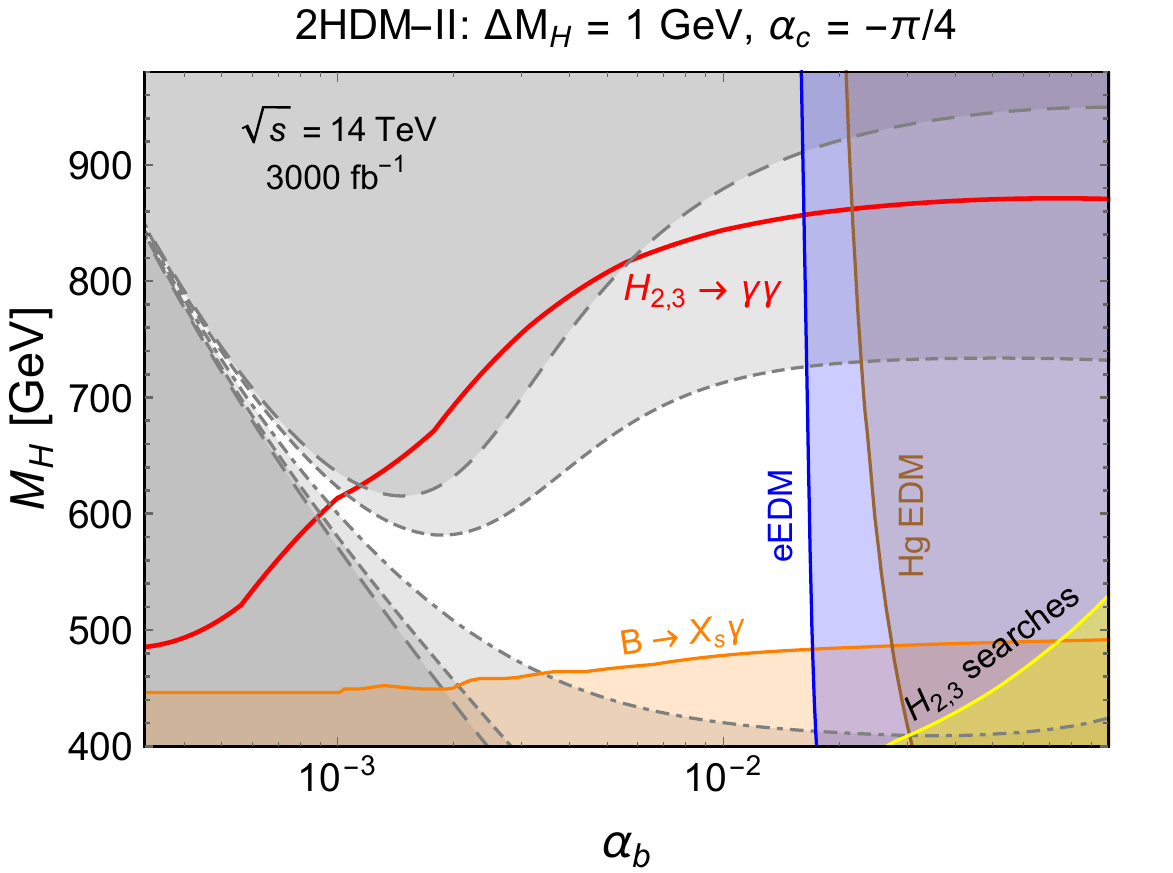} \\
  \includegraphics[width=0.4\textwidth]{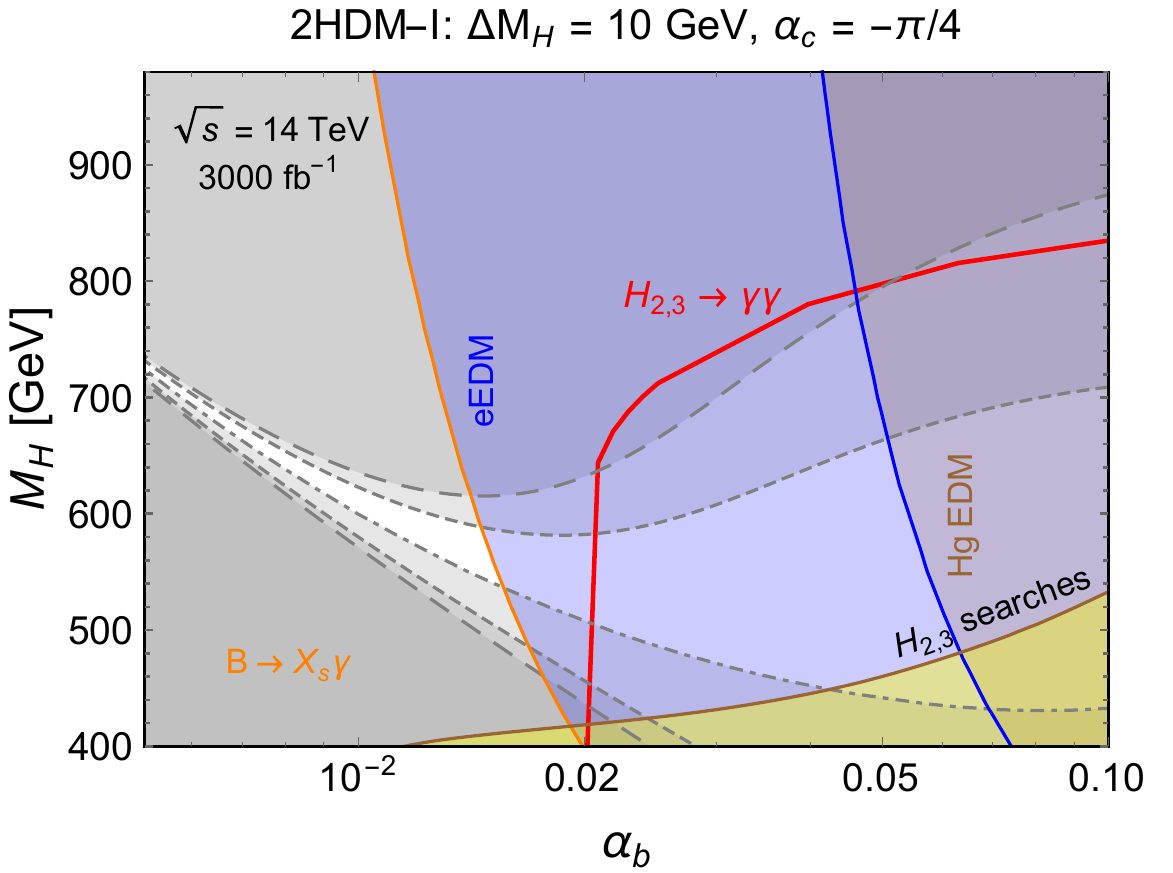}
  \caption{The same as in Figs.~\ref{fig:sensitivity1} and \ref{fig:sensitivity2} in the $M_H - \alpha_b$ plane, with $\alpha_c = -\pi/4$, $\Delta M_H  =1$ GeV (upper) and 10 GeV (lower). The type-II 2HDM with $\Delta M_H = 10$ GeV is excluded by the electron EDM measurements~\cite{Baron:2013eja}.}
  \label{fig:sensitivity5}
\end{figure*}

\begin{figure*}[!t]
  \centering
  \includegraphics[width=0.4\textwidth]{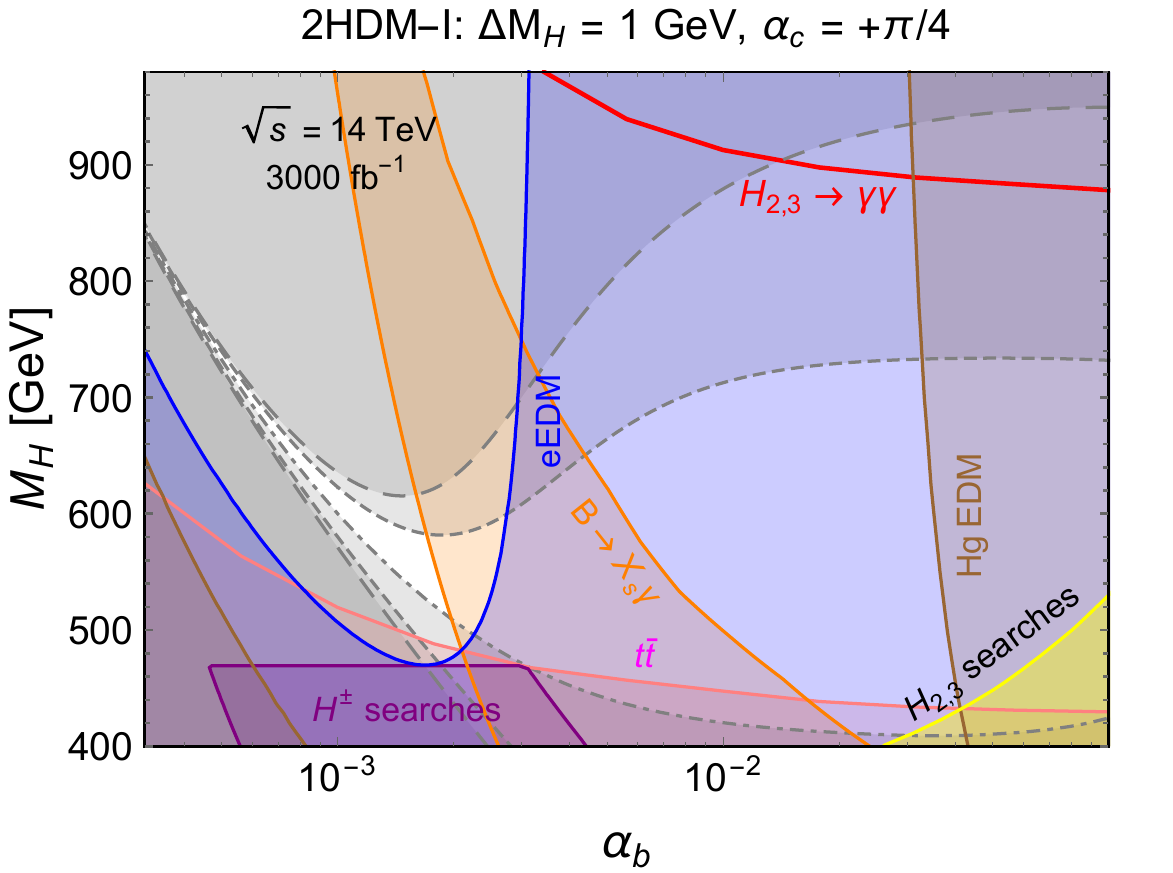} \hspace{-7.5pt}
  \includegraphics[width=0.4\textwidth]{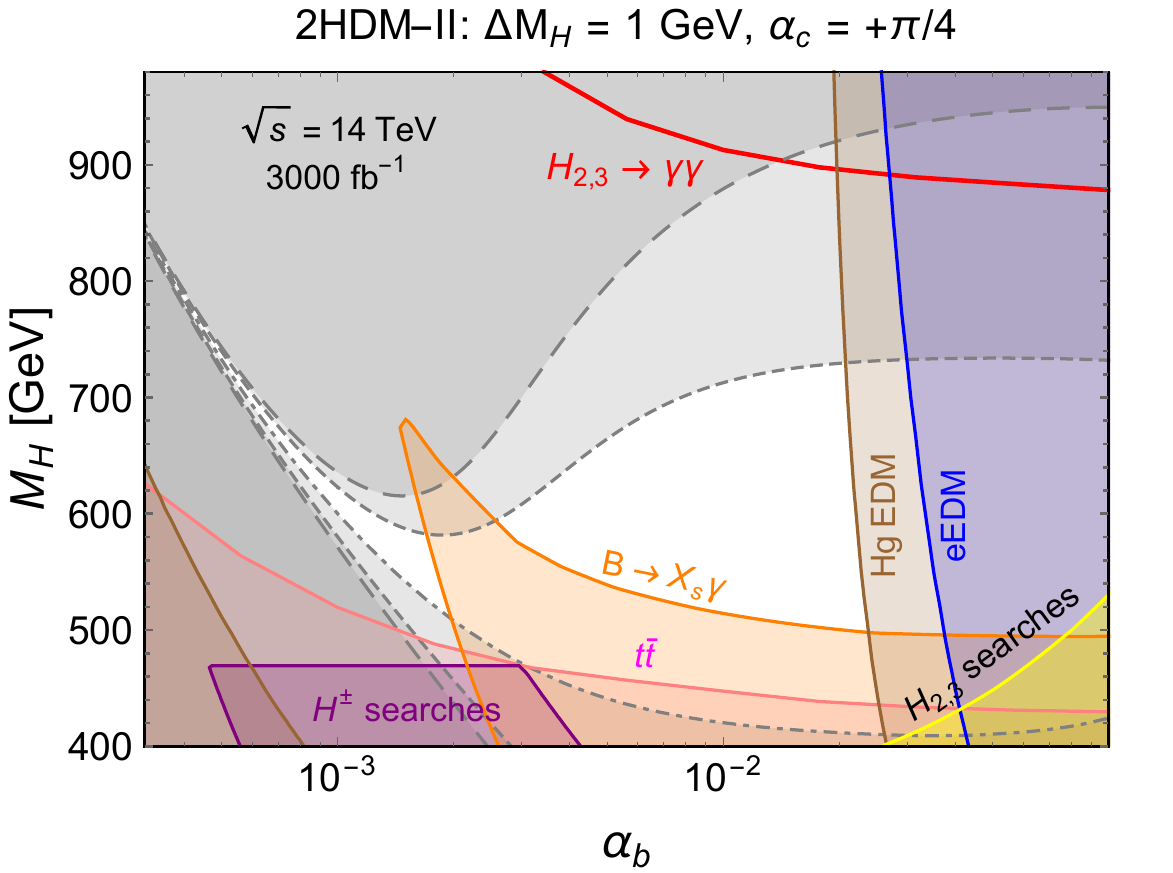}
  \caption{The same as in Figs.~\ref{fig:sensitivity1} and \ref{fig:sensitivity2} in the $M_H - \alpha_b$ plane, with $\alpha_c = +\pi/4$ and $\Delta M_H  =1$ GeV. The scenarios with $\Delta M_H = 10$ GeV is excluded by the electron and mercury EDM measurements~\cite{Baron:2013eja, Graner:2016ses}.}
  \label{fig:sensitivity6}
\end{figure*}

\begin{figure*}[!t]
  \centering
  \includegraphics[width=0.4\textwidth]{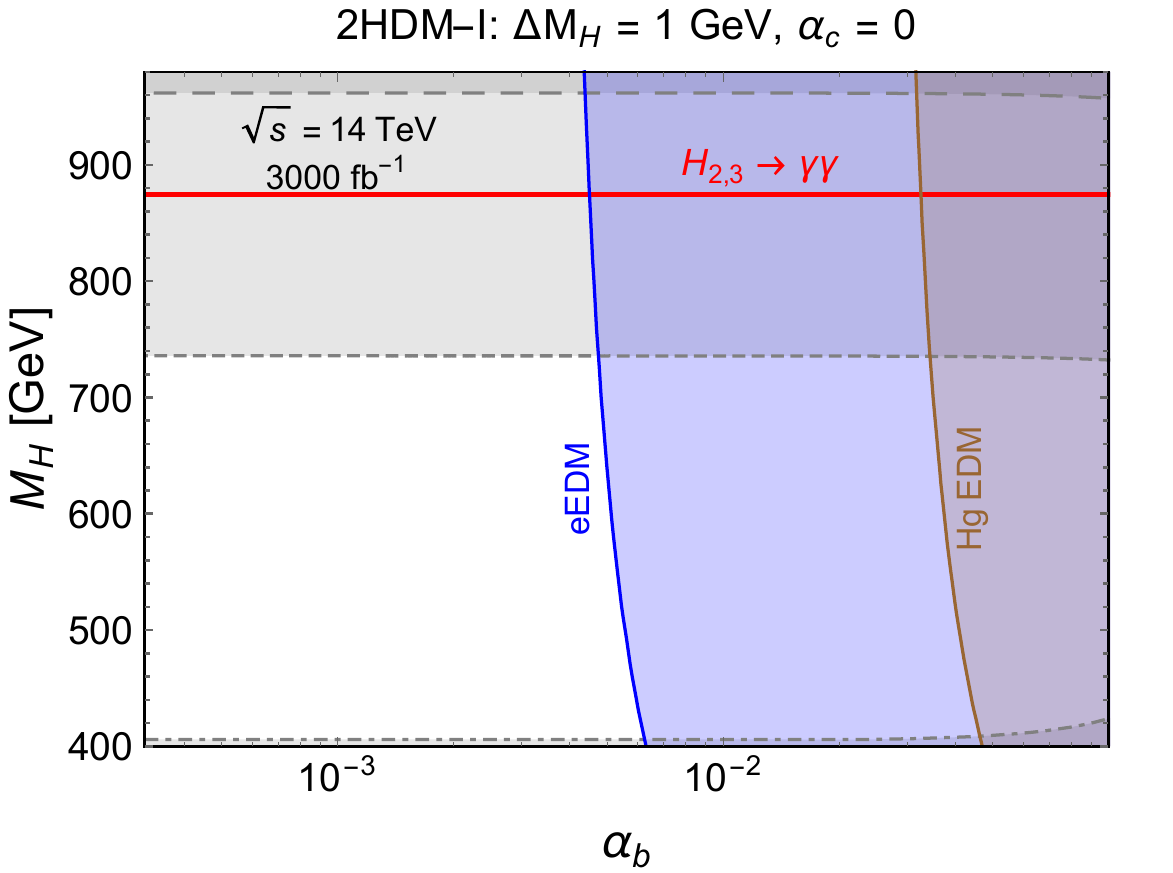} \hspace{-7.5pt}
  \includegraphics[width=0.4\textwidth]{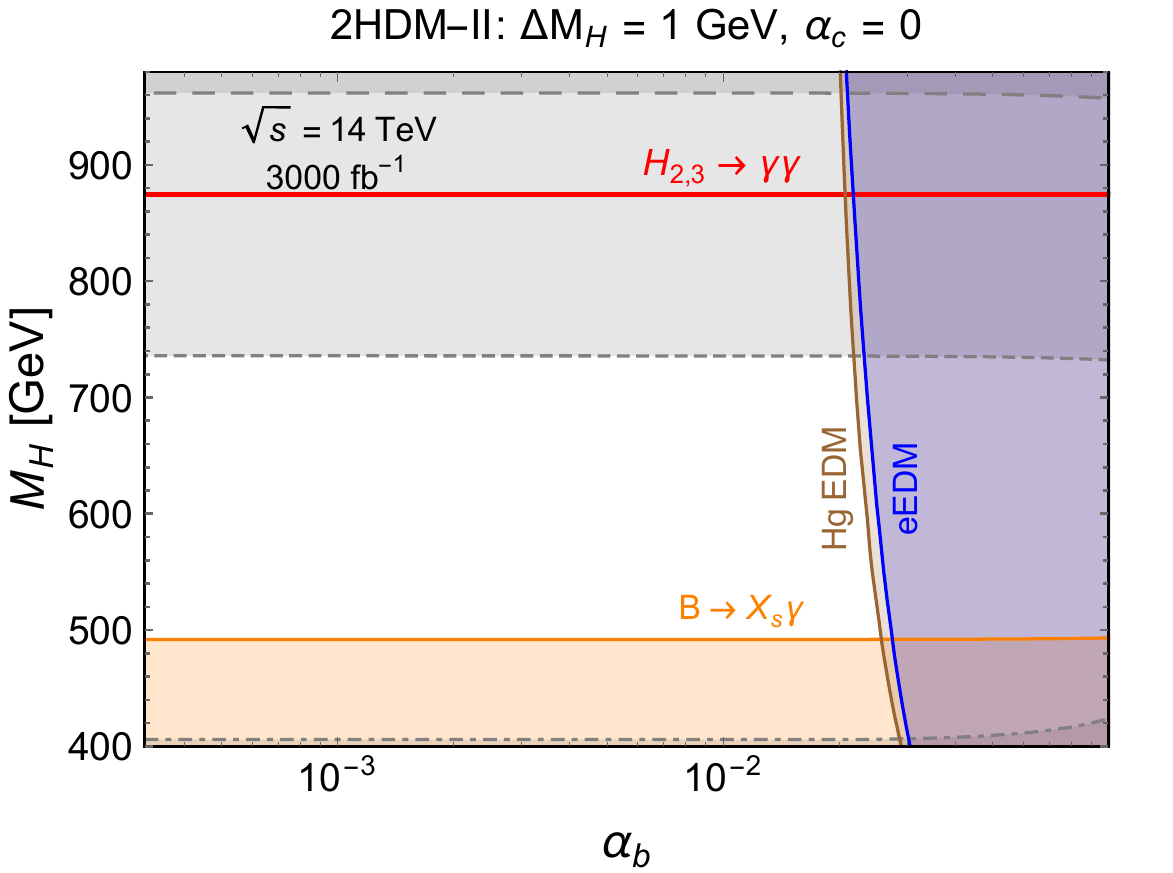} \\
  \includegraphics[width=0.4\textwidth]{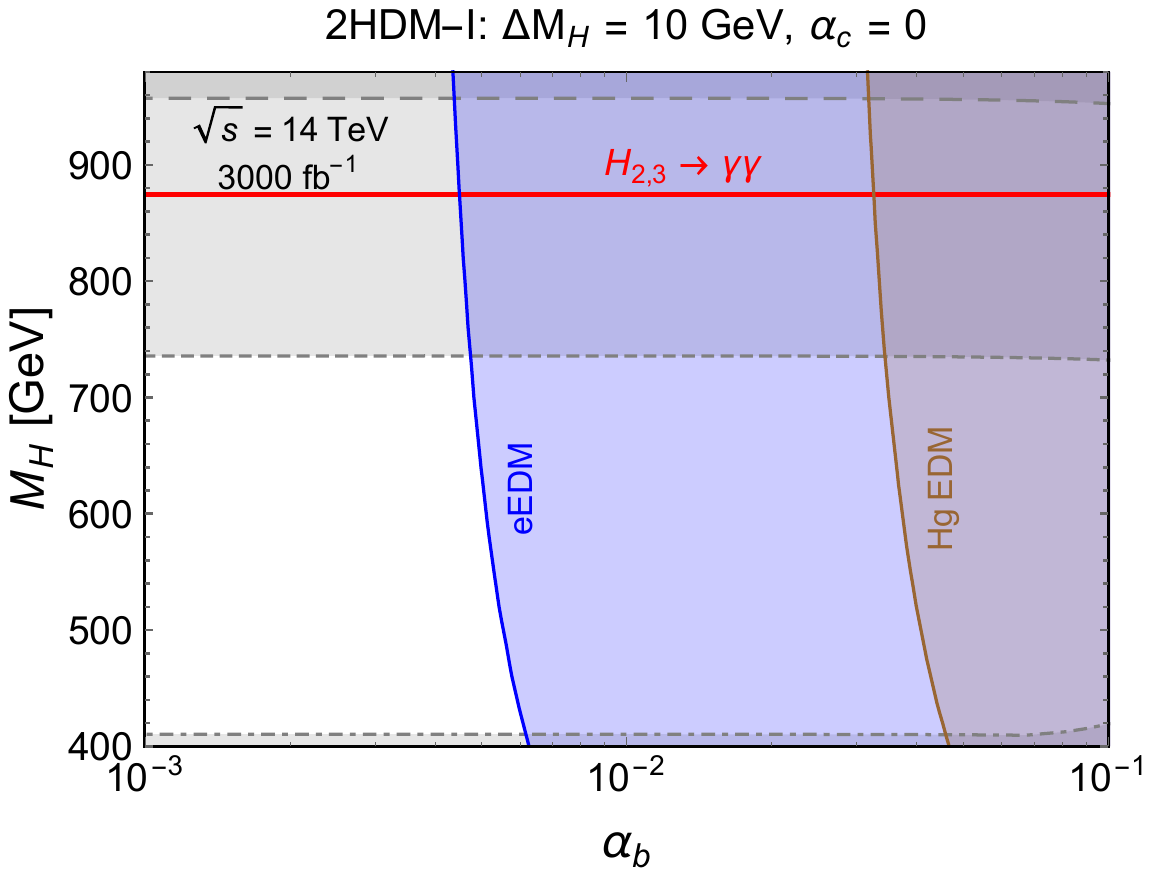} \hspace{-7.5pt}
  \includegraphics[width=0.4\textwidth]{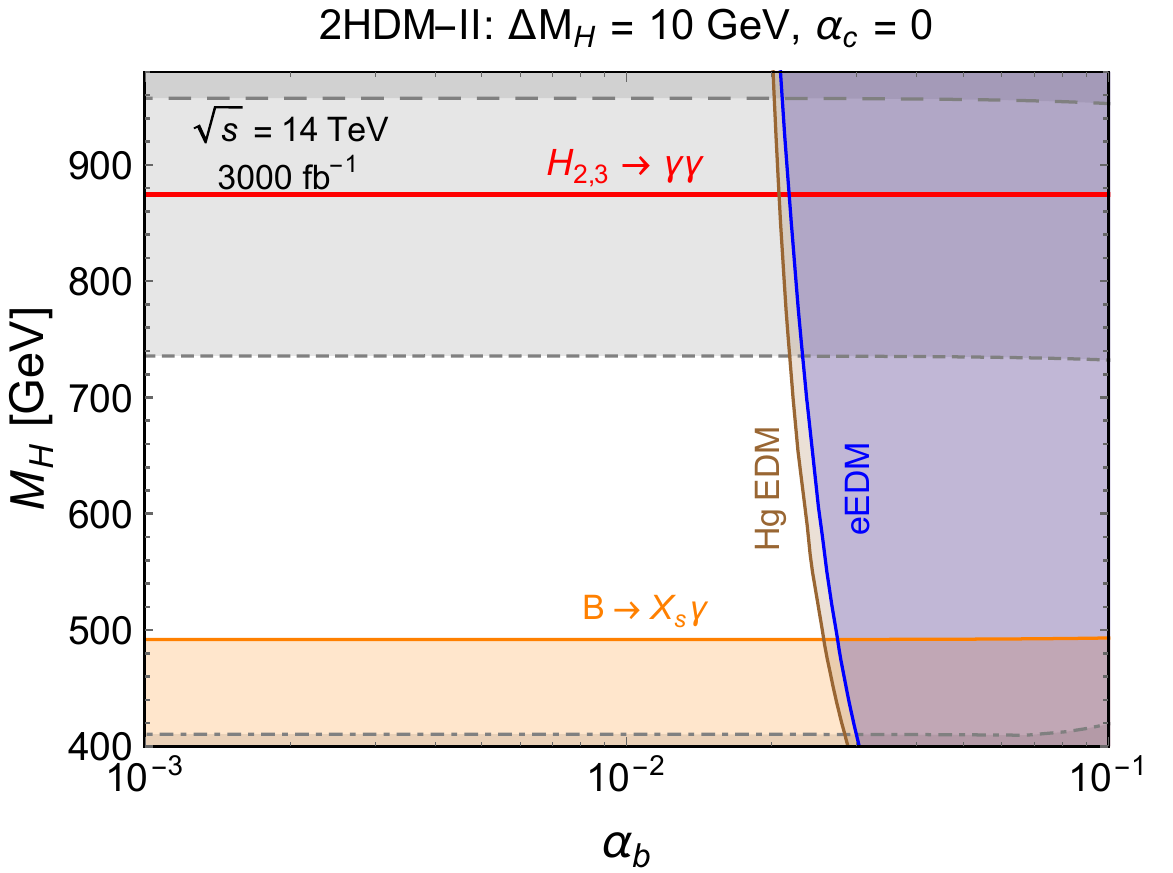}
  \caption{The same as in Figs.~\ref{fig:sensitivity1} and \ref{fig:sensitivity2} in the $M_H - \alpha_b$ plane, with $\alpha_c = 0$, $\Delta M_H  =1$ GeV (upper) and 10 GeV (lower).}
  \label{fig:sensitivity7}
\end{figure*}

All the diphoton prospects at the LHC are presented in the two-dimensional space of $M_H -\tan\beta$, $M_H - \alpha_c$ and $M_H - \alpha_b$ in Figs.~\ref{fig:sensitivity1} to \ref{fig:sensitivity7}, for both the type-I and type-II CPV 2HDM, with two typical values of small splitting $\Delta M_H = 1$ GeV and 10 GeV, as above, and some benchmark values of the mixing parameters $\alpha_b$ or $\alpha_c$. All the regions below the red lines are probable at the HL-LHC at the 95\% CL, with $\sqrt{s} = 14$ TeV and an integrated luminosity of 3000 fb$^{-1}$. The collider limits from differential $t\bar{t}$ data in the right panel of Fig.~\ref{fig:LHC}, direct $H^\pm$ searches and $B \to X_s \gamma$ data in Fig.~\ref{fig:Hpm} are also shown, respectively, as the pink, purple and orange shaded regions. The gray regions are excluded by the theoretical arguments of perutrbativity, stability and unitarity. In some of the plots the combined limits from direct searches of $H_{2,3}$ in the final state of $WW/ZZ$, $hh$ and $hZ$ are shaded in yellow, while the electron and mercury EDM limits are in blue and brown. Note that the perturbativity, unitarity and stability requirements on the theoretical side are rather stringent: Depending on $\tan\beta$ and the mass and mixing parameters, the quartic couplings $\lambda_i$ in the scalar potential are tightly constrained. We will not scan the full parameter space for the theoretical limits, but rather for the sake of concreteness, in all these plots we have taken the values of $M_{H^\pm} = M_H$ and $m_{\rm soft} = 300$ GeV. In the quasi-degenerate case, depending on $\Delta M_H$ and $\alpha_{b,c}$, the heavy scalar mass $M_H$ is required to be roughly within a range of 400 GeV to 1 TeV, unless some parameters in 2HDM are fine-tuned. Heavy scalars beyond 1 TeV push some of the quartic couplings $\lambda_i$ non-perturbative, while a smaller $M_H$ drives the stability conditions violated (depending on the parameter $m_{\rm soft}$).

The diphoton sensitivities as well as the theoretical and experimental constraints, in the parameter space of $M_H - \tan\beta$ are presented in Fig.~\ref{fig:sensitivity1}, where we have set $\alpha_c = 0$ (and resultantly $\alpha_b = 0$). In the CP conserving limit of $\alpha_{b,c} = 0$, we do not have the limits from direct searches of $H_{2,3} \to WW/ZZ$, $hh$ and $hZ$, and the purely CPV phenomena of EDMs. A large region is excluded by the theoretical arguments and the $B \to X_s \gamma$ data. With a high luminosity of 3000 fb$^{-1}$ at the LHC, almost all the allowed regions in Fig.~\ref{fig:sensitivity1} could be tested in the diphoton channel, though the ${\rm BR} (H_{2,3} \to \gamma\gamma)$ is rather small compared to other decay modes. One should note that the red lines go beyond 1 TeV when $\tan\beta \lesssim 1$, which are not shown explicitly in these plots; the small $\tan\beta$ regions are all covered implicitly.

The maximal CPV case with $\alpha_c = \pi/4$ is shown in Fig.~\ref{fig:sensitivity2}, where the EDM constraints becomes important. Depending on the Yukawa couplings and the mass splitting $\Delta M_H$, a sizable region in the $M_H - \tan\beta$ plane has been excluded by the EDM measurements. In the case of type-II 2HDM with $\Delta M_H = 10$ GeV, when the electron and mercury EDM constraints are combined together, the whole $M_H - \tan\beta$ plane is excluded, see the lower right panel in Fig.~\ref{fig:sensitivity2}. This excluded scenario could be confirmed or falsified at the LHC via the diphoton searches $gg \to H_{2,3} \to \gamma\gamma$. A positive signal in the excluded region would imply the incompleteness of CPV 2HDM at the TeV scale, which has to be further extended, or the experimental data should be interpreted in other beyond SM frameworks.

The projected diphoton sensitivities in the $M_H - \alpha_c$ plane are presented in Figs.~\ref{fig:sensitivity3} and \ref{fig:sensitivity4}, with respectively $\alpha_b = 10^{-3}$ and $10^{-2}$. For the small CPV angle $\alpha_b = 10^{-3}$ in Fig.~\ref{fig:sensitivity3}, the 2HDM contribution to the EDMs are highly suppressed, and could not provide any limits beyond the theoretical constraints. However, when $\alpha_b$ becomes larger, e.g. $10^{-2}$ in the plots of Fig.~\ref{fig:sensitivity4}, the electron and mercury EDMs could exclude large parameter space as in Fig.~\ref{fig:sensitivity2}. In particular, the type-I 2HDM with $\Delta M_H = 1$ GeV is all excluded by the electron EDM, see the caption of Fig.~\ref{fig:sensitivity4}. It is transparent in these plots that the searches of $H_{2,3} \to \gamma\gamma$ at the LHC could probe the whole allowed regions in the $M_{H} - \alpha_c$ plane, at least for the benchmark scenarios given here.

In the following Figs.~\ref{fig:sensitivity5} and \ref{fig:sensitivity6} we show the diphoton sensitivities and the constraints in the parameter space of $M_H$ and $\alpha_b$, with respectively $\alpha_c = - \pi/4$ and $+\pi/4$. For comparison, the $\alpha_c = 0$ case is shown in Fig.~\ref{fig:sensitivity7}, in which $\tan\beta = 1$, $\alpha_b \neq 0$ and the EDMs from CPV 2HDM are purely the $\alpha_b$-relevant contributions. Obviously the EDM measurements exclude large values of $\alpha_b$, depending on other parameters in the 2HDM. As in Figs.~\ref{fig:sensitivity3} and \ref{fig:sensitivity4}, some of the scenarios have been completely excluded by the EDM data, e.g. the type-II 2HDM with $\alpha_c = -\pi/4$ and $\Delta M_H = 10$ GeV, and both the type-I and type-II 2HDM with $\alpha_c = +\pi/4$ and $\Delta M_H = 10$ GeV.
For negative $\alpha_c$ in Fig.~\ref{fig:sensitivity5}, $\tan\beta >1$, and the production cross section of heavy scalars $\sigma (gg \to H_{2,3})$ is suppressed, when compared to the positive $\alpha_c$ case (and $\tan\beta < 1$) in Fig.~\ref{fig:sensitivity6}, therefore smaller regions could be probed in Fig.~\ref{fig:sensitivity5}, in particular when $\alpha_b$ is small and $\Delta M_H$ is large (see the lower panel in Fig.~\ref{fig:sensitivity5}). In contrast, in Fig.~\ref{fig:sensitivity7}, with $\alpha_c = 0$, $\tan\beta = 1$ is a constant, and thus the diphoton sensitivities are almost horizontal lines. Again, almost the whole parameter space could, in principle, be probed in the diphoton channel of heavy scalar decay in CPV 2HDM.

\begin{figure*}[!t]
  \centering
  \includegraphics[width=0.4\textwidth]{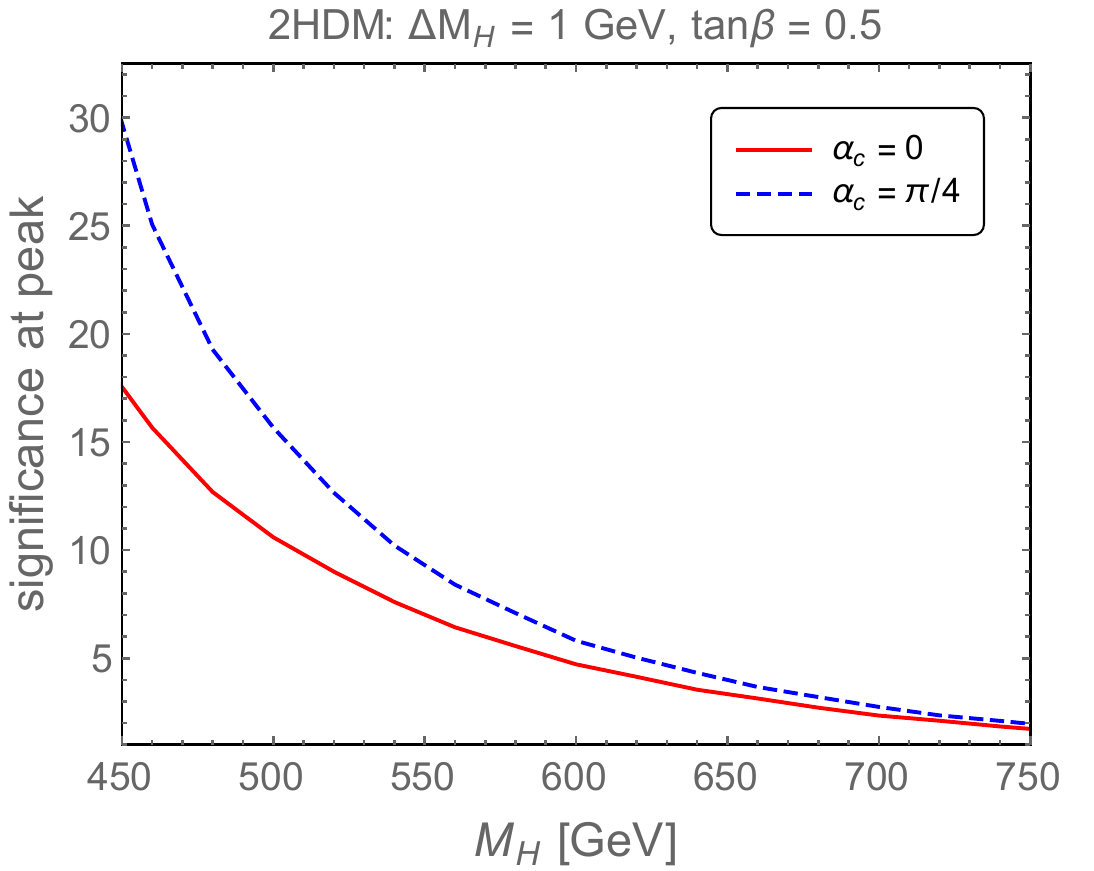} \hspace{-7.5pt}
  \includegraphics[width=0.4\textwidth]{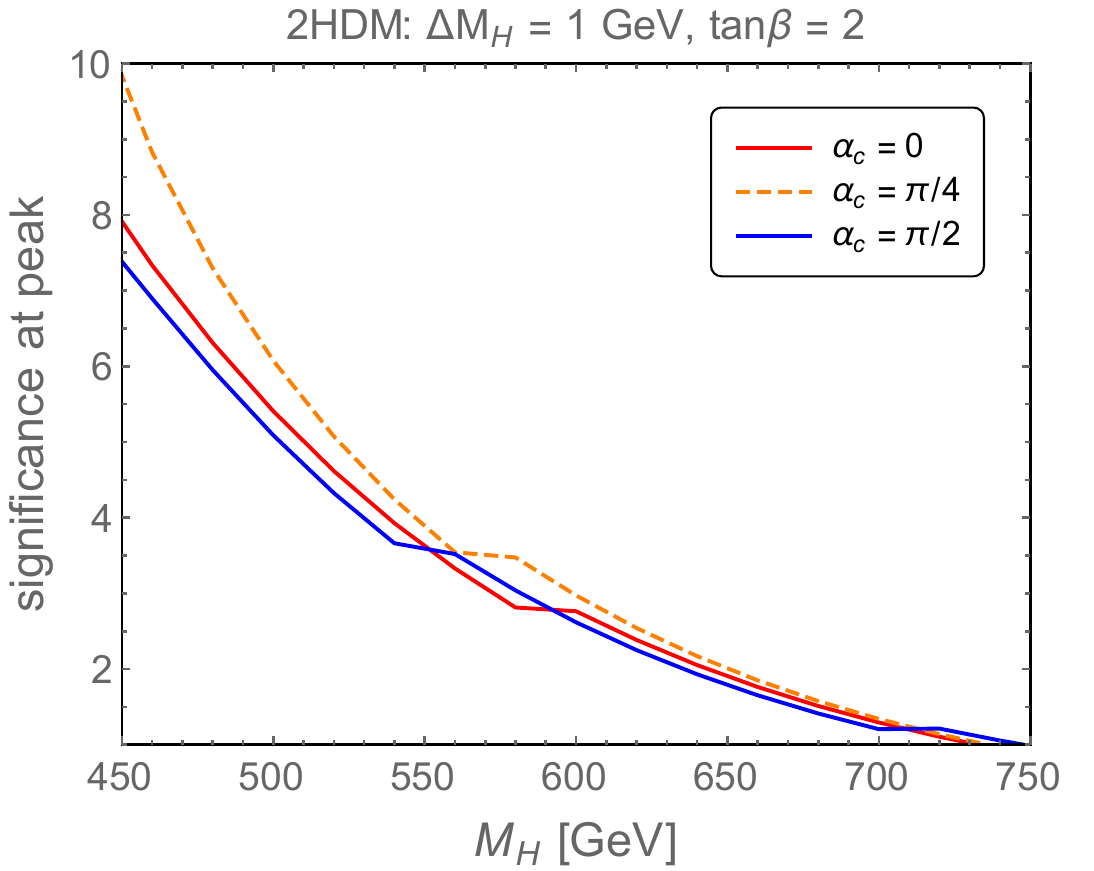}
  \caption{Significance at the resonance peak with a bin width of 10 GeV for the CPV 2HDM with $\Delta M_H = 1$ GeV, $\tan\beta = 0.5$ (left) and $2$ (right), and $\alpha_c = 0$, $\pi/4$ (and $\pi/2$ for $\tan\beta = 2$). The sensitivities for type-I and type-II 2HDM are almost the same.}
  \label{fig:sensitivity_CP}
\end{figure*}

To demonstrate the prospects of distinguishing the 2HDM scenarios with different CPVs in the scalar sector at the LHC, we compare the significance defined in Eq.~(\ref{eqn:chi2}) for the benchmark models given in Fig.~\ref{fig:example_CP}. As the CPV effects are most significant at the resonance, we count for simplicity only the single bin (with a bin width of 10 GeV) at the peak, which is rather conservative from this aspect. With a larger coupling to the top quark, the scenarios with $\tan\beta = 0.5$ have a larger cross section at the peak and thus higher significances in the left panel of Fig.~\ref{fig:sensitivity_CP}. As shown in Fig.~\ref{fig:example_CP}, a maximal CP violating mixing of the two heavy scalars, i.e. $\alpha_c = \pi/4$, could enhance significantly the cross sections at the peak, thus the lines in Fig.~\ref{fig:sensitivity_CP} with $\alpha_c = \pi/4$ have a larger significance, compared to the CP conserving limit of $\alpha_b = 0$ (and $\alpha_c = \pi/2$), especially when the scalars $H_{2,3}$ are not too heavy. Comparing the expected significances with different $\alpha_c$ in Fig.~\ref{fig:sensitivity_CP}, we could distinguish the maximal mixing case $\alpha_c = \pi/4$ from the CP conserving model at the HL-LHC, if the heavy scalar mass $M_H \lesssim 600$ (500) GeV for $\tan\beta = 0.5$ ($2$). Here we have considered only the peak bins, with more diphoton bins included, the distinguishing power could be further improved.

\subsection{Discussions}

All the theoretical and experimental limits and the diphoton prospects at HL-LHC in the framework of CPV 2HDM have been extensively studied in depth in the sections above. Before proceeding to the conclusion we comment in this section on the various effects of different mass, mixing and coupling parameters on these limits and the diphoton signal in both the type-I and type-II CPV 2HDMs, and grab some of the qualitative features.

One should be first aware of the constraints from the theoretical requirements of unitarity, perturbativity and stability in the scalar sector with two quasi-degenerate heavy scalars $H_{2,3}$, which prefer a $\tan\beta \sim 1$, as a sufficiently large $\tan\beta$ (or $\cot\beta$) would easily push the quartic couplings $\lambda_i$ non-perturbative or unitarity-violating: For instance, the prefactors $\sin^{-1}\beta$ ($\cos^{-1}\beta$) in Eqs.~(\ref{eqn:lambda1}) to (\ref{eqn:lambda3}) would be large in the limit of $\tan\beta \ll 1$ ($\tan\beta \gg 1$). In this sense, the theoretical limits are approximately ``invariant'' under the exchange $\tan\beta \leftrightarrow \cot\beta$, which is transparently demonstrated in the example shown in Figs.~\ref{fig:sensitivity1} to \ref{fig:sensitivity4}. Furthermore, these theoretical requirements also set both lower and upper limits on the quasi-degenerate heavy scalar masses $M_H$. An upper bound is easy to be understood, as, without the SM gauge symmetry extended, all the scalars in the 2HDM masses are proportional to the EW VEV $v$ through $M_i \sim \sqrt{\lambda} v$, therefore, with the perturbative couplings $|\lambda_i| < 4\pi$, one should expect all the heavy scalars in 2HDM are roughly below the TeV scale. In Eqs.~(\ref{eqn:lambda1}) and \ref{eqn:lambda2}, there are minus terms $-\nu$ which are proportional to the $Z_2$ soft breaking parameter $m_{\rm soft}^2$, then the lower bound on the heavy scalars depend largely on $m_{\rm soft}$. With the mass splitting $\Delta M_H$ gets larger, these theoretical constraints might be to some extent weakened, but the CPV phases $\alpha_{b,\,c}$ would ge generally much smaller, and we will lose the significant CPV effects at the ``double resonance''.

In a large region of parameter space, the production and decay processes $gg \to H_{2,3} \to \gamma\gamma$ are dominated by the top quark loop, thanks to the fact of $\tan\beta \sim 1$ (if $\tan\beta \sim m_t/m_b$, the bottom quark and $W^\pm$ loop would be very important), therefore the diphoton signal depend largely on the value of $\tan\beta$ (see Figs.~\ref{fig:sensitivity1} and \ref{fig:sensitivity2}) and could be largely enhanced by the CPV angle $\alpha_c$ ($\alpha_b$ is generally much smaller) at the resonance peak (see Figs.~\ref{fig:example_CP} and \ref{fig:sensitivity_CP}). The couplings to the down-type quarks and charged leptons, e.g. whether the couplings are of type-I or type-II, are important in the sense that they determine largely the limits from $B \to X_s \gamma$ and the EDMs of electron and mercury. The mass splitting $\Delta M_H$ is important when the scalar resonances becomes narrower, see the examples in Fig.~\ref{fig:example2} and plays also a important role in evaluating the EDMs, e.g. in Fig.~\ref{fig:sensitivity2}.


\section{Conclusion}
\label{section:conclusion}

In this paper, we have studied in detail the diphoton signal from the decay of two quasi-degenerate heavy scalars in the CPV 2HDM with both type-I and type-II Yukawa couplings. To simplify the scalar potential, we assume there is a soft-breaking $Z_2$ symmetry, under which there are only two CP violating terms in the potential: the soft-breaking mass parameter ${\rm Im}\, m_{12}^2$ and one of the quartic couplings ${\rm Im}\, \lambda_5$. With these CP violating terms, the three neutral scalars are no longer CP eigenstates but all mix with each other, with the lightest one being SM-like with mass of 125 GeV, leaving the other two heavier. The CP violating mixing angles $\alpha_{b,c}$ are linked intimately to the scalar masses, in particular depending non-trivially on the mass splitting of the two heavier states $H_{2,3}$. Roughly speaking, with the two heavy scalars approaching to be degenerate, their mixing tends to be larger, or even maximal, which is in general more important than their CP violating mixing with the SM Higgs which is somewhat suppressed by the large mass splitting of $M_H - m_h$.

Throughout this paper we have considered two benchmark values of small splitting of $\Delta M_H = 1$ GeV and $10$ GeV, and work in the simplified case of $\alpha = \beta - \pi/2$ with $\alpha_{b,c}\neq 0$ which is {consistent with} the current SM Higgs data. We have collected in Section~\ref{sec:limitsall} all the relevant theoretical and experimental limits on the CPV 2HDM, with some typical example shown in the two-dimensional parameter space of heavy scalar mass $M_H$ verses $\tan\beta$, $\alpha_c$ and $\alpha_b$, i.e. Figs.~\ref{fig:limits1} to \ref{fig:limits5} and Figs.~\ref{fig:sensitivity1} to \ref{fig:sensitivity7}. It turns out that the theoretical limits from the requirements of unitarity, perturbativity and stability of the scalar potential impose severe constraints on the parameter space in our model, demanding that the heavy scalar masses satisfy $400 \, {\rm GeV} \lesssim M_H \lesssim 1\, {\rm TeV}$. The direct searches of heavy neutral scalars $H_{2,3} \to WW/ZZ,\, hh,\, hZ$ performed at the LHC could hardly constrain the heavy scalars heavier than roughly 450 GeV, unless there are to some extent fine-tuning in the scalar sector, due to the small branching ratios of $H_{2,3}$ decay into the massive SM bosons. As the heavy scalars decay almost 100\% into the top quark pairs, the consistency of experimental differential $t\bar{t}$ data with the SM predictions could constrain more effectively the couplings of $H_{2,3}$, in particular when $\tan\beta$ is small. Benefitting from the oblique $T$ parameter constraints on the neutral-charged scalar splitting $|M_H - M_\pm|$, the direct search of charged scalars and the rare $B$ decay data of $B \to X_s \gamma$ provide additional limits on the neutral scalar sector. The electron and mercury EDM constraints on the CP violating couplings, e.g. those in Table~\ref{tab:couplings}, exclude also large regions in the parameter space.

Though the branching ratios to diphoton are generally very small, typically of order $10^{-5}$ in a large region of the parameter space, the clean SM background renders it one of the key channels to search for heavy neutral scalars, as for the SM Higgs. The full details of the (differential) diphoton cross section are given in Section~\ref{sec:diphoton}, for both the resonance and interference contributions. The SM background is expected to be much larger than the pure signal resonances, thus the continuum-resonance interference is crucially important for the heavy scalar searches. By na\"ively counting the numbers of diphoton events as functions of the invariant mass $M_{\gamma\gamma}$, we have estimated the expected sensitivities for the searches of $H_{2,3} \to \gamma\gamma$ in the CPV 2HDM at the $\sqrt{s} = 14$ TeV HL-LHC with an integrated luminosity of 3000 fb$^{-1}$, which are presented in Figs.~\ref{fig:sensitivity1} to \ref{fig:sensitivity7}. It turns out that almost all the allowed parameter space could be probed in the diphoton channel, at least at the 95\% CL, which is largely complementary to other direct searches at the LHC, e.g. in the final states of the SM $h$, $W$ and $Z$ bosons. A large mixing $\alpha_c$ of the two nearly-degenerate heavy scalars could enhance significantly the cross section at the resonance peak, see the examples in Fig.~\ref{fig:example_CP}. Therefore with sufficient events collected at the resonance peak, we could obtain some information of CPV in the scalar sector of 2HDM, e.g. the examples given in Fig.~\ref{fig:sensitivity_CP}, which is largely complementary to the low-energy probe of CPV in the EDM experiments.

In this paper we have focused only on the type-I and type-II 2HDM with CPV in the scalar sector, which could be generalized to the decay $H_{2,3} \to Z\gamma$ though the interference effects might be tinny there. The angular distributions of the leptons from $Z$ decay could, in principle, be used to suppress the SM background and provide more information of the couplings of the heavy scalars. In addition, we could do analogous studies in the framework of supersymmetric models with also two scalar doublets. The heavy super-particles might be important for the loop-level $H_{i}\gamma\gamma$ couplings, and leave the footprint in the diphoton signal. All these open questions will be pursued in future follow-up papers.


\section*{Acknowledgements}

We would like to thank Yanwen Liu, Zheng-Tian Lu for very useful discussions and communication.  {The work of LGB
is partially supported by the National Natural Science Foundation of China (under Grant No. 11605016), Korea Research Fellowship Program
through the National Research Foundation of Korea(NRF) funded by the Ministry of Science and ICT (2017H1D3A1A01014046), and Basic Science Research Program through the National Research Foundation of Korea (NRF) funded by the Ministry of Education, Science and Technology (NRF-2016R1A2B4008759).}
NC is partially supported by the National Natural Science Foundation of China (under Grant No. 11575176). NC would like to thank the Center for High Energy Physics at Peking University for their hospitalities where part of this work was prepared. YCZ would like to thank the IISN and Belgian Science Policy (IAP VII/37) for support.


\begin{appendix}

\section{Differential cross section for $gg \to H_{2,3} \to t\bar{t}$}
\label{sec:ttbar}

In the CPV 2HDM, the parton-level cross sections for the resonance and interference terms are respectively:
\begin{widetext}
\begin{eqnarray}
\label{eqn:ttbar1}
&& \frac{{\rm d} }{{\rm d}z} \hat{\sigma}^{\rm res} (gg \to H_{2,3} \to t\bar{t}) =
k_F \, \frac{3 G_F^2 \alpha_s^2 (\sqrt{\hat{s}}) \hat{s}^2 m_t^2 \beta_t}{2^{13}\pi^3} \nonumber \\
&& \quad \times \sum_{ij} \left(
\left| \sum_q c_{q,i} A_{1/2}^{H} (\tau_{q}) \right|^2 +
\left| \sum_q \tilde{c}_{q,i} A_{1/2}^{A} (\tau_{q}) \right|^2 \right) P_{ij}
\left( \beta_t^2 \left| c_{t,j} \right|^2 + \left| \tilde{c}_{t,j} \right|^2 \right) \,, \\
\label{eqn:ttbar2}
&& \frac{{\rm d} }{{\rm d}z} \hat{\sigma}^{\rm int} (gg \to H_{2,3} \to t\bar{t}) =
- k_F \, \frac{G_F \alpha_s^2(\sqrt{\hat{s}}) m_t^2}{2^{13/2}\pi}
\frac{\beta_t}{1-\beta_t^2 z^2} \nonumber \\
&& \quad \times \sum_{ij} \left[
\left( \sum_q c_{q,i} A_{1/2}^{H} (\tau_{q}) P_{ij} \right)^{\rm Re} \beta_t c_{t,j} +
\left( \sum_q \tilde{c}_{q,i} A_{1/2}^{A} (\tau_{q}) P_{ij} \right)^{\rm Re} \tilde{c}_{t,j} \right] \,,
\end{eqnarray}
\end{widetext}
where $z = \cos\theta$ is the scattering angle, $\beta_t = \sqrt{1- 4m_t^2/\hat{s}}$, $k_F = 2$ the $k-$factor for the high-order corrections, the propagator elements $P_{ij}$ given in Eq.~(\ref{eqn:propagator}), and ``Re'' takes only the real parts. The minus sign in Eq.~(\ref{eqn:ttbar2}) is from the fermion loops in the signal amplitude for the $H_i gg$ couplings. In numerical calculations, the strong coupling $\alpha_s$ is evaluated at the heavy scalar mass $M_H$. In both the resonance and interference terms, the contribution from the $H_{2,3}$ mediated processes (the diagonal terms $P_{ii}$) and the mixed ones (the off-diagonal terms of $P_{ij}$ with $i\neq j$) are all summed up. It is straightforward to obtain the differential cross sections with respect to the invariant top pair mass $M_{t\bar{t}}$, by integrating over the scattering angle $z$ and convoluting with the gluon distribution luminosity ${\cal L}_{gg}$ in proton:
\begin{eqnarray}
\frac{{\rm d} \sigma^{\rm res}}{{\rm d} M_{t\bar{t}}} &=&
\frac{2}{M_{t\bar{t}}} \hat{\sigma}^{\rm res} (\hat{s} = M_{t\bar{t}}^2) {\cal L}_{gg} \,, \\
\frac{{\rm d} \sigma^{\rm int}}{{\rm d} M_{t\bar{t}}} &=&
\frac{2}{M_{t\bar{t}}} \hat{\sigma}^{\rm int} (\hat{s} = M_{t\bar{t}}^2) {\cal L}_{gg} \,.
\end{eqnarray}

\section{Oblique parameters in the CPV 2HDM}
\label{sec:oblique}

In the limit of $\beta - \alpha = \pi/2$, the oblique parameters in 2HDM are~\cite{Chen:2015gaa}
\begin{widetext}
\begin{eqnarray}
\Delta S&=&\frac{1}{24\pi} \Big\{  c_{2W}^2\, G( M_\pm^2 \,, M_\pm^2\,, m_Z^2) + s_{\alpha_b}^2 \Big[  c_{\alpha_c}^2 G(M_1^2\,, M_2^2\,, m_Z^2) + s_{\alpha_c}^2 G(M_1^2\,, M_3^2\,, m_Z^2)\nonumber\\
& +& s_{\alpha_c}^2 \hat G(M_2^2\,, m_Z^2) + c_{\alpha_c}^2 \hat G(M_3\,, m_Z^2)   \Big]  + c_{\alpha_b}^2 \Big[  \hat G(M_1^2\,, m_Z^2) + G(M_2^2\,, M_3^2\,, m_Z^2)  \Big]\nonumber \\
&+&\log \Big(  \frac{M_1^2 M_2^2 M_3^2}{ M_\pm^6 } \Big) - \Big[  \hat G(M_{1}^2\,, m_Z^2)+ \log \Big( \frac{M_{1}^2}{M_\pm^2} \Big) \Big] \Big\} \,,\\
\label{eqn:oblique2}
\alpha \Delta T&=& \frac{1}{16\pi^2\, v^2} \Big\{  s_{\alpha_b}^2 F( M_\pm^2\,, M_1^2)+ ( 1 - s_{\alpha_b}^2 s_{\alpha_c}^2 ) F( M_\pm^2\,, M_2^2)+ ( 1 - s_{\alpha_b}^2 c_{\alpha_c}^2 ) F( M_\pm^2\,, M_3^2)\nonumber \\
&-& c_{\alpha_c}^2 s_{\alpha_b}^2 F(M_1^2\,, M_2^2) - s_{\alpha_c}^2 s_{\alpha_b}^2 F(M_1^2\,, M_3^2) - c_{\alpha_b}^2 F(M_2^2\,,M_3^2) \nonumber \\
&+& 3 c_{\alpha_b}^2 \Big[  F(m_Z^2\,, M_1^2) - F(m_W^2\,, M_1^2) \Big] + 3 s_{\alpha_b}^2 s_{\alpha_c}^2 \Big[ F(m_Z^2\,, M_2^2) - F(m_W^2\,, M_2^2)  \Big]\nonumber \\
&+& 3 s_{\alpha_b}^2 c_{\alpha_c}^2 \Big[ F(m_Z^2\,, M_3^2) - F(m_W^2\,, M_3^2)  \Big]\nonumber \\
&-& 3 \Big[ F(m_Z^2\,, M_{1}) - F(m_W^2\,, M_{1}) \Big] \Big\}\,,
\end{eqnarray}
\end{widetext}
with $M_{1}= m_h$ the SM Higgs mass, and the auxiliary functions are defined as
\begin{eqnarray}
F(x\,,y)&=& \left\{ \begin{array}{ll}
\frac{x+y}{2}- \frac{xy}{x-y} \log ( \frac{x}{y} ), & x\neq y \vspace{0.2cm} \\
0, & x=y \end{array}  \right. \\
G(x\,,y\,,z)&=& -\frac{16}{3} + \frac{5(x+y)}{z} - \frac{2(x-y)^2}{z^2} \nonumber \\
&&+ \frac{3}{z} \Big[ \frac{x^2+y^2}{x-y} - \frac{x^2 - y^2}{z} + \frac{ (x-y)^3 }{3z^2}   \Big]\log \frac{x}{y}\nonumber \\
&&+ \frac{z^2 - 2z(x+y) + (x-y)^2 }{z^3} \nonumber \\
&&\times f\Big( x+y-z\,, z^2 - 2z(x+y) + (x-y)^2 \Big)  \,,\nonumber \\ && \\
\tilde G(x\,,y\,,z)&=&
-2 + \Big( \frac{x-y}{z} - \frac{x+y}{x-y} \Big) \log \frac{x}{y} \nonumber \\
&& + \frac{1}{z} f(x+y-z\,, z^2 - 2z(x+y) + (x-y)^2 ) \,, \nonumber \\ && \\
\hat G(x\,,y)&=& G(x\,,y\,,y) + 12 \tilde G(x\,,y\,,y)\,,
\end{eqnarray}
with
\begin{eqnarray}
f(x\,,y)&=&\left\{ \begin{array}{ll}
\sqrt{y}\log \Big|  \frac{x- \sqrt{y}}{x + \sqrt{y} } \Big|, & y>0 \\
0, & y=0 \\
2\sqrt{-y} \tan^{-1} \frac{\sqrt{-y} }{x}, &  y<0 \end{array} \right.\,.
\end{eqnarray}

\section{EDMs in the CPV 2HDM}
\label{sec:EDMapp}

\subsection{All the separate contributions }


For light fermions, the dominant contributions to their EDMs and CEDMs come from the two-loop Barr-Zee type diagrams~\cite{Barr:1990vd}. In particular, the Wilson coefficient $\delta_e$ receive contributions from the following terms:
\begin{eqnarray}
\label{eq:eEDMOp_total}
\delta_e &=&
(\delta_e )_t^{H\gamma\gamma} +
(\delta_e )_t^{H Z \gamma} +
(\delta_e )_W^{H \gamma\gamma} +
(\delta_e )_W^{H Z \gamma} \nonumber \\
&& +(\delta_e )_{H^\pm}^{H \gamma\gamma} +
(\delta_e )_{H^\pm}^{H Z \gamma} +
(\delta_e)_H^{H^\pm W^\mp \gamma}\,,
\end{eqnarray}
where the diagrams with effective $H_i \gamma\gamma$ and $H_i Z\gamma$ couplings from integrating out a top quark loop are respectively
\begin{eqnarray}
\label{eq:haaOp_top}
\left(\delta_f \right)^{H\gamma\gamma}_{t} &=& -  \frac{N_c Q_f Q_{t}^2 e^2}{64\pi^4} \sum_{i=1}^3 \left[
f(z^i_{t}) \, c_{t,i} \tilde c_{f,i} +
g(z^i_{t}) \, \tilde c_{t,i} c_{f,i} \right] \,, \nonumber \\ && \\
\label{eq:haZOp_top}
\left(\delta_f \right)^{H Z\gamma}_{t} &=& -  \frac{N_c Q_f g_{Z\bar f f}^V g_{Z\bar tt}^V}{64\pi^4} \sum_{i=1}^3 \left[
\tilde f\left(z^i_{t}, \frac{m_{t}^2}{M_Z^2}\right) c_{t,i} \tilde c_{f,i} \right. \nonumber \\
&& + \left. \tilde g\left(z^i_{t}, \frac{m_{t}^2}{M_Z^2}\right) \tilde c_{t,i} c_{f,i} \right] \,,
\end{eqnarray}
where $z_X^i \equiv m_X^2 / M_i^2$, $g_{Z f\bar f}^V$ is the vector-current couplings of $Z$ boson to the fermions.
The loop integral functions are respectively
\begin{eqnarray}
f(z) &\equiv& \frac{z}{2} \int_0^1 dx \frac{1-2x(1-x)}{x(1-x)-z} \log\frac{x(1-x)}{z} \ , \\
g(z) &\equiv& \frac{z}{2} \int_0^1 dx \frac{1}{x(1-x)-z} \log\frac{x(1-x)}{z} \,, \\
\tilde f(x\,,y) &\equiv& \frac{y f(x) - x f(y) }{y - x } \,,\\
\tilde g(x\,,y) &\equiv& \frac{y g(x) - x g(y) }{y-x}\,.
\end{eqnarray}
There are also contributions from the $W$ boson and its Nambu-Goldstone bosons to the $H_i \gamma\gamma$ and $H_i Z\gamma$ operators, which were gauge-invariant and have been obtained in Refs.~\cite{Chang:1990sf,Leigh:1990kf, Abe:2013qla}.
\begin{widetext}
\begin{eqnarray}
\label{eq:haaOp_W}
\left(\delta_f \right)^{H\gamma\gamma}_W &=&
\frac{Q_f e^2}{256\pi^4} \sum_{i=1}^3 \left[
\Big( 6 + \frac{1}{z^i_W} \Big) f(z^i_w) +
\Big( 10 - \frac{1}{z^i_W} \Big) g(z^i_w)
+  \frac{3}{4}
\Big( g(z^i_W) +  h (z^i_W)  \Big)  \right] a_i \tilde c_{f,i}  \ ,  \\
\label{eq:haZOp_W}
\left(\delta_f \right)^{HZ\gamma}_W &=&  \frac{g_{Z\bar f f}^V g_{ZWW}}{256\pi^4} \sum_{i=1}^3 \left[
\left(6 -\sec^2\theta_W + \frac{2-\sec^2\theta_W}{2z^i_w} \right)\tilde f(z^i_W, c_W^2 ) \right.\non
&&\hspace{2cm}+ \left.
\left( 10- 3\sec^2\theta_W - \frac{2-\sec^2\theta_W}{2z^i_w}\right)\tilde g(z^i_W, c_W^2 ) + \frac{3}{2}
\Big( g( z^i_W ) + h(  z^i_W) \Big)  \right] a_i \tilde c_{f_i} \ ,
\end{eqnarray}
\end{widetext}
where the gauge coupling $g_{WWZ} = e/\tan\theta_W$, and the loop function
\begin{eqnarray}
h(z) &\equiv& \frac{z}{2} \int_0^1 dx \frac{1}{z-x(1-x)} \nonumber \\
&& \times \left( 1+ \frac{z}{z-x(1-x)} \log\frac{x(1-x)}{z} \right) \,.
\end{eqnarray}
The contributions by integrating out the charged Higgs bosons loops read
\begin{eqnarray}
\left(\delta_f \right)^{H\gamma\gamma}_{H^\pm} &=& \frac{Q_f e^2}{256 \pi^4} \sum_i  \Big[ f(z_\pm^i ) - g( z_\pm^i )  \Big]  \bar \lambda_i \tilde c_{f\,,i} \,,
\label{eq:haaOp_Hpm}\\
\left(\delta_f \right)^{H Z \gamma}_{H^\pm} &=& \frac{g_{Z\bar f f}^V g_{ZH^+ H^- } }{256 \pi^4 } \Big( \frac{v}{M_\pm } \Big)^2 \sum_i  \left[
\tilde f\left(z_\pm^i\,, \frac{M_\pm^2}{m_Z^2}\right) \right. \nonumber \\
&& \left. - \tilde g\left(z_\pm^i\,, \frac{M_\pm^2}{m_Z^2}\right) \right] \bar\lambda_{i} \tilde c_{f\,,i}\,,
\label{eq:haZOp_Hpm}
\end{eqnarray}
with $z_\pm^i = M_\pm^2/ M_i^2$, $g_{ZH^+ H^- }= e(1 - \tan\theta_W^2)/(2 \tan\theta_W)$, and $\bar\lambda_i = -\lambda_{i+-}/v$ the effective trilinear scalar couplings given in Eqs.~(\ref{eqn:lambda+-1}) to (\ref{eqn:lambda+-3}). Additional contributions are from the $H^\pm W^\mp \gamma$ operators~\cite{Abe:2013qla}, which read
\begin{eqnarray}
\label{eq:HpmWaOp_H}
(\delta_f)_H^{H^\pm W^\mp \gamma}&=& \frac{s_f}{512\pi^4} \sum_i \left[
\frac{e^2}{ 2 s_W^2} \mI_4(M_i^2\,, M_\pm^2) \, a_i \tilde c_{f\,,i} \right. \nonumber \\
&& \left. - \mI_5 (M_i^2\,, M_\pm^2) \, \bar \lambda_i  \tilde c_{f\,,i} \right]\,,
\end{eqnarray}
where $s_f=+1$ for the down-type quarks and charged leptons, and $-1$ for the up-type quarks, and the two-loop integral functions are defined as
\begin{eqnarray}
\mI_{4\,,5} (M_1^2\,,M_2^2) &\equiv&
\frac{m_W^2}{ M_\pm^2 - m_W^2 } [ I_{4\,,5}(m_W\,, M_1) \nonumber \\
&& - I_{4\,,5} (M_2\,,M_1)  ]\,,\\
I_4(M_1\,, M_2)&\equiv& \int_0^1 dz\, (1-z)^2 \left( z-4 + z \frac{ M_\pm^2 - M_2^2}{m_W^2}  \right)\non
&&\times \frac{M_1^2}{ m_W^2 (1-z) + M_2^2 z - M_1^2 z (1-z)} \nonumber \\
&& \times \log \left( \frac{ m_W^2 (1-z) + M_2^2 z }{M_1^2 z (1-z) } \right) \,,\\
I_5(M_1\,, M_2) &\equiv&
\int_0^1 dz\, \frac{M_1^2 z(1-z)^2 }{ m_W^2 (1-z) + M_2^2 z - M_1^2 z(1-z)}\non
&\times& \log \left( \frac{ m_W^2 (1-z) + M_2^2 z }{ M_1^2 z (1-z) }\right) \,.
\end{eqnarray}



For the CEDM, the top quark loop is integrated out to obtain the effective $h_iGG$ or $h_iG\tilde G$ operators, which leads to the following CEDM operators~\cite{Gunion:1990iv},
\begin{eqnarray}
\label{eq:CEDM}
\tilde{\delta}_q(\Lambda) &\equiv& \left(\tilde{\delta}_q \right)^{hgg}_{t} = - \frac{g_s^2}{128\pi^4}
\sum_{i=1}^3 \left[ f(z^i_{t}) \, c_{t,i} \tilde c_{q,i} \right. \nonumber \\
&& \left. +g(z^i_{t}) \, \tilde c_{t,i} c_{q,i} \right] \,.
\end{eqnarray}


The contribution to the dimension-6 Weinberg operator arises predominantly from the top loop~\cite{Weinberg:1989dx}, which gives
\begin{eqnarray}
\label{eq:Weinberg}
C_{\tilde{G}} (\Lambda) \equiv (C_{\tilde{G}})_t 
= - \frac{g_s^2}{3} \frac{1}{128 \pi^4} \sum_{i=1}^3 h_0(z_t^i) \, c_{t,i} \tilde c_{t,i} \,,
\end{eqnarray}
with the two-loop integral function
\begin{eqnarray}
h_0(z) \equiv \frac{z^4}{2} \int_0^1 \, dx \int_0^1 \, dy \, \frac{  x^3 y^3 (1-x) }{ \Big[ z^2 x(1-xy) + (1-x)(1-y)  \Big]^2} \,. \nonumber \\
\end{eqnarray}

\subsection{RG running and mixing effects}

During the RG running from the new physics scale down to the hadronic scale, the
nontrivial corrections
to the Wilson coefficients of the CEDM and Weinberg operators induced by
 flavor-conserving CP-odd four-fermion operators need to be take into account. The complete Lagrangian for the calculation of mercury EDMs should be
\begin{eqnarray}
{\cal L}_\text{CPV} &=& {\cal L}_{\rm (C)EDM}+
 \sum_q \frac{C_4^q}{\Lambda^2} {\mathcal O}_4^q
+\sum_{q'\ne q} \frac{\widetilde{C}_1^{q'q} }{\Lambda^2}\widetilde{\mathcal O}_1^{q'q} \nonumber \\
&& + \frac{1}{2}\sum_{q'\ne q}\frac{\widetilde{C}_4^{q'q}}{\Lambda^2}\widetilde{\mathcal O}_4^{q'q} \ .
\label{effop}
\end{eqnarray}
Here, the first two CP-odd four-fermion operators
\begin{eqnarray}
{\mathcal O}_4^q
&=&\, (\overline{q} q) (\overline{q} \,i\gamma_5 q) \,,
 \\
\widetilde{\mathcal O}_1^{q'q} &=&\, (\overline{q'} q') (\overline{q} \,i\gamma_5 q) \,,%
\label{4foperator}
\end{eqnarray}
can be generated through the CPV Yukawa threshold corrections and the CPV neutral Higgs-boson mixing in the $t$-channel. The corresponding CP-odd coefficients are given respectively as
\begin{eqnarray}
  \label{eq:cff}
C_4^q\ &=&\ g_q\, g_{q}\,
\frac{c_{q}\,\tilde c_{q}}{M_{H}^2} \,, \\
\widetilde{C}_1^{q'q}\ &=&\ g_{q'}\, g_q\,
\frac{c_{q'}\,\tilde c_{q}}{M_{H}^2}\; ,
\end{eqnarray}
with $g_{q(q')}=m_{q(q')}/v$. On the other hand, the last CP-odd four-fermion operator
\begin{eqnarray}
\widetilde{\mathcal O}_4^{q'q}
&=& (\overline{q'_\alpha} \sigma^{\mu\nu} q'_\beta)
(\overline{q_\beta} \,i\sigma_{\mu\nu}\gamma_5 q_\alpha) \,,
\label{add4foperator}
\end{eqnarray}
 is generated from the operator mixing effects of $\widetilde{C}_1^{q'q}$ and $\widetilde{C}_1^{qq'}$ which follows the Eq.~(\ref{Eq:Gamma}) below. To obtain the value of the Wilson coefficients $\left(\delta_q, \tilde{\delta}_q, - \frac{3 C_{\tilde{G}}}{2} \right)$ at a GeV scale,
we need to take an evolution for
\begin{eqnarray}
{\bf C} = \left( \delta_q,\, \tilde{\delta}_q,\,
- \frac{3 C_{\tilde{G}}}{2},\,
C_4^q,\, \widetilde{C}_1^{q'q},\,
\widetilde{C}_1^{qq'},\, \widetilde{C}_4^{q'q} \right)
\end{eqnarray}
from the 2HDM scale $v$ down to the GeV scale, based on the Renormalization Group Equations (RGE)~\cite{Degrassi:2005zd,Hisano:2012cc,Dekens:2013zca} :
\begin{eqnarray}
\label{RGE}
\frac{d}{d \ln \mu}\bold{C} =
\bold{C}\cdot \bold{\Gamma} \,.
\end{eqnarray}
Here, the one-loop anomalous dimension matrix is given by
\begin{align}
{\bf \Gamma} = \begin{bmatrix}
\frac{\alpha_s}{4\pi} \gamma_s & {\bf 0}                        & {\bf 0} \\
\frac1{(4\pi)^2} \gamma_{sf}   & \frac{\alpha_s}{4\pi} \gamma_f & {\bf 0} \\
\frac1{(4\pi)^2} \gamma'_{sf}  & {\bf 0}                        & \frac{\alpha_s}{4\pi} \gamma'_f
\end{bmatrix}, \label{Eq:Gamma}
\end{align}
with
\begin{align}
\gamma_s &=
\begin{bmatrix}
+ 8 C_F & 0         & 0 \\
+8C_F & +16C_F-4N & 0 \\
0     & +2N       & N+2n_f+\beta_0
\end{bmatrix}, \label{Eq:Gamma_s} \\
\gamma_f &=
\begin{bmatrix}
-12C_F+6     \end{bmatrix}, \label{Eq:Gamma_f} \\
\gamma'_f &=
\begin{bmatrix}
-12C_F                      & 0       & -1      \\
0                           & -12C_F   &-1\\
-12 &-12&-8C_F-\frac{6}{N}        \end{bmatrix}, \label{Eq:Gamma_f'} \\
\gamma'_{sf} &=
\begin{bmatrix}
0&0&0\\
0 & 0  &0    \\
-8\frac{m_q'}{m_q}\frac{Q_q'}{Q_q}&-8\frac{m_q'}{m_q}&0
\end{bmatrix}, \label{Eq:Gamma'_sf} \\
\gamma_{sf} &=
\begin{bmatrix}
+2 & +2 &0
\end{bmatrix}. \label{Eq:Gamma_sf}
\end{align}
Where $q$ runs over $u, d, b$, $N=3$, $C_F = (N^2-1)/(2N)=4/3$, $\beta_0 = (11N-2n_f)/3$, and $n_f$ is the flavor number.

For the RGE running from the 2HDM scale down to the scale of $m_b$, we assume a five-flavor scheme.
Keeping only the leading logarithmic terms that make additional contributions to  the CEDMs of bottom and
light quarks at the matching scale $\mu=m_b$, we have
\begin{eqnarray}
\Delta \tilde{\delta}_b (m_b) &\approx& \frac{1}{8\pi^2}  C_4^b
\log\left(\frac{M_H}{m_b} \right) \ ,\label{shiftC}
\end{eqnarray}
where $\Delta\tilde{\delta}_b (m_b)$ could be figured out from Eq.~(\ref{Eq:Gamma}) and is from integrating out the bottom quark at one-loop level. After the bottom quark being integrated out, its CEDM makes a shift to the Weinberg operator~\cite{Boyd:1990bx, Hisano:2012cc},
\begin{eqnarray}
\Delta C_{\tilde{G}} (m_b) = \frac{\alpha_S(m_b)}{12\pi} \, \tilde{\delta}_b(m_b) \ .
\end{eqnarray}
Here the the two-loop Barr-Zee graphs generated CEDMs $\tilde{\delta}^0_b(m_b)$ has been modified to be $\tilde{\delta}_b(m_b)   =  \tilde{\delta}^0_b(m_b)  +  \Delta \tilde{\delta}_b(m_b)$ to obtain the whole $b$-quark CEDM at the $m_b$ scale.
  The shift to CEDMs of quarks are given by,
\begin{eqnarray}
\Delta \tilde{\delta}_q(m_b) &\approx&  \frac{g_s^2}{64\pi^4} \frac{m_b}{m_q} ( \tilde{C}_1^{bq}+\tilde{ C}_1^{qb}) \left(\log\frac{M_H}{m_b}\right)^2  \,.
\end{eqnarray}
We would like to mention that $\Delta\tilde{\delta}_q(m_b)$ is nontrivial, which is induced by $\widetilde{C}_4^{qq'}$ through integrating out the bottom quark at two-loop level. Below the $m_b$ scale,  we assume a four-flavor scheme for the RGE running of the Wilson coefficients $\delta_q$, $\tilde{\delta}_q$ and $C_{\tilde{G}}$ between $m_b$ and $m_c$, and a three-flavor scheme below $m_c$.

\end{appendix}

\end{document}